\documentclass[letterpaper]{elsarticle}

\usepackage{lineno,hyperref}
\modulolinenumbers[5]

\usepackage[margin=2.5cm]{geometry}

\usepackage{placeins}

\usepackage{bm}  

\journal{Astroparticle Physics}

\begin{document}

\begin{frontmatter}
  \title{The Cosmic-Ray Energy Spectrum between 2 PeV and 2 EeV Observed 
    with the TALE detector in monocular mode}

  \author{\par\noindent
R.U.~Abbasi$^{1}$,
M.~Abe$^{2}$,
T.~Abu-Zayyad$^{1}$,
M.~Allen$^{1}$,
R.~Azuma$^{3}$,
E.~Barcikowski$^{1}$,
J.W.~Belz$^{1}$,
D.R.~Bergman$^{1}$,
S.A.~Blake$^{1}$,
R.~Cady$^{1}$
% ...
\par\noindent
{\footnotesize\it
$^{1}$ High Energy Astrophysics Institute and Department of Physics and Astronomy, University of Utah, Salt Lake City, Utah, USA \\
$^{2}$ The Graduate School of Science and Engineering, Saitama University, Saitama, Saitama, Japan \\
$^{3}$ Graduate School of Science and Engineering, Tokyo Institute of Technology, Meguro, Tokyo, Japan \\
% ...
\addtocounter{footnote}{-1}\let\thefootnote\svthefootnote
}
\par\noindent
}
  \corref{correspondingauthor}
  \cortext[correspondingauthor]{Corresponding author\\
    E-mail: tareq@cosmic.utah.edu}

  \begin{abstract}
    We report on a measurement of the cosmic ray energy spectrum by the 
    {\bf T}elescope {\bf A}rray {\bf L}ow-{\bf E}nergy Extension (TALE) air
    fluorescence detector. The TALE air fluorescence detector is also sensitive
    to the Cherenkov light produced by shower particles. Low energy cosmic
    rays, in the PeV energy range, are detectable by TALE as ``Cherenkov
    Events''. Using these events, we measure the energy spectrum from a low
    energy of $\sim 2$ PeV to an energy greater than 100 PeV.  Above 100 PeV 
    TALE can detect cosmic rays using air fluorescence.  This allows for the
    extension of the measurement to energies greater than a few EeV.  In
    this paper, we will describe the detector, explain the technique, and
    present results from a measurement of the spectrum using $\sim 1000$
    hours of observation.  The observed spectrum shows a clear steepening 
    near $10^{17.1}$~eV, along with an ankle-like structure at $10^{16.2}$~eV.
    These features present important constraints on galactic cosmic rays origin 
    and propagation models.  The feature at $10^{17.1}$~eV may also mark the end
    of the galactic cosmic rays flux and the start of the transition to 
    extra-galactic sources.
  \end{abstract}
\end{frontmatter}


\section{Introduction}
\label{sec:introduction}

The TALE detector was designed to look for structure in the
energy spectrum and associated change in composition of cosmic rays
below the “ankle” structure at $10^{18.6}$ eV.  The ankle structure has
been observed now by a number of experiments, though the
interpretation of the structure is still a matter of open debate.  A
possible explanation, one that would be consistent with a changing
composition, is that the ankle represents the transition from galactic
flux (below the ankle) to that of cosmic rays of extragalactic origin
\cite{Hill:1983mk}.  An alternative interpretation for the ankle is that 
it is a "dip" in the cosmic rays flux caused by the pair-production 
interaction of extragalactic protons with the cosmic microwave 
background (CMB) photons~\cite{Berezinsky:1988wi,Berezinsky:2005qe}.
  The latter scenario would be consistent with the predominately protonic,
composition reported by HiRes~\cite{Abbasi:2004nz} and
Telescope Array (TA)~\cite{Abbasi:2014sfa}, and one would then
expect a galactic-extragalactic transition to occur at a lower energy.
Yet another interpretation of the ankle as the maximum energy at the source
of extragalactic protons ($E_{max} \sim 5Z\times 10^{18}$~eV) 
has been proposed in light of Auger data~\cite{Aloisio:2013hya}.  This 
scenario also requires that the galactic-extragalactic transition to occur 
at an energy below that of the ankle.

\begin{figure}[htb]
\centering
\includegraphics[height=2.4in]{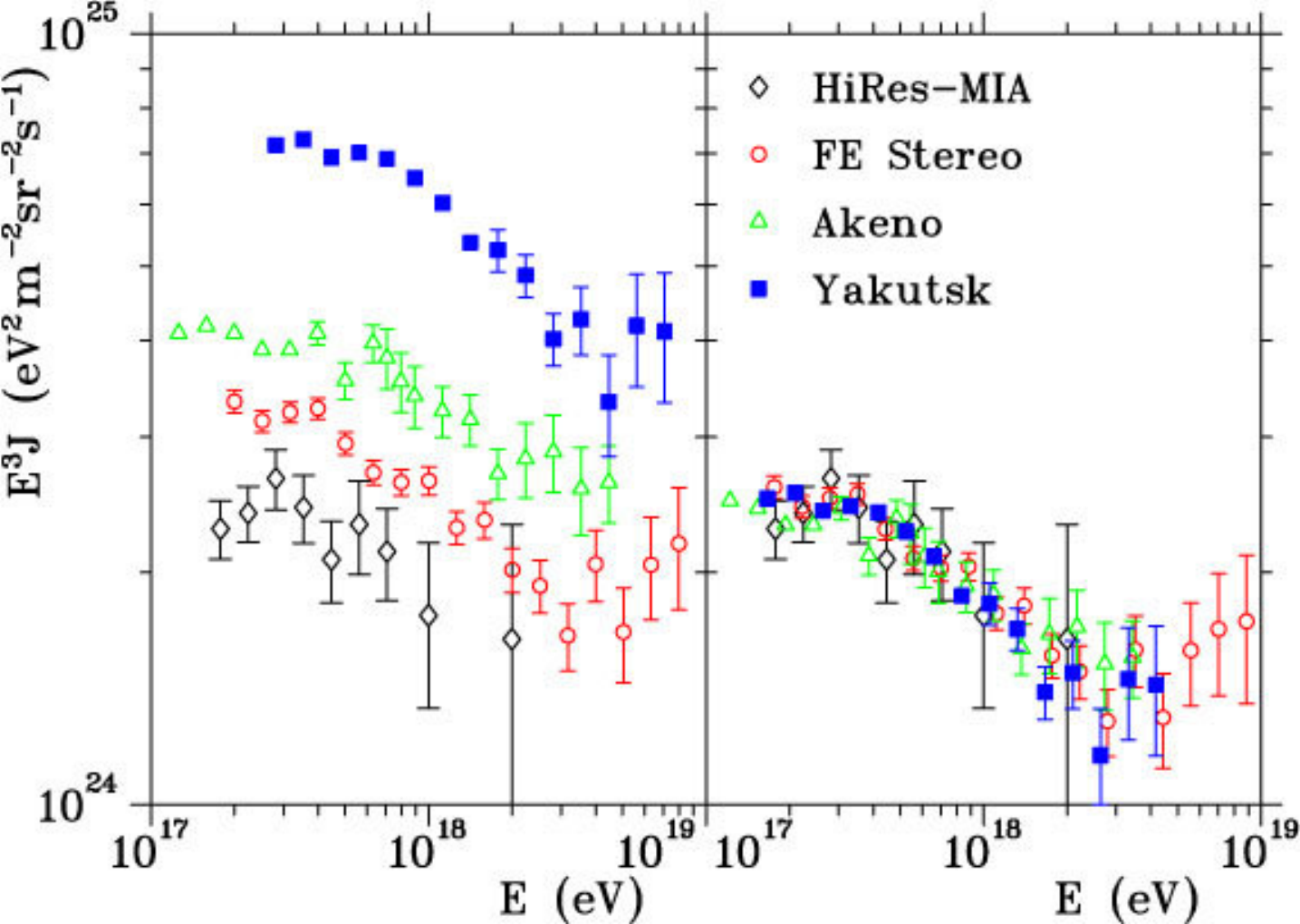}
\caption{Left: Cosmic ray spectra measured by the Fly’s
  Eye~\cite{Bird:1994wp}, HiRes/MIA~\cite{AbuZayyad:2000ay},
  Akeno~\cite{Nagano:1991jz}, and Yakutsk~\cite{Egorova:2004cm} experiments. 
  Each shows a flat part, a break, and a falling part (on an $E^3J$ plot,
  where $J$ denotes the differential flux).  Right: Aligning the flat parts of
  the spectra from the four experiments, the spectral feature known as
  the second knee appears at the same energy, about $10^{17.5}$ eV.  Below
  this energy the spectrum is flat (on an $E^3J$ plot).}
\label{fig:2nd_knee}
\end{figure}

Indications of a “knee”-like structure in the $10^{17}$ decade have been 
seen in the spectra reported by several previous experiments ``e.g.''  
\cite{Bird:1994wp,AbuZayyad:2000ay,Nagano:1991jz,Egorova:2004cm}.  As
shown in Figure~\ref{fig:2nd_knee}, the energy scales of these detectors
differed by about a factor of two, so the energy at which this
spectral break occurs is quite uncertain.  However, each data set
shows a definite softening (break) in the power law spectral index,
that we call the {\it second knee}.  When the energy scales are adjusted in
the right half of Figure~\ref{fig:2nd_knee}, the four spectra can be
made to align simultaneously in normalization (flux) and in the
location of the break.  Thus, while we know the break in the spectrum
exists, its energy is uncertain, except that it occurs somewhere in
the $10^{17}$ eV decade.  No real progress on understanding this
important feature can occur until one single experiment measures all
three spectral features of the ultra-high energy (UHE) cosmic ray 
regime: the high energy suppression (GZK cut-off), the ankle, and 
the second-knee.

Using both Fluorescence and Cherenkov light, the TALE fluorescence 
detector (FD) is able to push the minimum energy of the spectrum 
measurement to well below $10^{16}$~eV.  This gives plenty of lever-arm 
for fitting the spectrum around the second knee feature to power laws, 
and also allow us to probe for additional spectral features down to the 
``first knee'' near $10^{15.5}$~eV.  At the same time, TALE measurements 
overlap that of the original TA detectors, and with a common energy 
scale.  Hence TA/TALE will observe all three spectral features indicated 
above within the framework of a single experiment.

Further composition measurements below $10^{18.5}$~eV, from hybrid
analysis (determining shower geometries using both TALE FD and TALE
SD), will answer the question of the origin of the ankle and the
second knee.  With reliable and redundant measurements of the
composition, one can also selectively analyze the heavy or light parts
of the event sample to see where the second knee presents itself.  Its
presence in the heavy part of the data would indicate that the second
knee is a galactic effect.  Conversely, a second knee with light
composition would suggest an extra-galactic origin for the effect.

This paper represents the first step in the TALE program, where we
measure and present the cosmic ray spectrum from $10^{15.3}$~eV to $10^{18.3}$~eV.
Only the high elevation ($31^{\circ}$-$55^{\circ}$) telescopes of 
TALE FD, see section~\ref{sec:tale_detector},  are used, as the majority 
of interesting events in this range, especially toward the region of the 
second knee and below, register only in the upper elevation detector.

In the remainder of the paper, we will describe
the detector and data collection in section~\ref{sec:tale_detector},
and explain the event selection procedure and event reconstruction
method in section~\ref{sec:data_process}. Section~\ref{sec:data_process}
concludes with an overview of the spectrum measurement procedure.  
In section~\ref{sec:MC} we will describe the Monte Carlo (MC) simulation, 
and show its result for the detector aperture calculation. We will also 
present the results of MC studies of the reconstruction performance, in 
order to understand the detector resolution and validate the reconstruction 
technique. Finally, key distributions related to the aperture 
of the detector from the simulation will be compared to those from actual 
data in order to verify the accuracy of the detector simulation and the 
aperture calculation.  Following a discussion of the systematic uncertainties 
in section~\ref{sec:systematics}, the measured spectrum will be shown in 
section~\ref{sec:results_and_discussion}, along with a brief discussion
of the measured results.  The paper will conclude with a summary 
in section~\ref{sec:summary}.

\FloatBarrier

\section{TALE Detector and Operation}
\label{sec:tale_detector}

\begin{figure}[htb]
\centering
\includegraphics[height=2.4in]{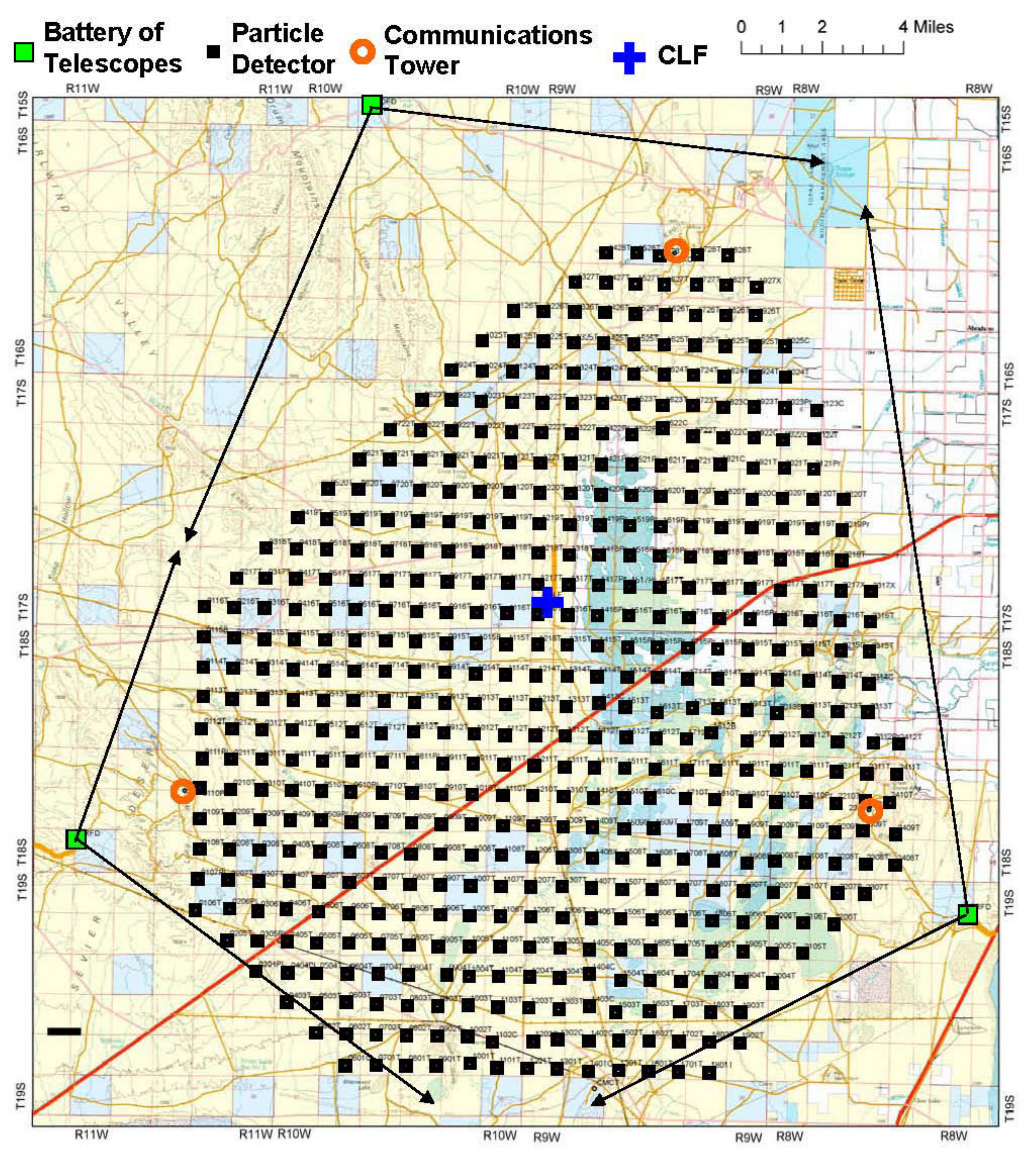}
\caption{Map of the Telescope Array surface detector and the three
  fluorescence detectors overlooking the array.  The MD site is at the
  green square at the top of the map.}
\label{fig:ta_config}
\end{figure}

TA is an international collaboration with members
from Japan, U.S., South Korea, Russia, and Belgium. The TA experiment
is located in the West Desert of Utah, about 150 miles southwest of
Salt Lake City, and is the largest cosmic ray detector in the northern
hemisphere.  In operation since 2008, TA consists of 507 surface
detectors (SD), arranged in a square grid of 1.20 km 
spacing~\cite{tasd-nim-a}.  A total of 38 telescopes are distributed 
among three FD stations at the periphery of the SD
array~\cite{AbuZayyad:2012qk,tafd-nim-a}. The FD telescopes observe
the airspace above the SD array.  This arrangement of detectors is
shown in Figure~\ref{fig:ta_config}. TA is the direct successor to
both the Akeno Giant Air Shower Array (AGASA) and High resolution
Fly’s Eye (HiRes) experiments \cite{Teshima:1985vs,Sokolsky:2011zz},
incorporating the scintillation counter design of the former for the
SD, and the fluorescence detector technique of the latter.  The goal
of TA is to clarify the origin of ultra-high energy cosmic 
rays~(UHECR) and related extremely high energy phenomena in the 
universe.  The results of measuring the energy spectrum, composition, 
and anisotropy in the arrival direction distribution for energies
above $10^{18.2}$~eV have been 
published~\cite{AbuZayyad:2012ru,Abbasi:2014sfa,Abbasi:2014lda}

\begin{figure}[htb]
\centering
\includegraphics[height=2.4in]{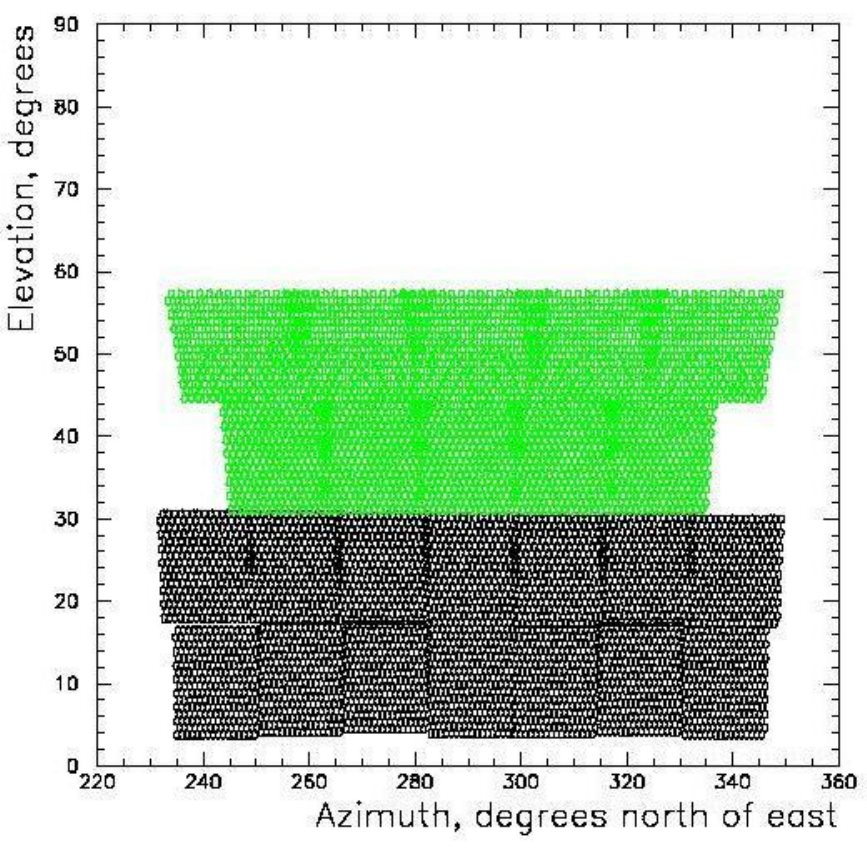}
\caption{Schematic of TALE/MD mirrors showing azimuthal and elevation
  coverage.  The aperture of the TA FD is shown in black, 
  and TALE is shown in green.}
\label{fig:tale_fov}
\end{figure}

At the northern FD station (Middle Drum), a TA low-energy extension
(TALE) detector \cite{tale_icrc2011} has been under construction since
2011.  Ten new telescopes were added to the 14 of the TA FD at the
site.  All 24 were refurbished from components previously used by
HiRes, and updated with new communications hardware.  Whereas the
original 14 TA FD telescopes, distributed in two “rings”, view between
$3^{\circ}$ to $31^{\circ}$ in elevation, the new TALE FD telescopes
cover between $31^{\circ}$ to $59^{\circ}$.  Both the TA and TALE
telescopes view approximately southeast, over the part of the sky
above the SD array.  This arrangement is illustrated in
Figure~\ref{fig:tale_fov}. The new telescopes of the TALE FD were
completed in 2013, and have been taking data since the fall of that
year.

\begin{figure}[htb]
\centering
\includegraphics[height=2.4in]{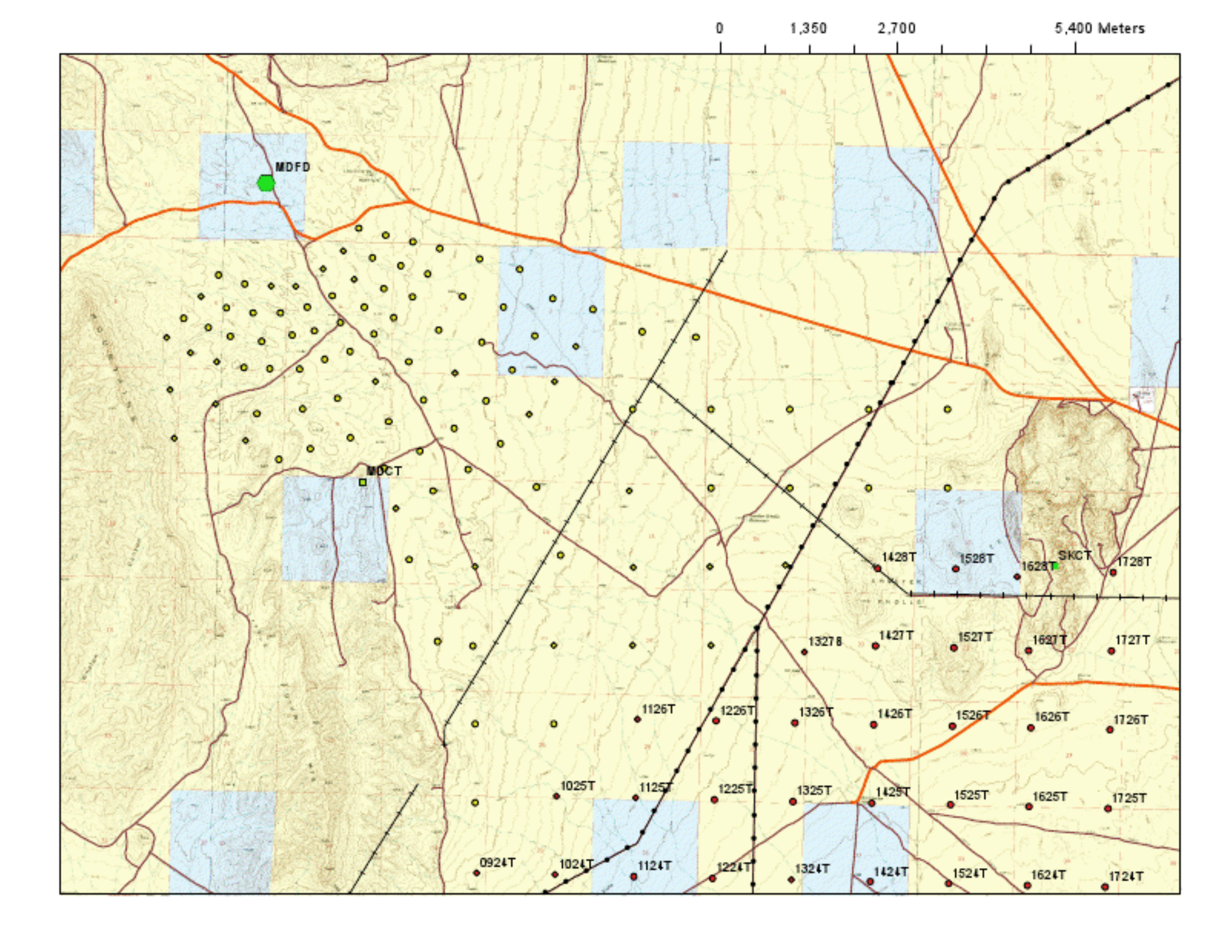}
\caption{Map of the TALE surface detector. The TALE FD site can be 
  seen near the top left of the figure}
\label{fig:tale_sd_config}
\end{figure}

In addition to the ten new, high-elevation angle, FD telescopes, TALE also
incorporates 103 new SD counters arranged in a graded spacing array.
The arrangement of the new counters is shown in
Figure~\ref{fig:tale_sd_config}. As of the summer of 2017, all TALE SD counters
are in place, and 80 are operational.

The TALE FD telescopes were assembled from refurbished HiRes-II
mirrors, cameras, and readout electronics~\cite{Boyer:2002zz}.  The
mirrors are each made from four spherical segments arranged in a
clover-leaf pattern.  The unobscured viewing area of each mirror is
approximately $3.7$~m$^{2}$.  A camera of 256 pixels is placed at the
focal plane of the mirror.  The pixels consist of 2-inch hexagonal
XP3062 photo-multiplier tubes (PMTs) made by Photonis. They are
arranged in a $16\times16$ hexagonal close-packed arrangement.  Each
pixel covers a one degree cone in the sky, and each camera has a
field-of-view (FOV) of $16^{\circ}$ in azimuth by $14^{\circ}$ in
elevation.

The PMT signals are recorded by a 10 MHz FADC readout system with an
8-bit resolution. Analog sums over rows and columns of pixels, also
sampled at 8-bits, allow recovery of saturated PMTs in most cases.
The summed signals are also used as the input to the trigger logic of
the system, which looks for a three-fold coincidence in rows or in
columns.  The communications and timing systems were upgraded for TALE to
include a LINUX-based link module that transfers telescope data to a
central data acquisition (DAQ) computer over standard 100 Mbps
Ethernet~\cite{Zundel_thesis_ch4}.  Inter-mirror triggers, GPS
timing, and clock signals are distributed over a new custom serial
system tied to a central timing module~\cite{hires_gpsy}.  All of
these replaced a token-ring like, proprietary fiber system of hires, and
improved the data throughput from about 10 Mbytes/s to 100 Mbytes/s.

TALE FD construction began in 2011.  The first observation runs with
all 10 telescopes began in the September of that year.  Over the
following year, we continued to tune the gain and trigger of the
system to achieve maximum data sensitivity while maintaining detector
stability. These changes resulted in a number of distinct “epochs” of
data that differ primarily in the separate settings of gain of the
individual PMTs and those of the trigger analog row and column sums.
Also, a set of filters, named ``Moments and Clusters'', originally
designed to remove Cherenkov-dominated “blasts”, were disabled at the
beginning of 2015. In these events the detector sees an elliptical
pattern left by the Cherenkov cone of an event coming toward the
telescopes at small incident angles.  During HiRes operation, such
events were not written out.  The removal of these filters was
motivated in part by an attempt to run TALE FD in coincidence with a
prototype non-imaging Cherenkov detector (NICHE)~\cite{jNICHE}, and 
this change also improved TALE FD acceptance at energies below a 
few PeV.  Table \ref{table_epochs} summarizes the changes.

\begin{table}[htb]
\centering
\caption{TALE operation Epochs.  Note that Moments/Clusters refer to
  high level online trigger routines which keep or reject telescope
  triggers based on tube trigger pattern (main use is to reject 
  Cherenkov blasts.)}.
\label{table_epochs}
\begin{tabular}{|l|l|l|l|l|l|l|}
\hline
Epoch & Start   & End     & PMT gain & Trigger Gain & Moments/Clusters  & Nightly UVLED Calib. \\ \hline
1     & 09/2013 & 01/2014 & 0.63     & 1.0          & Yes               & No                   \\ \hline
2     & 06/2014 & 01/2015 & 1.0      & 0.63         & Yes               & Yes                  \\ \hline
3     & 01/2015 & current & 1.0      & 0.63         & No                & Yes                  \\ \hline
\end{tabular}
\end{table}

A measurement of the cosmic-ray energy spectrum using TALE epoch 1
data has already been reported~\cite{Abuzayyad:2015fvm}.  This 
paper deals with data collected during epochs 2 and 3.

Useful events recorded by TALE appear as tracks for which the observed
signal comprises a combination of direct Cherenkov light (CL) and
fluorescence light (FL), with some possible contribution from
scattered CL.  Contributions of light generated by these mechanisms are
both proportional to the number of charged particles in the extensive
air shower (EAS) at any point along its development.  Thus CL signals
can be analyzed in a manner analogous to that for FL to determine the
energy of the cosmic ray, and to determine the depth of shower maximum
($X_{\rm max}$) which is related to the mass of the primary particle.

There are some important differences between CL and FL measurements.
First, FL is emitted isotropically from the shower particles. In
contrast CL is strongly peaked in the forward direction of the shower
axis, and falls off rapidly as the incident angle of the shower to the 
detector increases.  CL also accumulates along the shower and increases in
overall intensity as the shower develops.  Both types of light also
undergo scattering in the atmosphere, from both molecules and
particulate aerosols.

\begin{figure}[htb]
\centering
\includegraphics[height=2.0in]{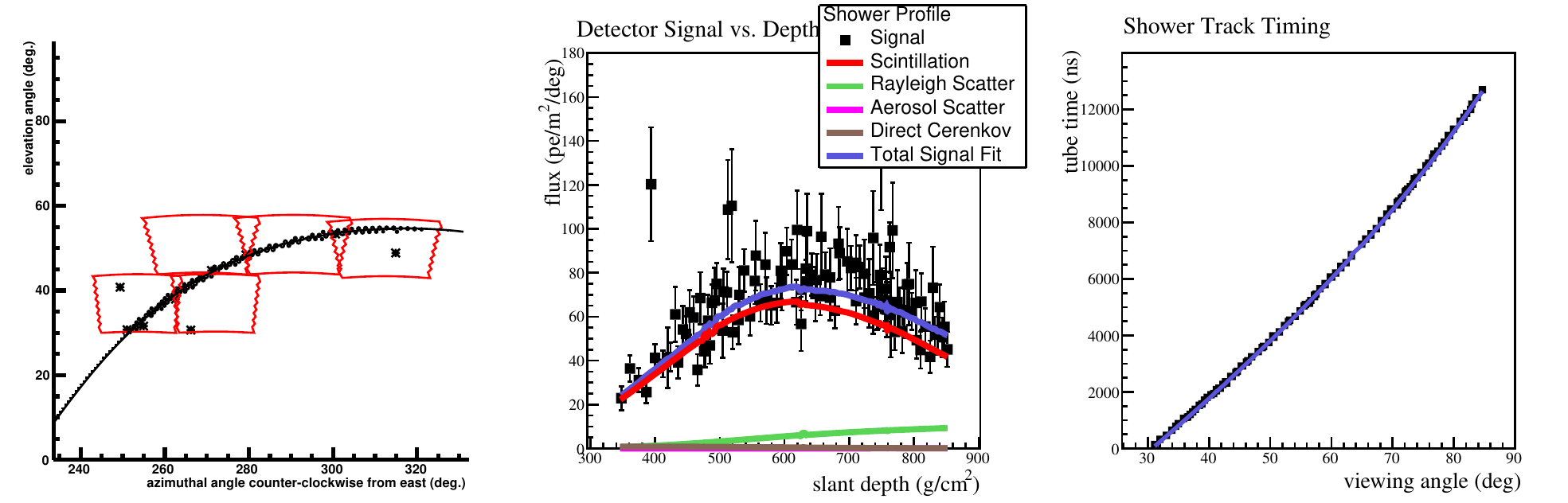}
\caption{A five-telescope fluorescence event.  The display panels show
  the event image (PMT trigger pattern), the reconstructed shower
  profile with relative contributions of FL/CL and scattered CL, and
  the time progression of triggered PMTs.}
\label{fig:qualitative_events_scin}
\end{figure}

Figure~\ref{fig:qualitative_events_scin} shows one extreme case of an
event observed by the TALE FD where the shower axis lies more or less
at right angle to the line of sight to the detector.  The observed
signal fitted to shower simulation shows that most of the signal seen
is scintillation light (FL) observed directly (i.e. not scattered
FL).  This event is somewhat unusual in that it spans $55^ {\circ}$ in
angle, and spread over five telescopes.  It also has a very long time
duration of about $13{\mu}$s.

\begin{figure}[htb]
\centering
\includegraphics[height=2.0in]{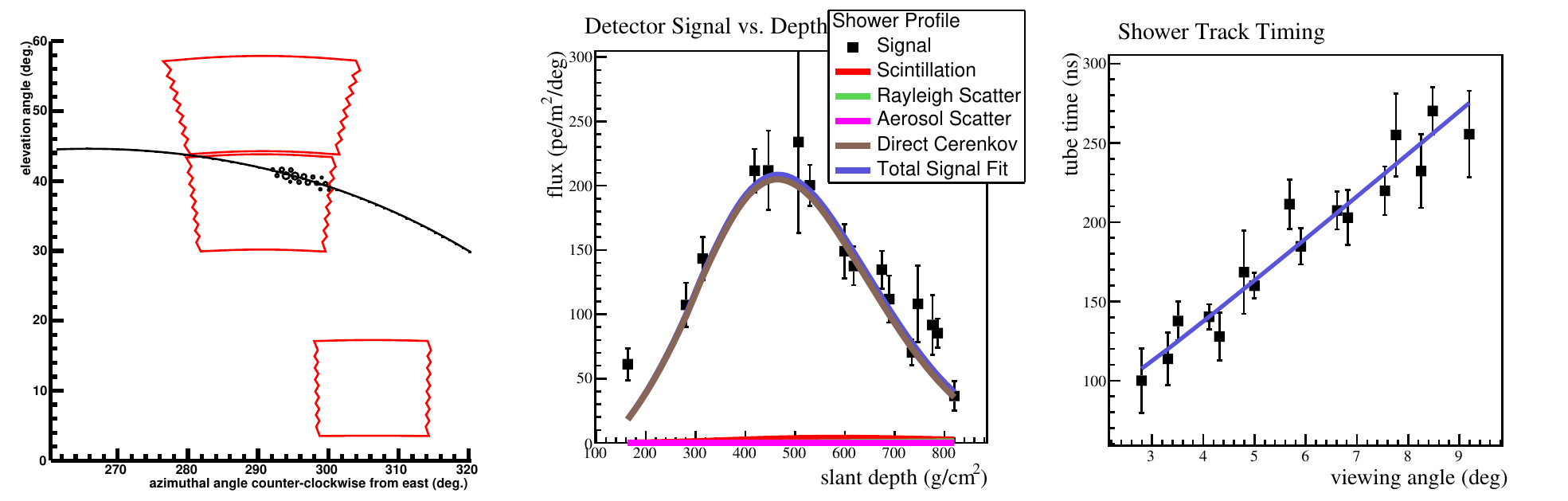}
\caption{A one-telescope Cherenkov event.  The display panels show the
  event image (PMT trigger pattern), the reconstructed shower profile
  with relative contributions of FL/CL and scattered CL, and the time
  progression of triggered PMTs.}
\label{fig:qualitative_events_ckov}
\end{figure}

An illustration of an event of the opposite extreme is shown in
Figure~\ref{fig:qualitative_events_ckov}.  This is a much lower energy
event for which the shower axis points towards the detector at a small
incident angle.  Almost all of the signal registered comes from direct
CL.  As is typical of CL-dominated events, the event is short: it is
contained in a single mirror, spanning just under $7^{\circ}$ in
angle, and only about 200ns in time.

The event set of TALE FD used in this paper contains a mix of CL and
FL.  However, at the lowest energies, events are dominated by CL.  Because
of the forward-beamed signal, those events pointing toward the
detector are of course preferentially recorded.  The cores of these
showers fall within a small area near the FD.  At the
higher energies, however, the FD becomes more sensitive to the
isotropically emitted FL that can land over a much larger area around
the detector site.  In the decade between $10^{16.5}$eV and
$10^{17.5}$eV, the events are typically recorded with a mix of both CL
and FL.  The very different nature of the signals makes for somewhat 
complicated evolution of detector aperture and resolutions with
energy, as will be discussed later in this paper.

\FloatBarrier

\section{Event Processing and Reconstruction}
\label{sec:data_process}

Nearly all of nightly data events from the TALE FD are from accidental
coincident hits in PMTs from night-sky brightness fluctuations, and
very low-energy Cherenkov blasts that cannot be used in the analysis.
In order to reduce the data to contain only usable events, a
processing chain was devised that analyzes the events and removes
those which are of low quality.  A brief description of the main steps 
and the data selection criteria used follows:

\begin{description}
\item[Step 1] Individual telescope data packets are combined to 
  form ``events'', and PMT signal calibration is applied.
\item[Step 2] A shower-detector plane (SDP) is reconstructed from the 
  pattern and pointing direction of the hit PMT pixels, as
  displayed in the left panels of figures~\ref{fig:qualitative_events_scin} 
  and~\ref{fig:qualitative_events_ckov}. Directional 
  correlation to the fitted SDP and expected time order are used to exclude
  accidental hits not related to the event.  This procedure involves 
  performing a {\it timing fit} of the hit pixels ``time vs. angle'', as
  displayed in the right panels of figures~\ref{fig:qualitative_events_scin} 
  and~\ref{fig:qualitative_events_ckov}.
  The following filtering criteria are then applied to select shower-induced,
  track-like events.
  \begin{enumerate}
  \item Arc length of track $>$ 4.5$^{\circ}$
  \item Inverse angular speed of track $>$ 0.012 $\mu s/deg$
  \item Event duration $>$ 0.1 $\mu s$
  \item Average number of photons/PMT $>$ 50
  \end{enumerate}
  The vast majority of the events removed at this stage actually failed the 
  duration cut, mainly very low-energy Cherenkov blasts.
\item[Step 3] Shower track reconstruction.
  The following selection criteria are applied to events at this stage:
  \begin{enumerate}
  \item Number of good-tubes (tubes associated with event) / deg of track $> 1.1$
  \item timing fit normalized $\chi^{2} <$ 5.0 
  \end{enumerate}
\item[Step 4] Shower profile and energy reconstruction.
\end{description}

After the filtering cuts, the remaining events are subjected to full
shower reconstruction, which includes the determination of shower
geometry, energy, and the depth of shower maximum.  In the original
technique developed by the Fly’s Eye experiment~\cite{Baltrusaitis:1985mx}, 
this process is divided into two steps.  Shower geometry is determined by 
fitting the pattern of hit pixels to a shower detector plane (SDP), and then
fitting the arrival time of light at the detector (in each hit pixel) 
as a function of the viewing angle of the pixel in the SDP:
\begin{equation}
  \label{eq:time_fit_flyseye}
  t_{i} = t_{0} + \frac{R_{p}}{c}\tan\left(\frac{\pi - \psi - \chi_{i}}{2}\right)
\end{equation}
where $R_{p}$ is the distance from the detector to the shower at the point 
of closest approach, $\psi$ is the incline angle of the track in the SDP, 
$t_{0}$ is a time offset, and $\chi_{i}$ is the viewing angle of the i-th pixel.

Once the shower geometry is determined, the pixel directions are converted 
to shower depths, and the signal size is fitted to the light profile expected 
for a given energy and shower $X_{\rm max}$ according to the Gaisser-Hillas
parameterization:
\begin{equation}
  \label{eq:time_gaisser_hillas}
  N(x) = N_{\rm max} \times \left(\frac{x -
    X_{0}}{X_{\rm max}-X_{0}}\right)^{(X_{\rm max}-X_{0})/\lambda}
    \exp\left(\frac{X_{\rm max}-x}{\lambda}\right)
\end{equation}
The parametrization gives the number of charged particles $N$ at
atmospheric depth $x$ along the shower track. $N_{\rm max}$, $X_{\rm max}$,
$X_{0}$, $\lambda$ are parameters.  Here, $N_{\rm max}$ is the number of 
shower particles at the point of maximum development, $X_{\rm max}$.  
$X_{0}$ represents the first point of interaction of the cosmic ray particle, 
and $\lambda = 70$~g/cm$^{2}$, in combination with $X_{0}$, sets the width of 
the shower profile curve.

The very low energy events observed by TALE, however, have track
lengths that are too short for independent geometry reconstruction.  So
in this paper we use a simultaneous fit of the geometry and shower
profile, but with holding the $X_{\rm max}$ fit parameter fixed to some 
nominal value.  This profile constrained geometry fit 
(PCGF)~\cite{AbuZayyad_thesis} was used successfully in the
analysis of HiRes-I data, which made the first observation of the GZK
suppression~\cite{Abbasi:2007sv}.

When applied to TALE events with significant CL signal, we found PCGF
to give very good geometry resolutions.  The sharply
peaked CL in the forward direction constrains most of the light to 
within a few degrees from the true shower direction, which made 
the shower profile fit very sensitive to obtaining the correct pointing 
direction: After examining the results from multiple fits with different 
``trial'' $X_{\rm max}$ parameters we noticed that the reconstructed geometry 
was essentially independent from the assumed trial values.  The resulting 
improvement of geometrical resolution (over that of monocular PCGF for FL
dominated events) made it possible to measure the $X_{\rm max}$ by adding a
step to the reconstruction procedure in which the PCGF-determined geometry 
can be fixed and a shower profile fit following standard
techniques can be performed.

The light signal recorded by a fluorescence detector contains a
contribution from Cherenkov light generated by the shower particles.
In event reconstruction, we distinguish among four contributions to
the total observed light signal:
\begin{enumerate}
  \item Direct Air Fluorescence light (FL).
  \item Direct Cherenkov light (CL).
  \item Rayleigh Scattered Cherenkov light
  \item Aerosols Scattered Cherenkov light
\end{enumerate}

These are the contributions to the total light signal shown in
figures~\ref{fig:qualitative_events_scin}
and~\ref{fig:qualitative_events_ckov} by the different colors.  All
four contributions are included in the PCGF and in the MC
simulation. The sum of the direct and scattered Cerenkov light make up
the total CL signal.

The TALE FD monocular data set spans over three decades of energy.
For optimizing the detector reconstruction algorithm and event
selection at different energy ranges, we classified events into three
categories:

\begin{enumerate}
  \item Cherenkov event: Fractional contribution to total signal 
    of CL $> 0.75$ and of direct CL $> 0.55$
  \item Fluorescence events: Fractional contribution to total signal
    of FL $> 0.75$
  \item Mixed events: those that do not belong to either (1) or (2) above.
\end{enumerate}

As will be shown below, fluorescence events dominated at energies
above $10^{17.5}$~eV, the Cherenkov 
events at below $10^{17}$ eV, and Mixed events in between.  FL and CL 
events have very different characteristics.  Mixed events may be more 
similar to CL or FL events depending on the direct CL contribution to 
the total signal.  Different event selection criteria were chosen for 
the different types of events, as summarized in Table~\ref{table:quality_cuts},
in order to remove poorly reconstructed events and assure good detector
resolution.

\begin{table}[htb]
\centering
\caption{Quality cuts applied to the three event categories.  Note
  that some cut values are the same across the three subsets while
  others have different numerical values.  Mixed events with direct CL
  fraction $>0.35$ are subject to the same cuts as Cherenkov events.}
\label{table:quality_cuts}
\begin{tabular}{l|c|c|c}
\hline
{\bf Variable}                              & {\bf CL}                        & {\bf Mixed} (dCL frac. $<0.35$) & {\bf FL}      \\ \hline
Angular Track-length [deg]                  & $trk > 4.5^{\circ}$               & $trk > 10.0^{\circ}$        & $trk > 10.0^{\circ}$      \\ 
Inverse angular speed [$\mu$s/deg]          & $0.014<1/\omega<0.1$            & $0.06<1/\omega<1.0$        & $0.06<1/\omega<1.0$     \\ 
Shower Impact parameter [km]                & $0.4<R_{p}<5.0$                  & $0.4<R_{p}<8.0$            & $1.0<R_{p}<20.0$         \\ 
Shower Zenith Angle [deg]                   & $28^{\circ} < \theta < 65^{\circ}$  & $\theta < 65^{\circ}$       & $\theta < 65^{\circ}$     \\ 
Shower $X_{\rm max}$ [g/cm$^2$]               & $475<X_{\rm max}<815$              & $525<X_{\rm max}<815$        & $465<X_{\rm max}<815$      \\ \hline
Estimated fit errors on energy               & \multicolumn{3}{c}{$\delta E/E < 0.6$}                  \\ 
Estimated fit errors on $X_{\rm max}$ [g/cm2]  & \multicolumn{3}{c}{$\delta X_{\rm max} < 200$}             \\ 
Timing fit $\chi^{2} / dof$                  & \multicolumn{3}{c}{$\chi^{2}_{tim}<4.5$}                   \\ 
Profile fit $\chi^{2} / dof$                 & \multicolumn{3}{c}{$\chi^{2}_{pfl}<12$}                    \\ \hline
\end{tabular}
\end{table}
The same PCGF reconstruction code and event selection is applied to
both data and simulated events.  The performance of this technique
will be discussed in a later section.  Table~\ref{table_event_summary}
summarizes the number of events which survive the application of the
quality cuts.  The two time epochs comprising the total data set are 
listed separately as well.

\begin{table}[htb]
\centering
\caption{Number of TALE events and numbers in each category.  
  The ``After cuts'' numbers include a requirement that the 
  shower energy is greater than $10^{15.3}$~eV.}
\label{table_event_summary}
\begin{tabular}{ l|r r}
\hline
{\bf Data Set}  & {\bf Events} &  {\bf After Cuts} \\ \hline
{\bf CL}        & 883637       & 392758            \\
{\bf Mixed}     & 21102        & 3726              \\
{\bf FL}        & 8060         & 3578              \\
{\bf All}       & 912799       & 400062            \\ \hline
\end{tabular}
\end{table}

The primary goal of the study described in this paper is to measure
the energy spectrum of cosmic rays in the range $10^{15.3}$ to
$10^{18.3}$ eV.  Energy spectrum refers to the differential flux of
cosmic rays, per unit energy, measured as a function of energy. We 
divide events in the data into bins
of fixed width $\Delta{x}=\Delta\log_{10}E[\rm eV]=0.05$.  For a given bin
at energy $E$, the differential flux is then approximated by the
formula:
\begin{equation}
J(E) = \frac{d^{3}N}{dE d(A\Omega)dt} \approx
\frac{\Delta{N}}{\Delta{E}\Delta(A\Omega)\Delta{t}}
\label{diff-flux}
\end{equation}
where $\Delta{N}$ is the number of events in the bin centered at energy $E$,
$\Delta{E}=(\ln{10})E\Delta{x}$ is the width of the bin at energy $E$,
$\Delta(A\Omega)$ is the aperture of the detector at energy $E$, and
$\Delta{t}$ the total run time.  The product
$\Delta(A\Omega)\Delta{t}$ is often referred to as the detector
exposure.

\begin{figure}[htb]
\centering
\includegraphics[height=2.4in]{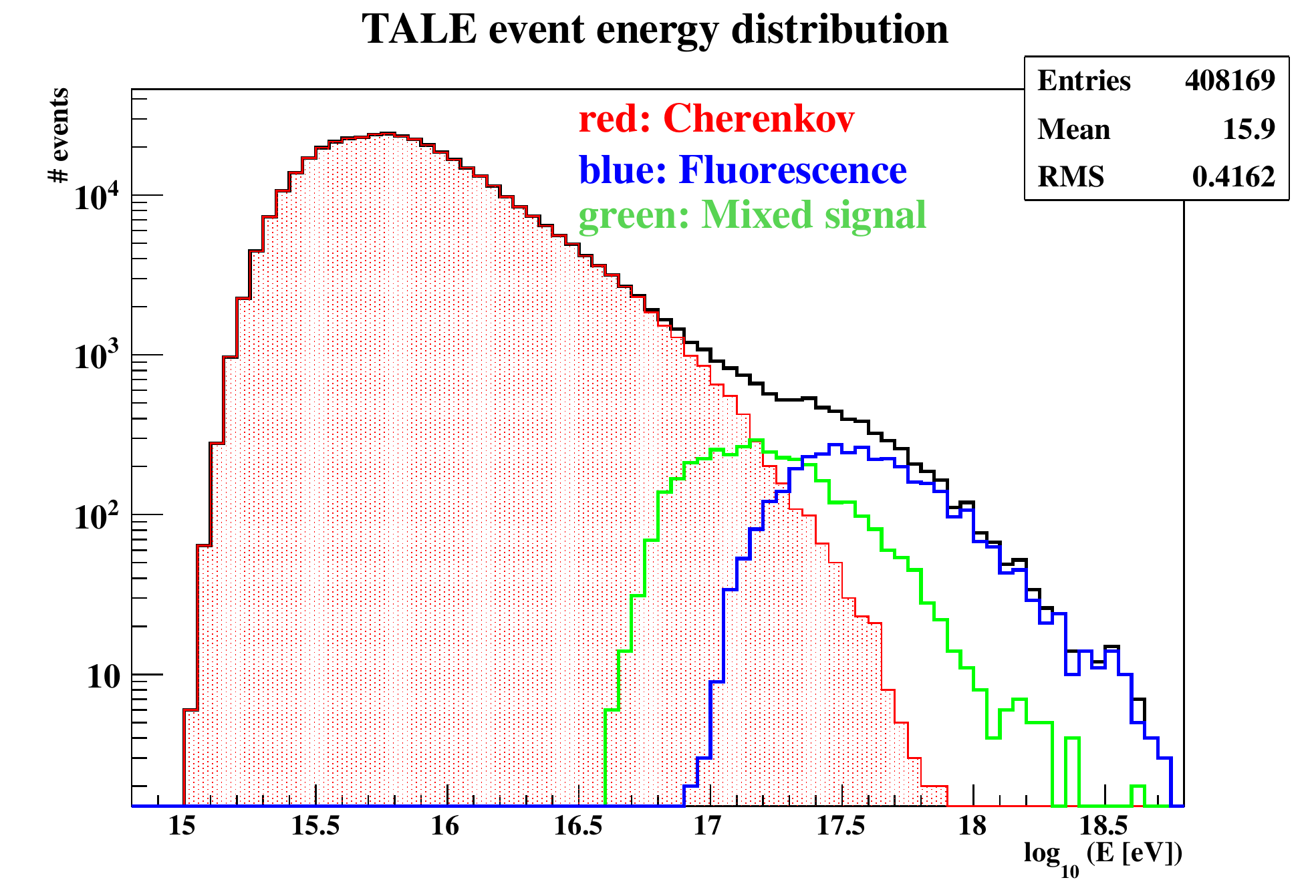}
\caption{Histogram of the energies of events from TALE FD data.  The
  distribution for Cherenkov, mixed, and fluorescence events are shown
  in red, green, and blue, respectively.  The distribution for the
  inclusive set is shown in black.  Note that the fluorescence events
  are restricted to energies above $\sim 10^{17}$ eV, the mixed events
  lowers the TALE FD threshold to about $10^{16.7}$ eV, and the
  Cherenkov events extends the reach by another 1.5 decades of energy
  to below $10^{15.5}$eV.  We should note that distributions obtained
  from the two data epochs look very similar to each other and to the 
  combined histogram.}
\label{fig:energy_distribution}
\end{figure}

Figure~\ref{fig:energy_distribution} shows the histogram of the
energies of events in the data using the selection, reconstruction
procedures, and quality cuts described in the preceding sections.
These bin values then provide the numerator for equation
\ref{diff-flux}.  These data contains a total of 400,062 events with
reconstructed energy greater than $10^{15.3}$~eV.
These events were
collected over a total observation time of 1080 hours, collected
between June, 2014 and March, 2016. This observation time then provides
one part of the denominator, $\Delta{t}$, of equation \ref{diff-flux}.
$\Delta{E}$ is clearly defined by the binning.  The only remaining
quantity needed for evaluating the differential flux is the aperture.

As the symbol $(A\Omega)$ implies, the aperture is the product of the
effective area and fiducial solid angle of the detector, which varies
with energy.  The effective area of the detector actually changes with
the angular configuration of the events, so that $(A\Omega)$ actually
represents a rather complicated convolution integral of the effective
area multiplied by the solid angle element $d\Omega$ over all angular
configurations.  For this analysis, we evaluate $(A\Omega)$ using
the Monte Carlo method, which we describe in the next section.

\FloatBarrier

\section {Simulation}
\label{sec:MC}

The TALE simulation program is an updated version of the TA Middle
Drum simulation program with added support for TALE. Much of the code
is inherited from the HiRes-1 simulation program, with the exception
of the TALE electronics simulation which has its origins in the
HiRes-2 simulation package. The core of this code has been refined and
tested against real data over a period of two decades.  For TALE FD,
the simulation uses actual run conditions: telescope live times, PMT
gains, trigger settings, and atmospheric parameters, for each night of
data collection.

The main components of the simulation are:
\begin{enumerate}
\item shower development and light production
\item light transmission through the atmosphere
\item detector optics ray-tracing 
\item electronics signal processing
\item detector trigger logic
\end{enumerate}

The longitudinal shower development is simulated using a library of
showers created using the CONEX~\cite{Bergmann:2006yz} package.  MC
events generated from this library are used to calculate the detector
aperture.  A smaller set of simulations generated from
CORSIKA~\cite{Heck:1998vt} is used to cross-check and validate the
CONEX simulation.  For instance, comparison between the two simulation
sets are used to verify the missing energy correction.

A 4-parameter Gaisser-Hillas function is used to describe the
longitudinal development of the shower. The lateral spread of the
shower electrons is simulated according to a modified (NKG-like)
parametrization from~\cite{Lafebre:2009en}.
Simulation of fluorescence light
production assumes the Kakimoto air fluorescence yield and the FLASH
wavelength spectrum as in the case of previous TA
papers~\cite{Kakimoto:1995pr,Belz:2005pv}.  We also used a Cherenkov
light angular distribution following the parametrization given
in~\cite{Nerling:2005fj}.  The amount of Cherenkov light production
follows the Fly's Eye formula~\cite{Baltrusaitis:1985mx} but with
electron energy distributions updated to those given
in~\cite{Nerling:2005fj}.

Propagation of light from the shower to the detector are performed
using photon-by-photon ray-tracing.  The atmospheric profiles are
provided from a GDAS database (one entry per three hour
interval)~\cite{GDAS}. Atmospheric aerosols density is treated as 
constant with an average density corresponding to a vertical 
aerosols optical depth (VAOD) of 0.04.  Scattering and attenuation 
from Rayleigh, aerosol, and ozone scattering are all included.

 The actual detector layout and geometry are reproduced in
the MC.  The shape and size of the telescope mirror, including
obscurations from the PMT camera box, its mechanical supports, and
cables are all included in the ray-tracing.  Pixel response is modeled
according to measured PMT cathode response profiles.  Both analog and
digital readout electronics, including the trigger logic, are simulated
in functional detail, with time development, including the effect of
analog filters, and night-sky background light.

The simulation used for the detector aperture calculation is run as
follows: A full energy range, $10^{15}$-$10^{18.5}$ eV, set and a 
high energy, $10^{16.4}$-$10^{18.5}$ eV, set were generated for the
aperture calculation.  The generated shower energies follow a
$E^{-2.92}$ spectrum. The simulated showers are drawn from a shower
library created using CONEX version 4.36 with QGSJet II-3 hadronic
interaction model. Five primaries: H, He, N, Mg, and Fe are included
in the simulations with a relative abundance based on the 
H4a model \cite{Gaisser:2012cc}.

In addition to the above model, we fit the $X_{\rm max}$ (data) distributions  
observed by the TALE FD to a mixture of four primaries (H, He, N, Fe), 
as a function of energy, and use the resulting abundances in our 
simulation.  The fit is performed by calculating a weighted sum of four MC 
histograms representing the reconstructed $X_{max}$ distributions of the four 
primaries~\cite{TFractionFitter, Barlow:1993dm}.  A "thrown fraction" is then 
determined taking into account the relative detection and reconstruction 
efficiencies for each primary type.  We refer to this primary mixture as 
``TXF'', for {\bf T}ALE $\bm{X}_{max}$ distributions {\bf F}its.
 
\begin{figure}[htb]
\centering
\includegraphics[height=3.2in]{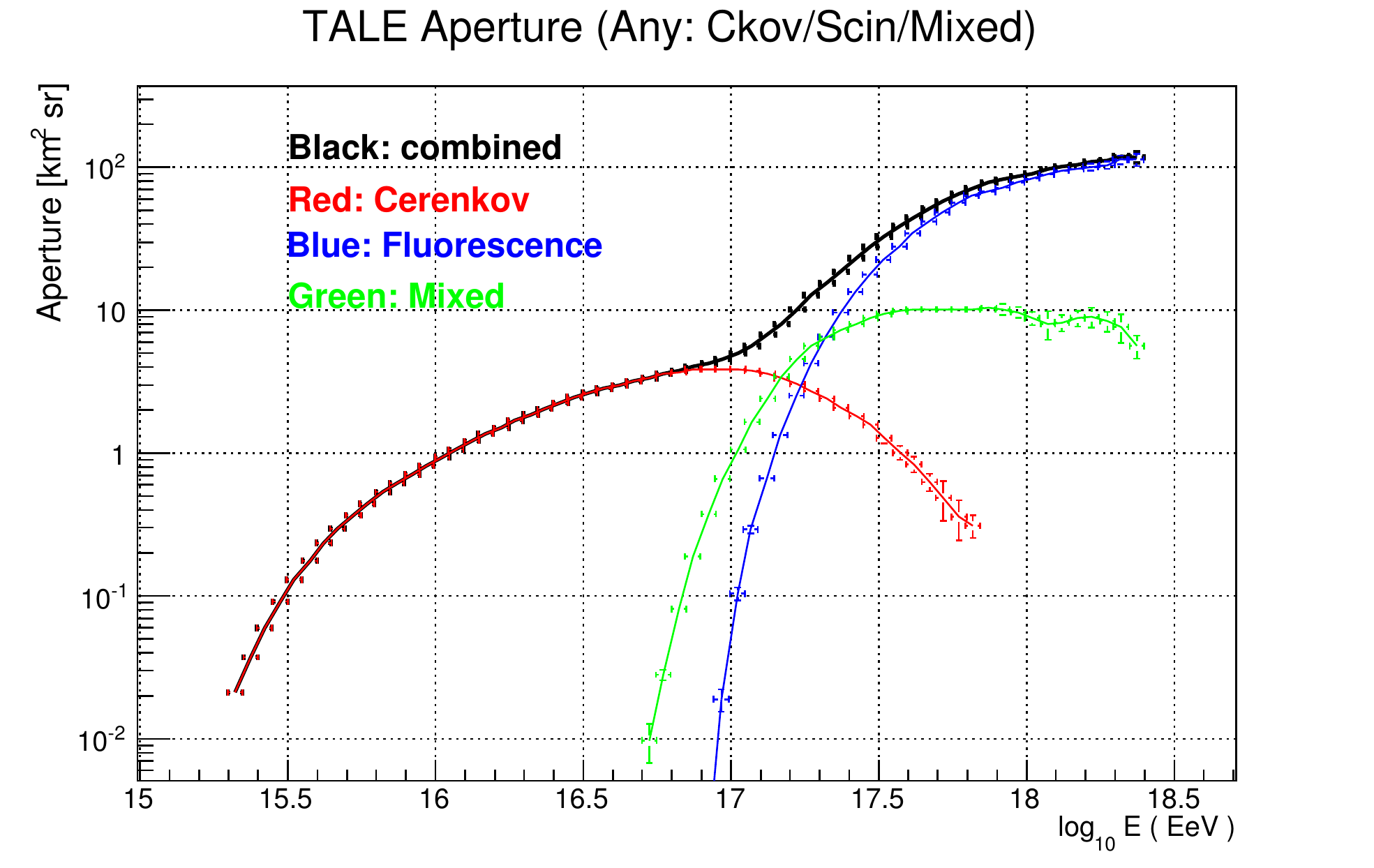}
\caption{TALE aperture using the combined set of events.  Aperture functions
  for each of the three subsets also shown.}
\label{fig:tale_aperture_all}
\end{figure}

This process of fitting the observed $X_{\rm max}$ distributions puts 
simulated showers in the actual positions in the sky as the real 
showers in the data, and is independent of hadronic interaction model.  
As will be seen below, our reconstruction
of $X_{\rm max}$ has excellent resolution below about $10^{17.5}$ eV, so the
fitting process successfully simulates the data.  Above this energy,
although the resolution is not as good, the aperture is quite insensitive
to composition, as will be shown in section~\ref{sec:results_and_discussion}, 
and so this process is robust. 
Figure~\ref{fig:tale_aperture_all} shows the aperture for this mix of primaries.

Simulated showers are reconstructed using the same procedure applied
to real data, and event selection is done in the same way.  A missing
energy correction is applied to the reconstructed data and MC showers
based on the same composition assumption, with the correction for each
primary type and energy being estimated from the CONEX generated showers
(cross checked against CORSIKA predictions).
  We use QGSJet II-3 because, at higher energies, we have measured\cite{Fujii:2017} 
the missing energy using a technique invented by the Auger collaboration, and found 
that it agrees well with that predicted by QGSJet II-3.

In addition to calculating the aperture, the simulation is also used
to verify the accuracy of the event reconstruction procedure.  In
particular we examine the “reconstruction resolution” of the following
parameters:

\begin{enumerate}
\item The angle in the shower-detector plane, $\psi$
\item The shower impact parameter to the detector, $R_{p}$
\item The depth of shower maximum, $X_{\rm max}$.  
\item The shower energy, $E$
\end{enumerate}

\begin{figure}[htb]
\centering
\includegraphics[height=1.45in]{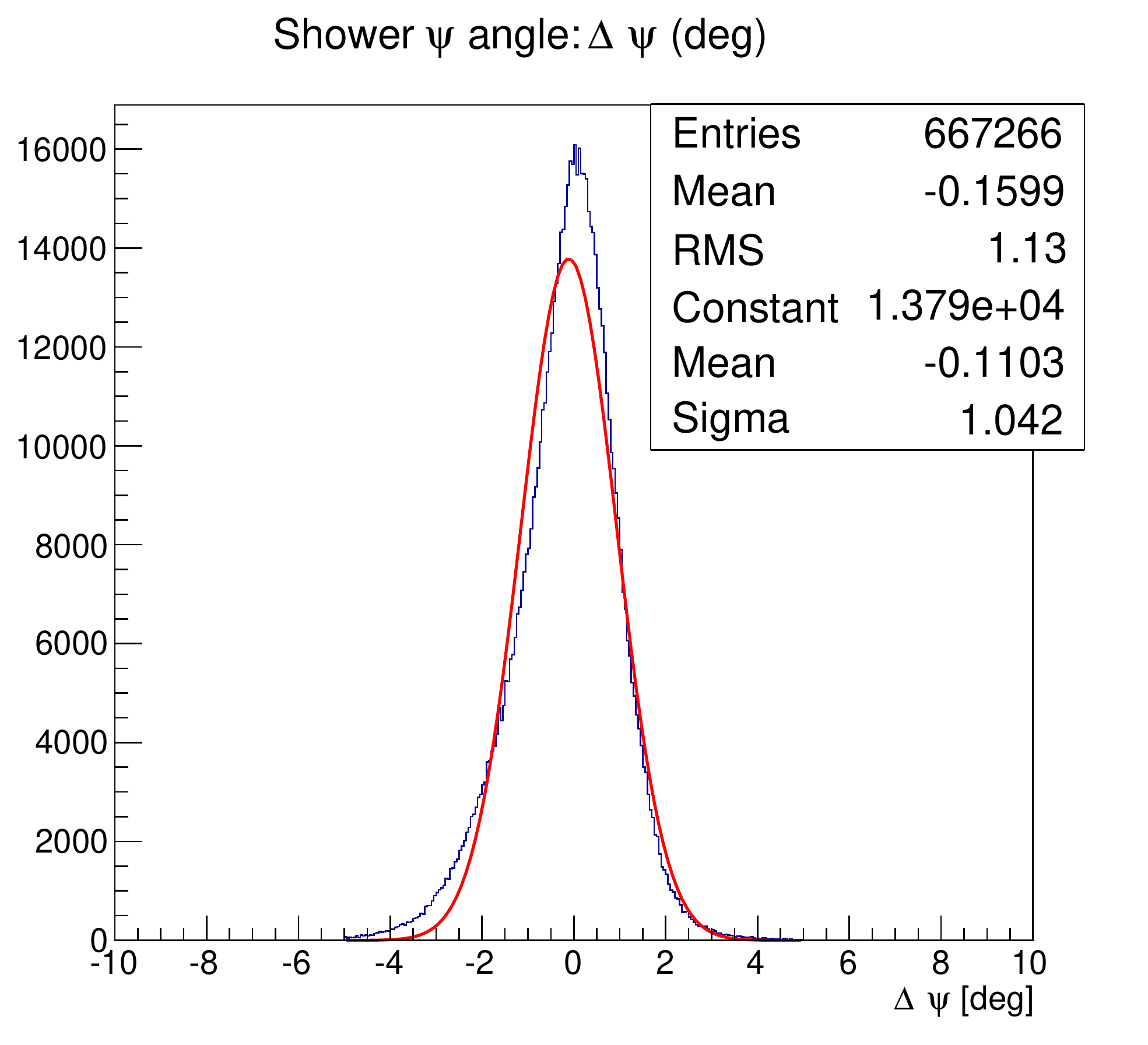}
\includegraphics[height=1.45in]{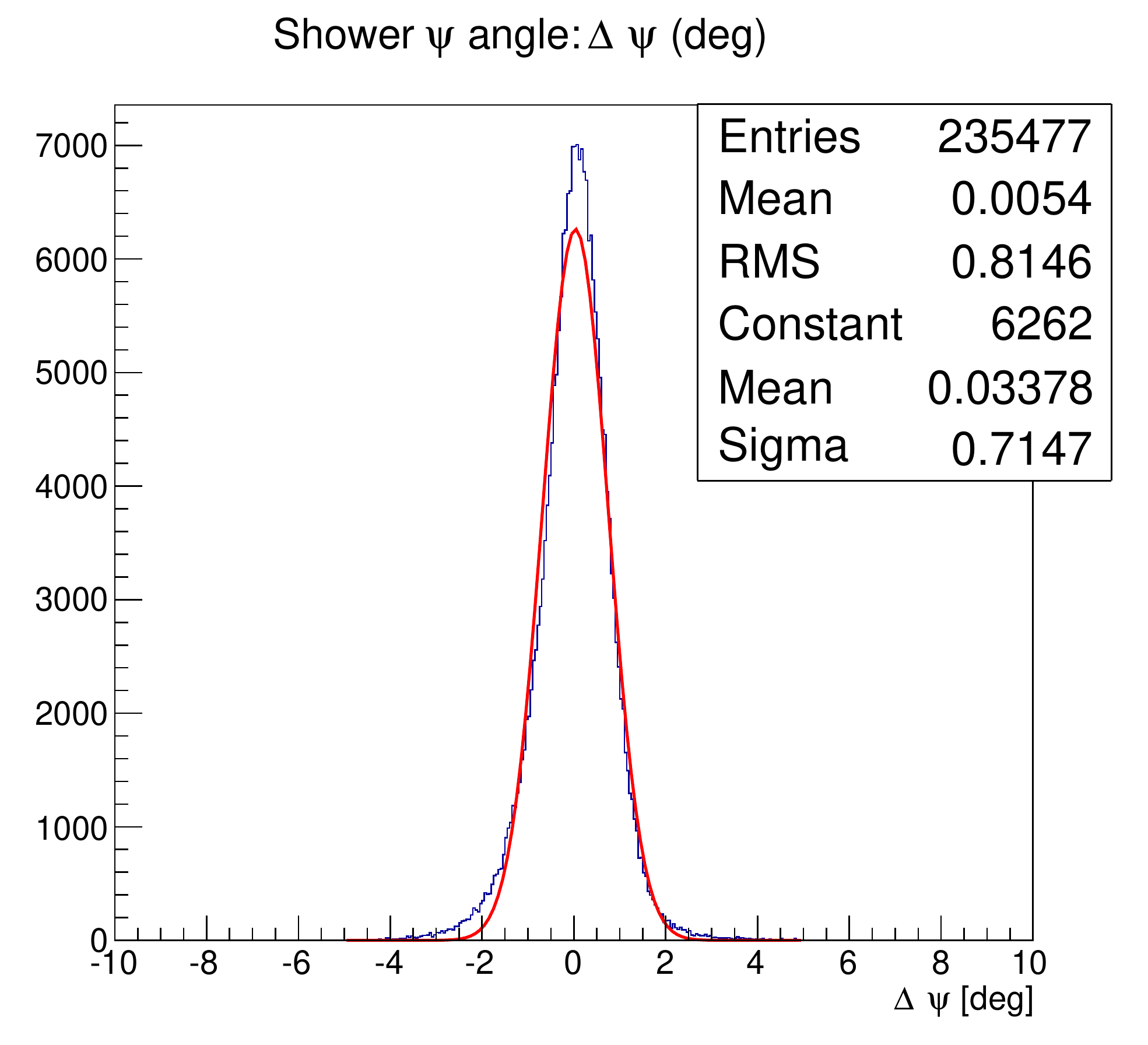}
\includegraphics[height=1.45in]{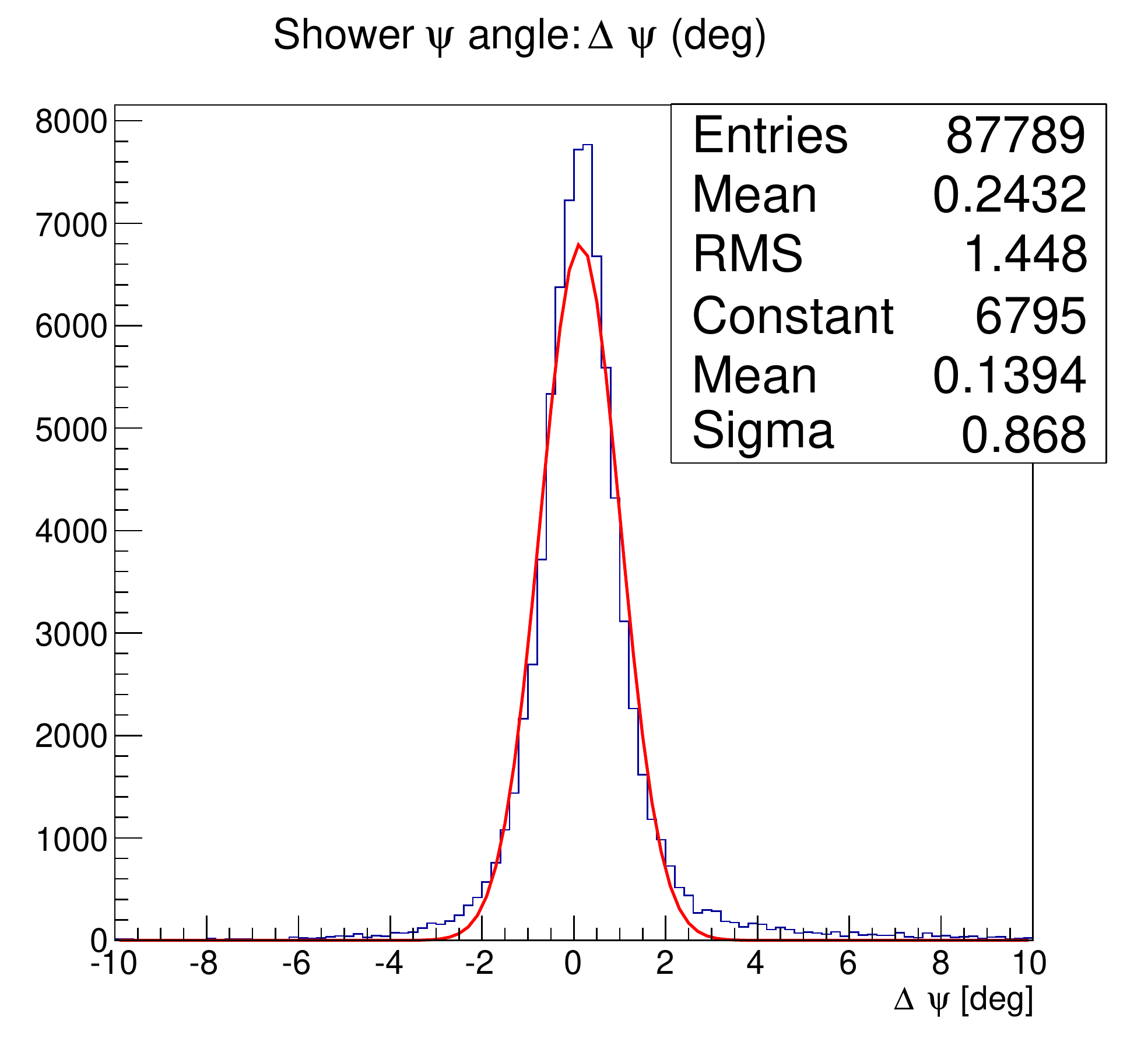}
\includegraphics[height=1.45in]{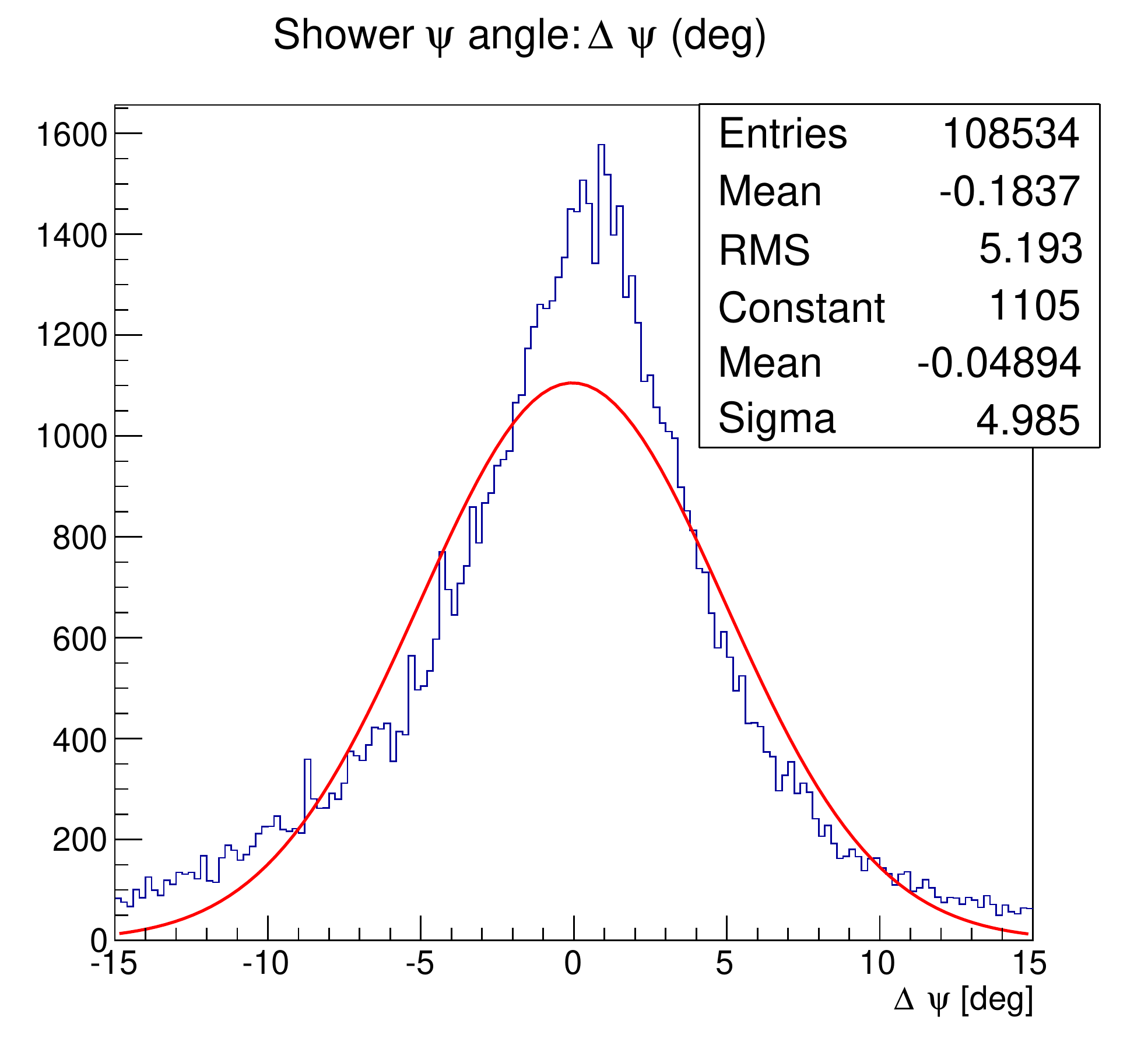}
\caption{Shower angle in the plane $\psi$ resolution after all quality cuts are applied.
  From the left: Cherenkov (all energies), Cherenkov ($E > 10^{16.7}$~eV), Mixed, 
  and fluorescence events subsets.  
  MC using a mixed composition, matching TXF results,} 
\label{fig:res_plots_spectral_set_1}
\end{figure}

\begin{figure}[htb]
\centering
\includegraphics[height=1.45in]{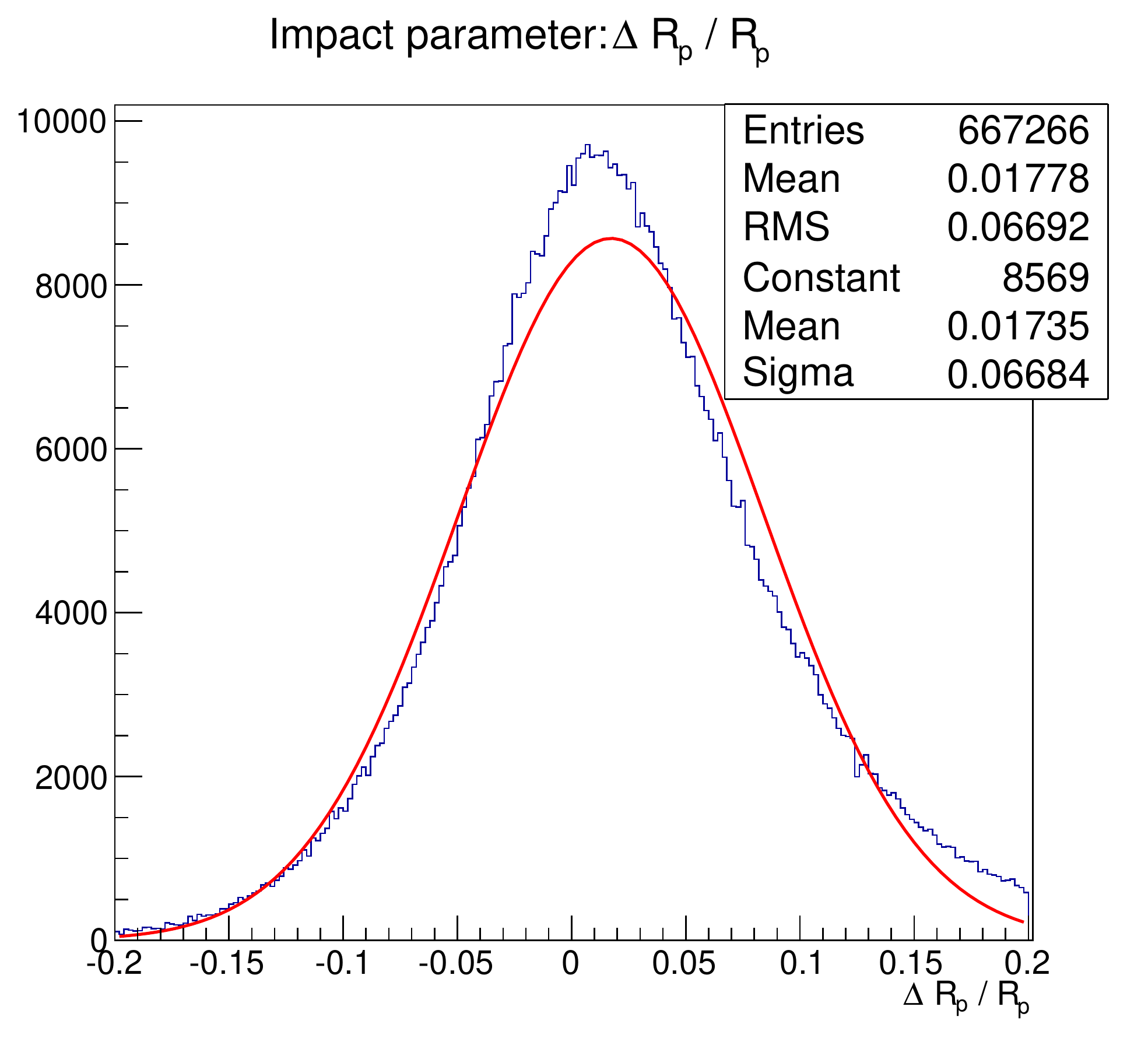}
\includegraphics[height=1.45in]{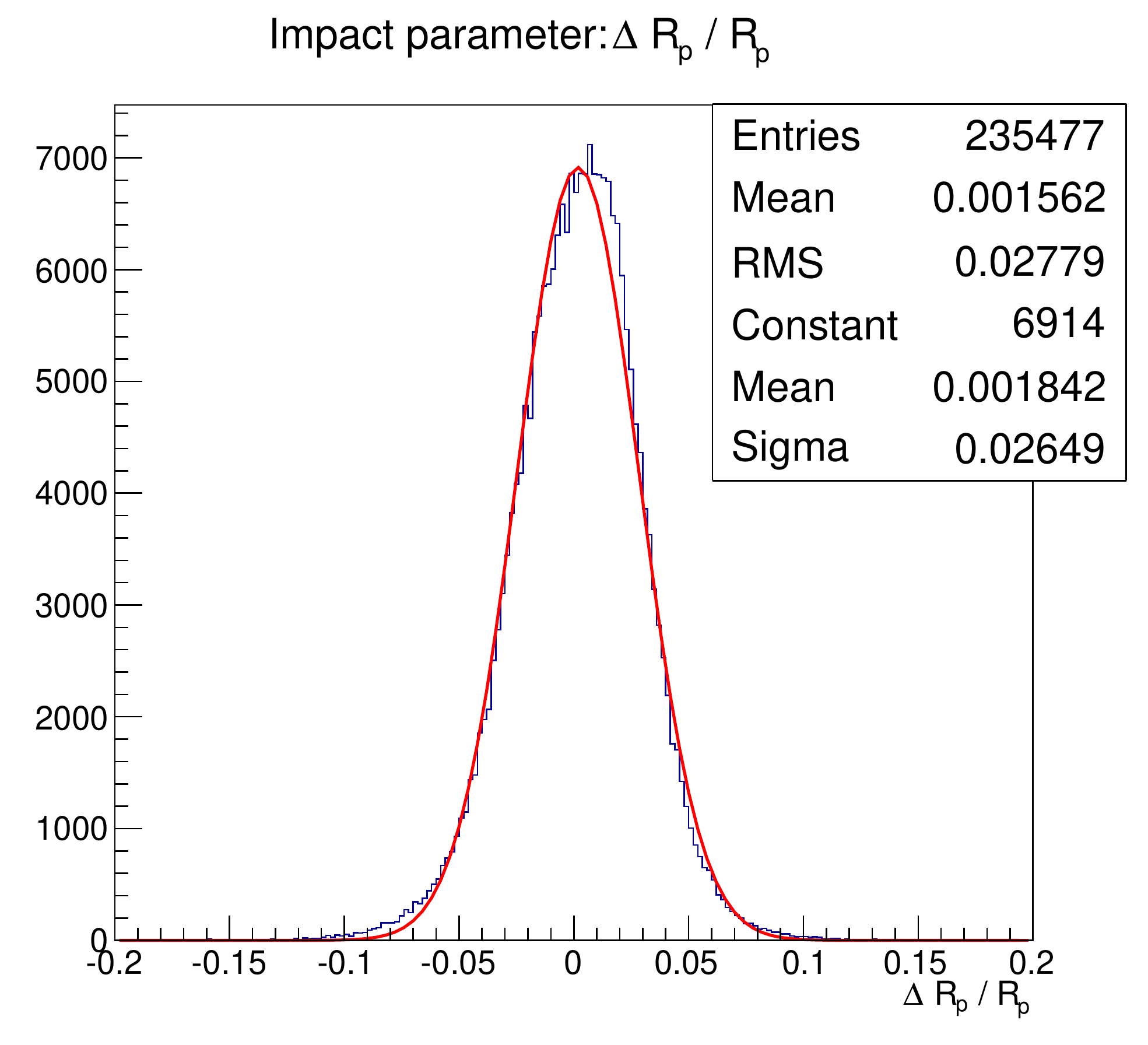}
\includegraphics[height=1.45in]{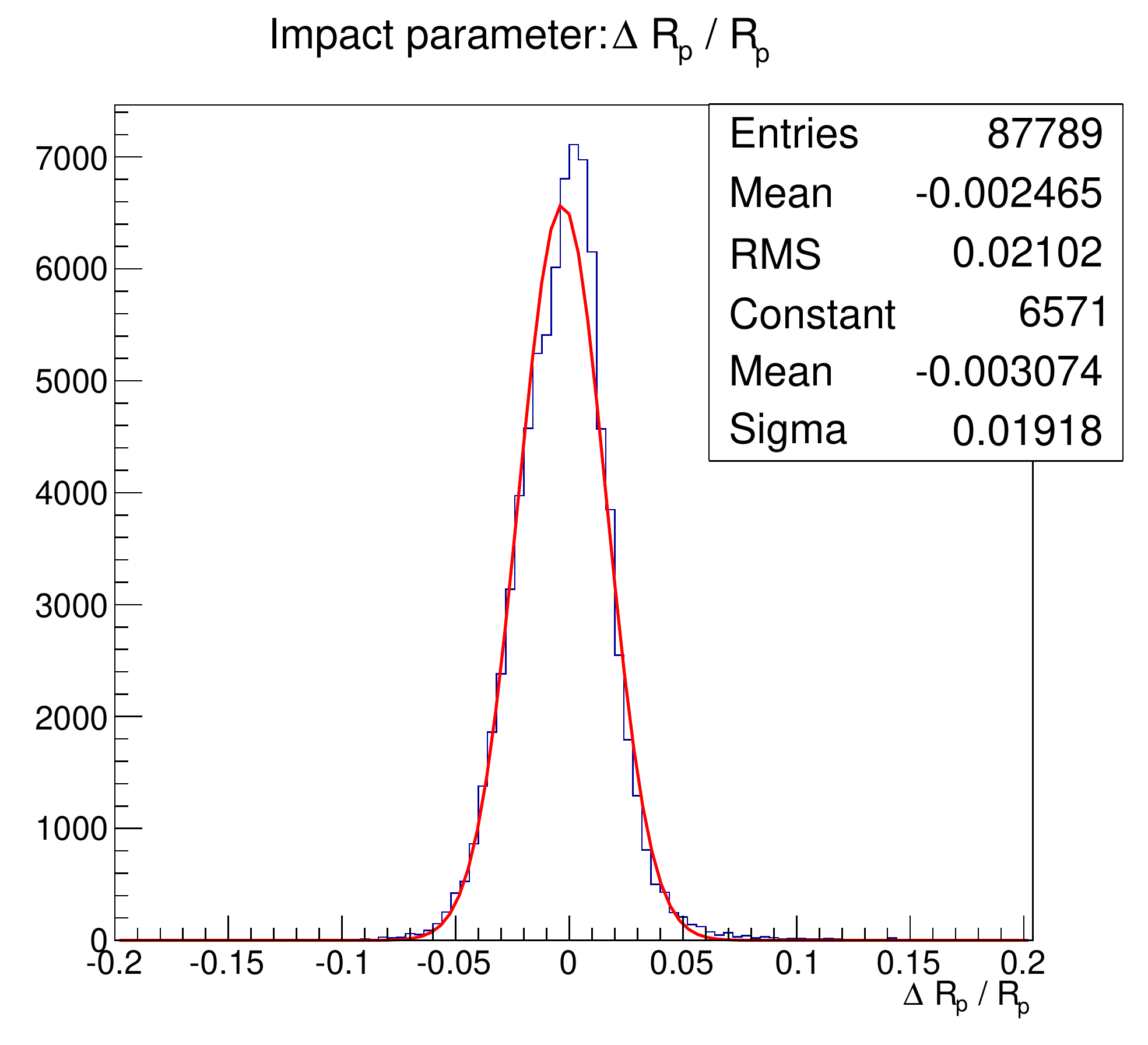}
\includegraphics[height=1.45in]{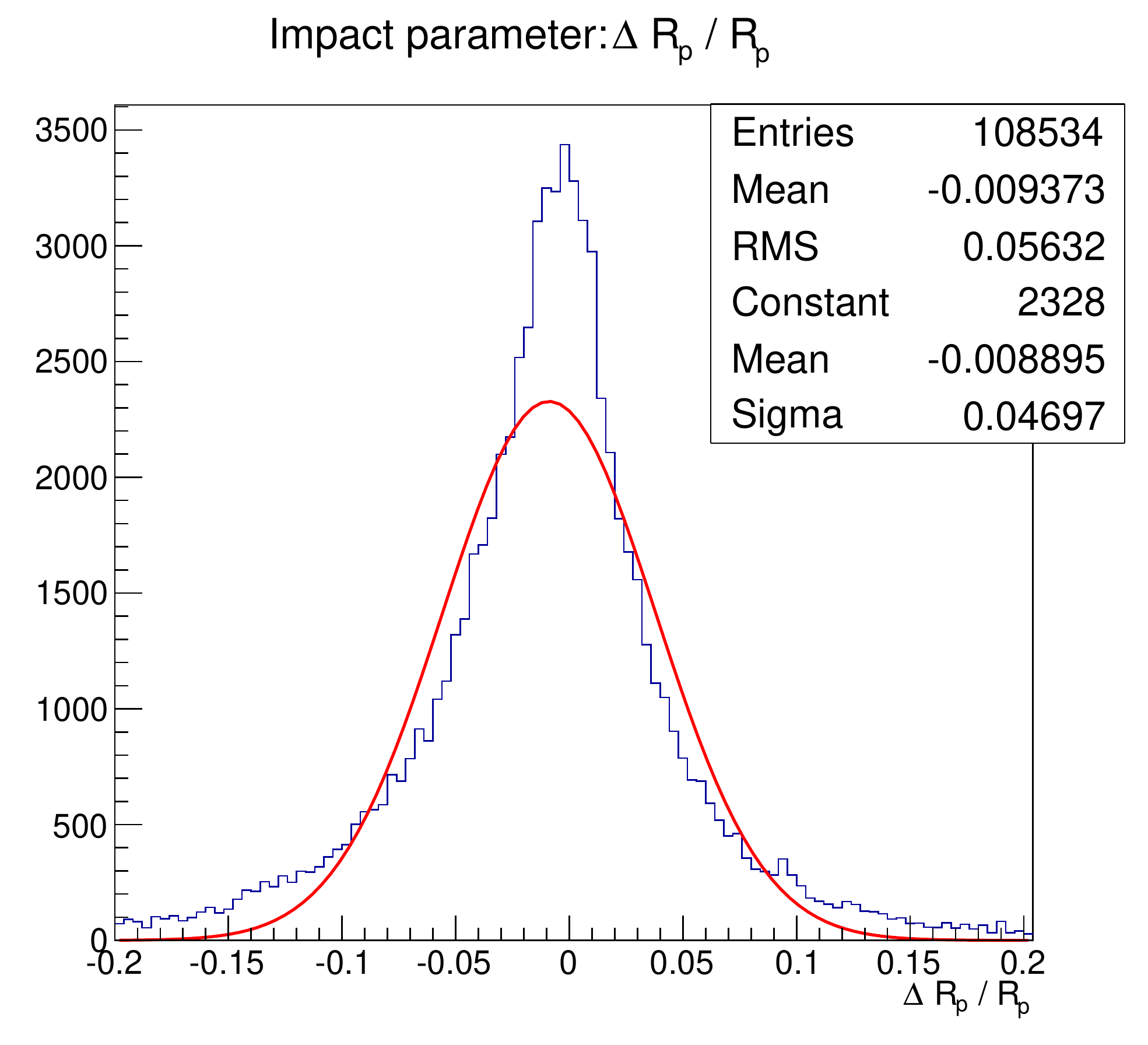}
\caption{Shower impact parameter to the detector, $R_{p}$, resolution after all quality 
  cuts are applied. From the left: Cherenkov (all energies), Cherenkov ($E > 10^{16.7}$~eV), Mixed, 
  and fluorescence events subsets. 
  MC using a mixed composition, matching TXF results,} 
\label{fig:res_plots_spectral_set_2}
\end{figure}

\begin{figure}[htb]
\centering
\includegraphics[height=1.45in]{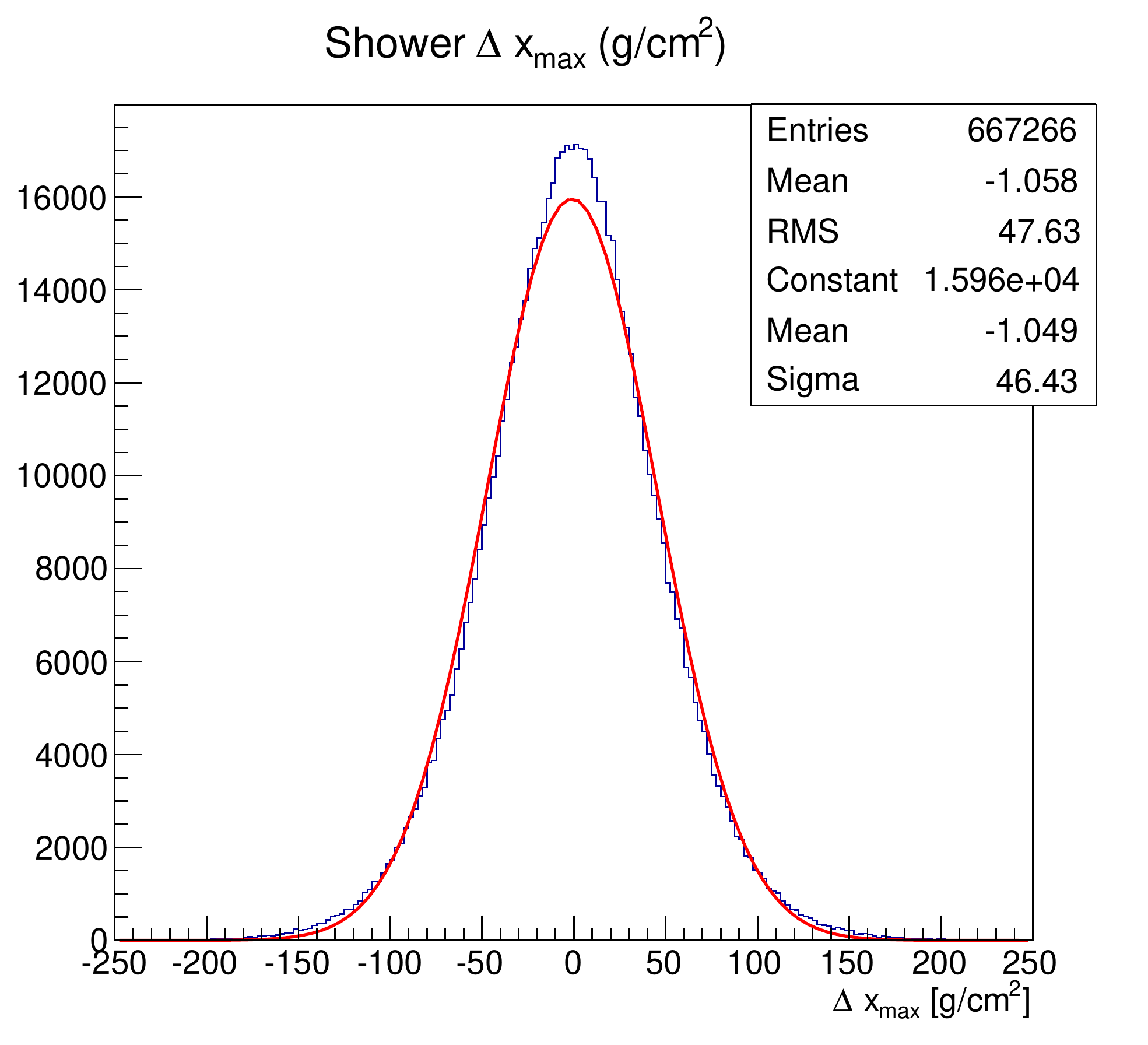}
\includegraphics[height=1.45in]{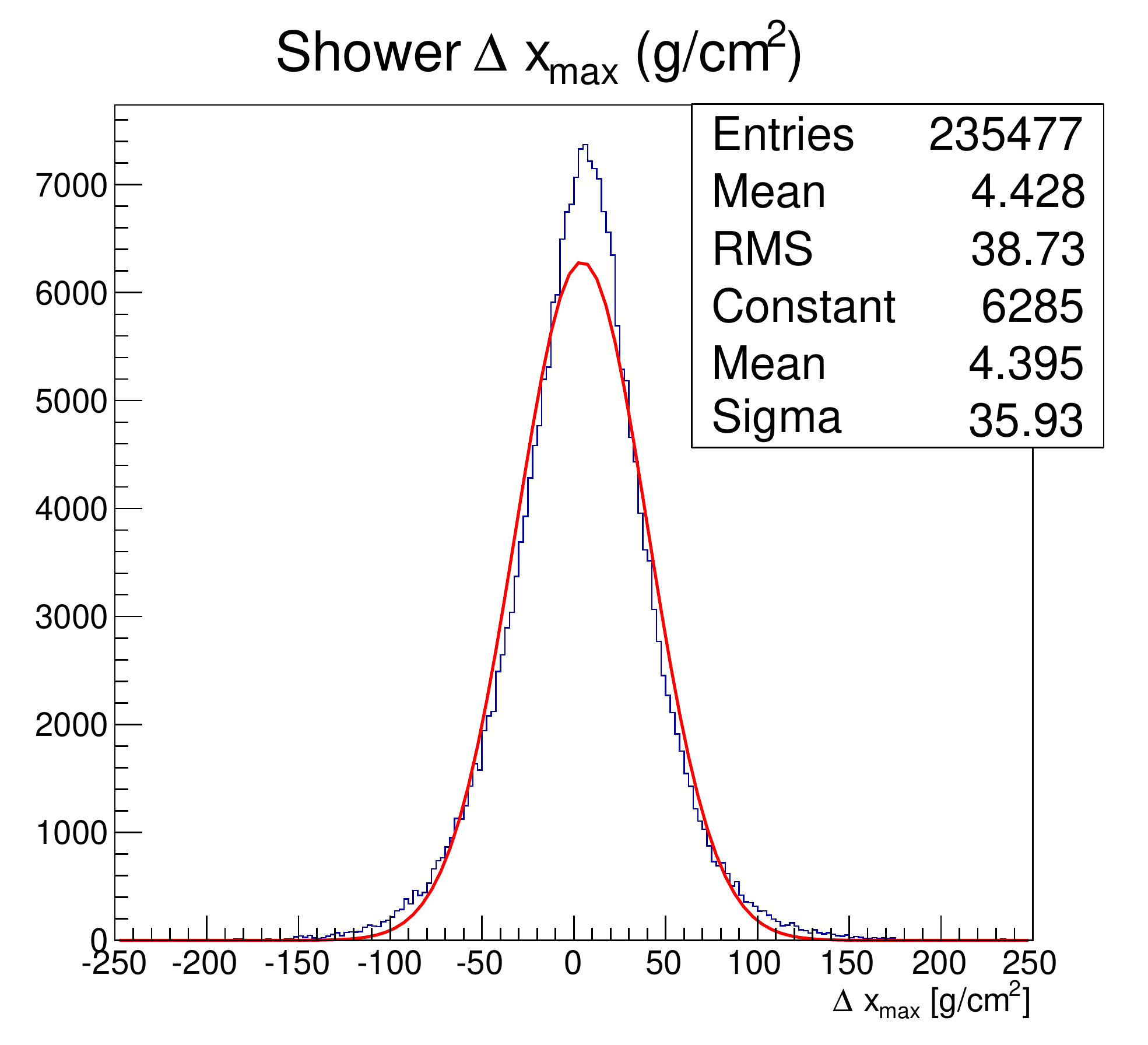}
\includegraphics[height=1.45in]{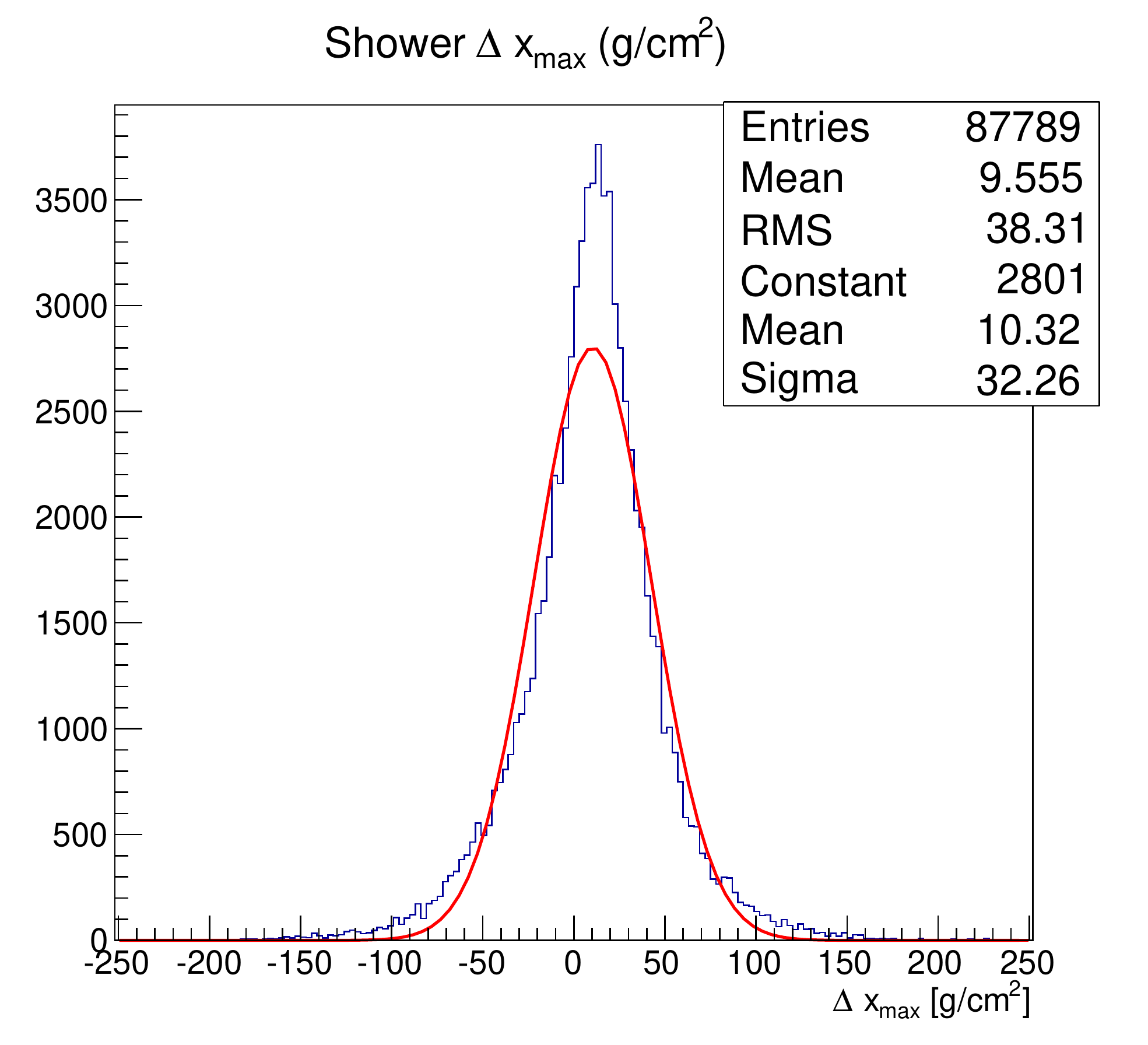}
\includegraphics[height=1.45in]{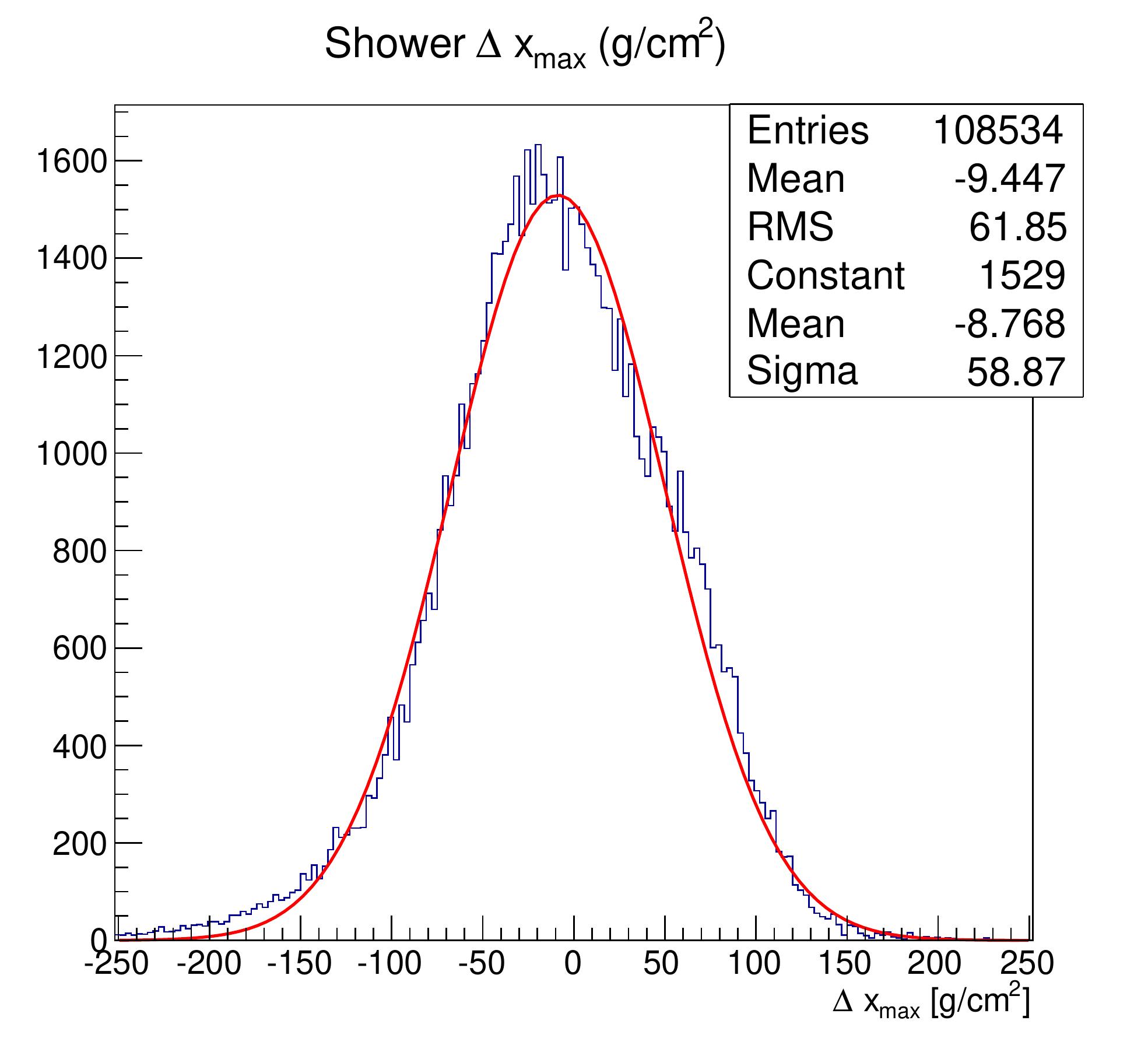}
\caption{Shower $X_{\rm max}$ resolution after all quality cuts are applied.  
  From the left: Cherenkov (all energies), Cherenkov ($E > 10^{16.7}$~eV), Mixed, 
  and fluorescence events subsets.  
  MC using a mixed composition, matching TXF results,} 
\label{fig:res_plots_spectral_set_3}
\end{figure}

\begin{figure}[htb]
\centering
\includegraphics[height=1.45in]{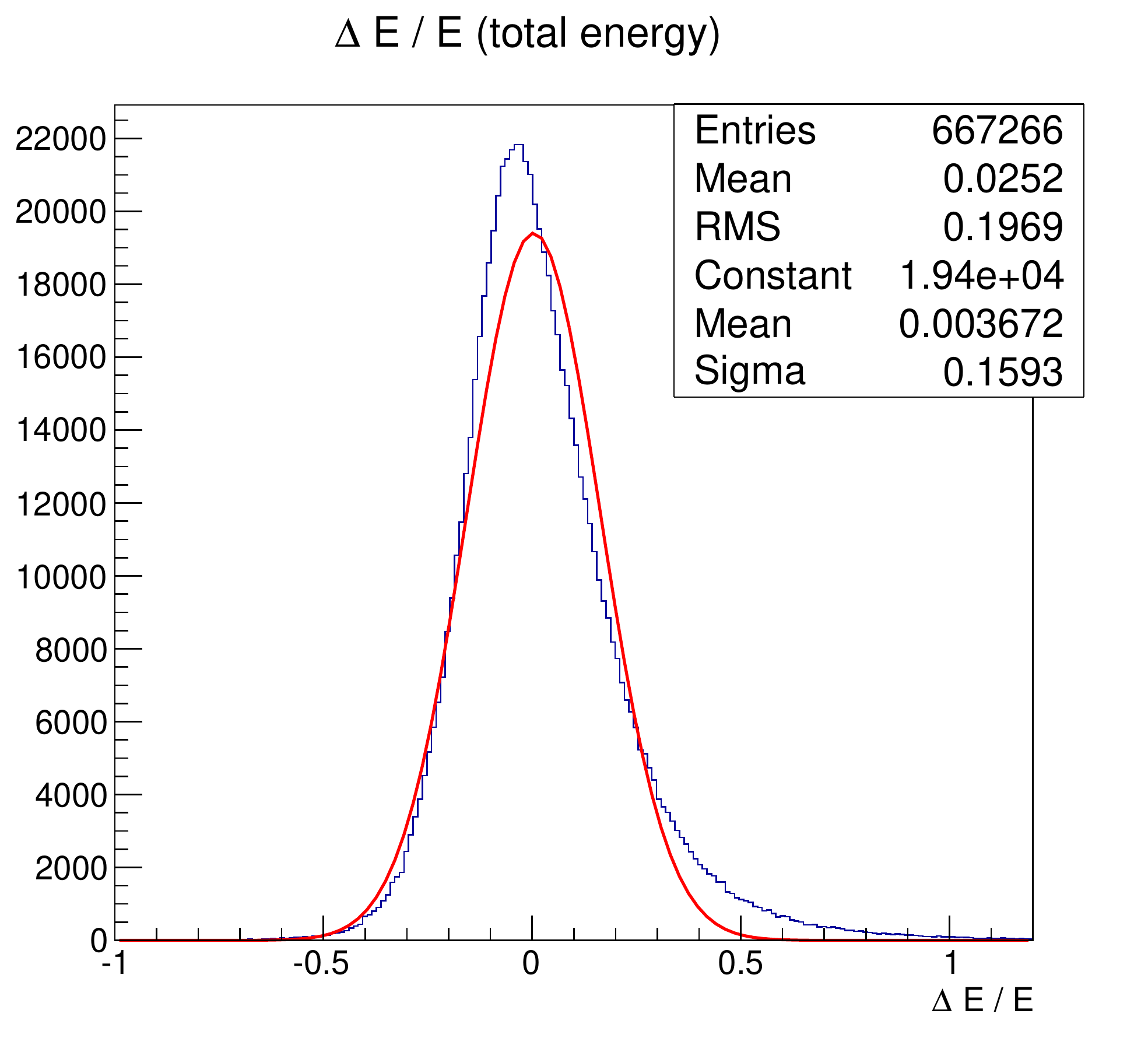}
\includegraphics[height=1.45in]{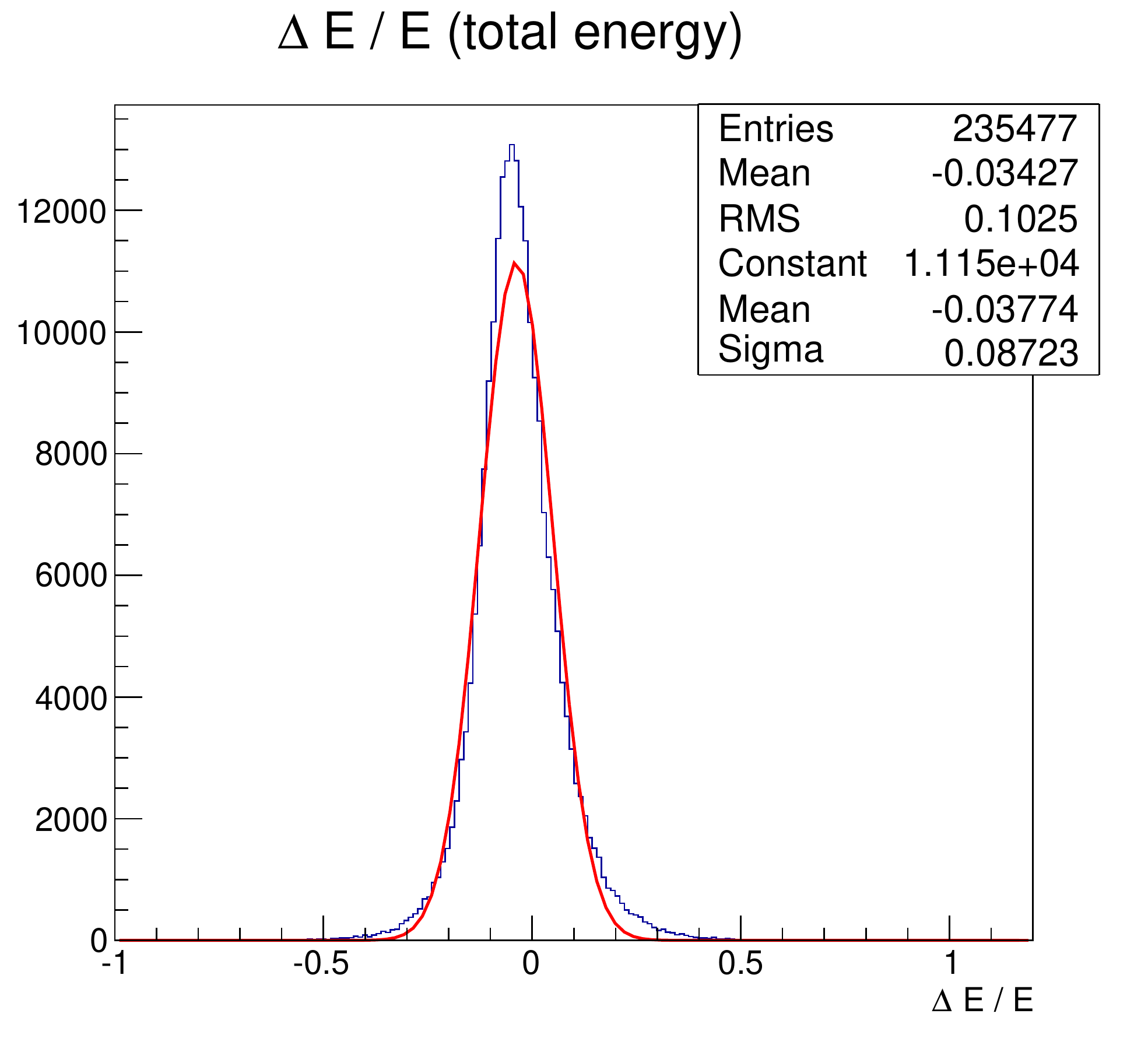}
\includegraphics[height=1.45in]{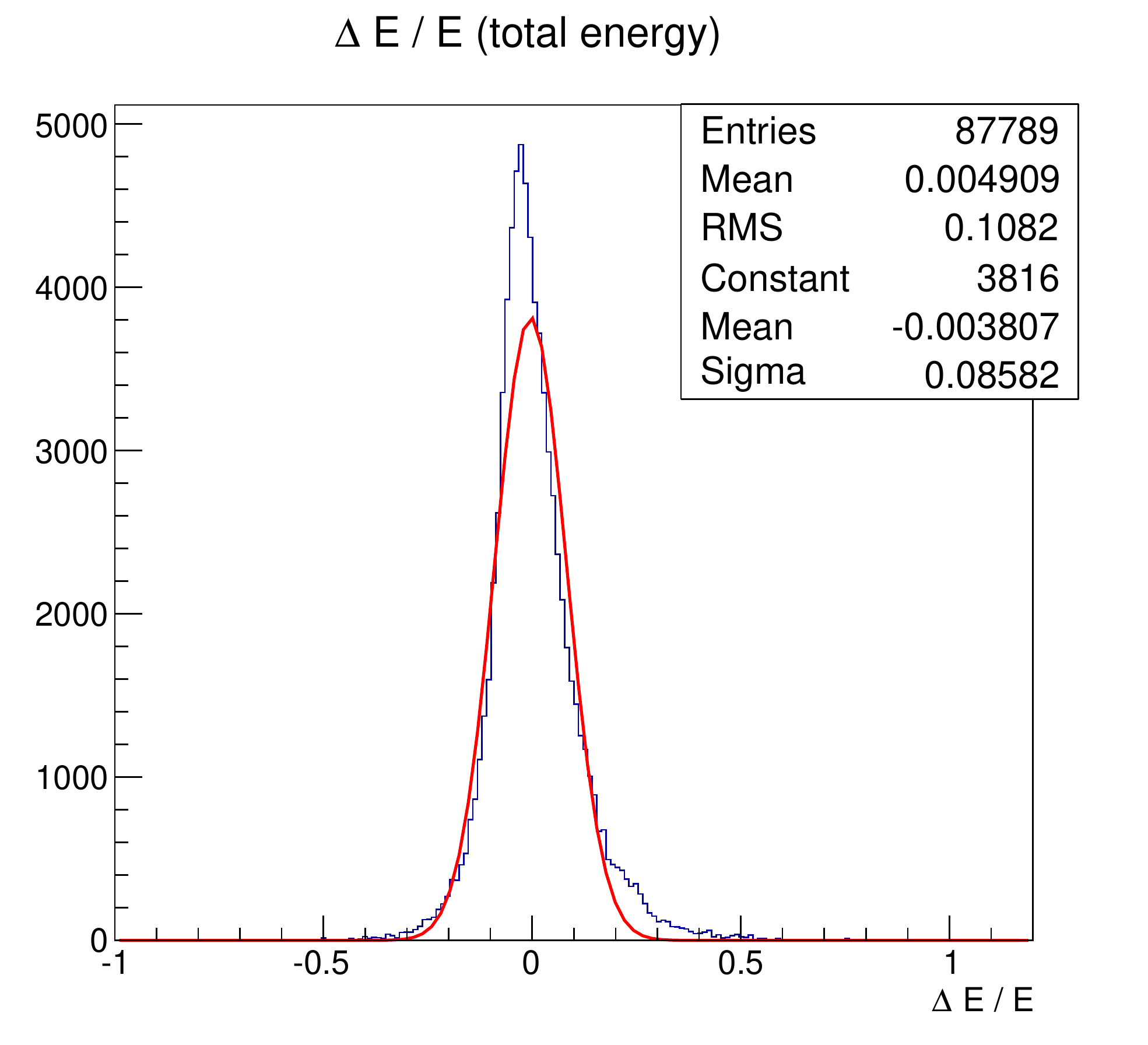}
\includegraphics[height=1.45in]{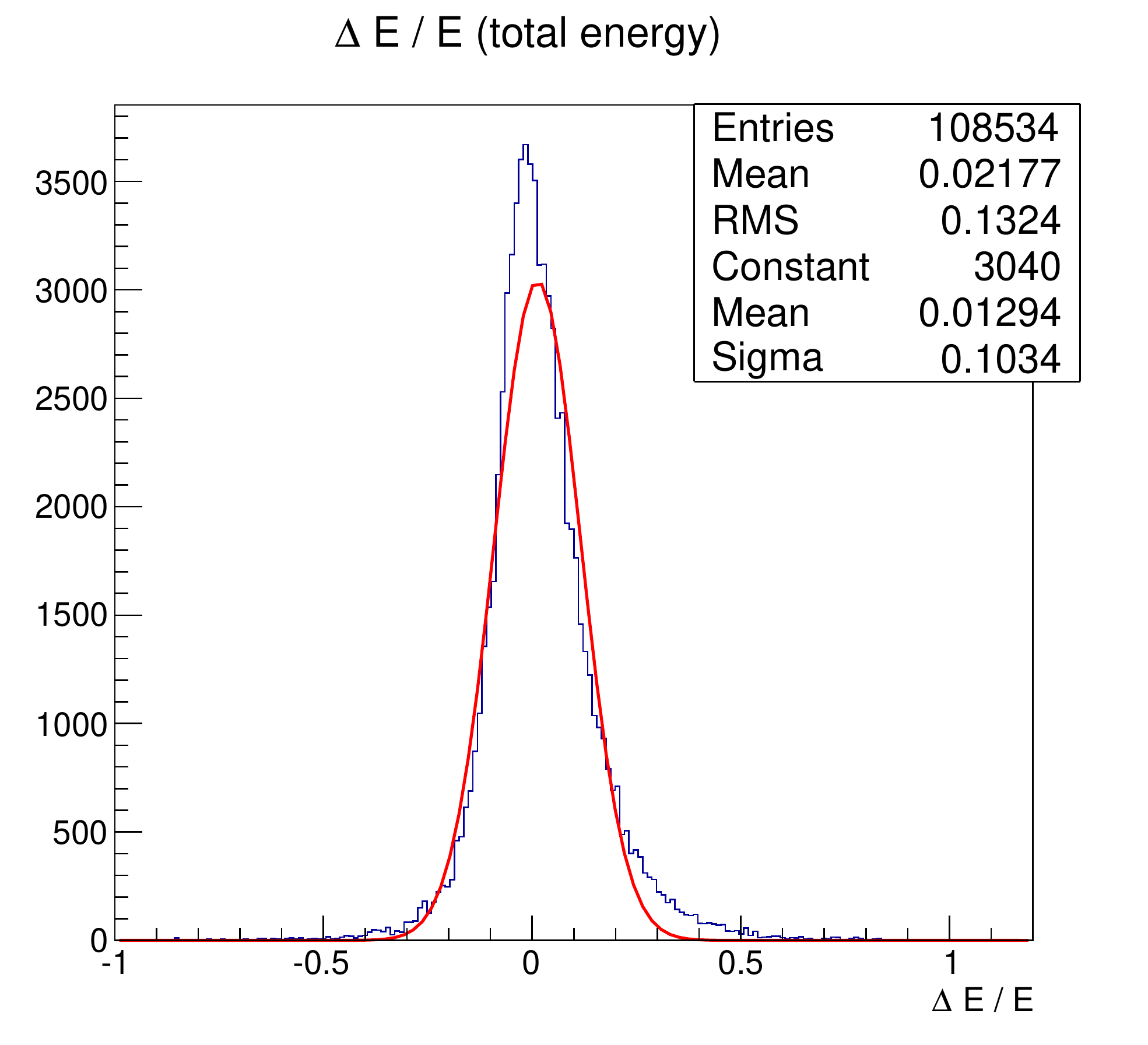}
\caption{Shower energy resolution after all quality cuts are applied.
  From the left: Cherenkov (all energies), Cherenkov ($E > 10^{16.7}$~eV), Mixed, 
  and fluorescence events subsets.  
  MC using a mixed composition, matching TXF results,} 
\label{fig:res_plots_spectral_set_4}
\end{figure}

Figures~\ref{fig:res_plots_spectral_set_1}~-~\ref{fig:res_plots_spectral_set_4}
show the difference between the reconstructed and thrown values of
simulated events, for these four key variables.  To understand how the
reconstruction and quality cuts affect the different types of events,
we show for each figure four resolution plots: Cherenkov (all energies), 
Cherenkov ($E > 10^{16.7}$~eV), Mixed signal, and fluorescence events separately.

For the fluorescence events in 
figures~\ref{fig:res_plots_spectral_set_1}~-~\ref{fig:res_plots_spectral_set_4}, 
we obtain resolutions of $\sim5^{\circ}$ for $d\psi$, $\sim5\%$ for 
$dR_p/R_p$, and $\sim60$~g/cm$^2$ for $dX_{\rm max}$.  These resolutions 
are comparable to those obtained for profile-constrained monocular 
reconstruction in previous studies~\cite{Abbasi:2002ta}.  However, the 
corresponding resolutions for mixed events and Cherenkov events 
($E>10^{16.7}$~eV) are much better than for fluorescence: $\sim1^{\circ}$ for 
$d\psi$, 2-3\% for $dR_p/R_p$, and $\sim35$~g/cm$^2$ for $dX_{\rm max}$.  The 
improvement is primarily due to the tight constraint in the reconstructed 
orientation of the shower from the strong forward peaking of direct 
Cherenkov light.  For the whole Cherenkov set, which is dominated by 
those events below $10^{16.7}$~eV, we see that $d\psi$, $dR_p/R_p$ and 
$dX_{\rm max}$ increase to $\sim1.6^{\circ}$, $\sim0.10$ and $\sim45$~g/cm$^2$, 
mostly as the result of shorter track lengths and event durations at 
the lowest energies.  In all cases, the bias in the reconstruction 
are negligible compared to the resolutions.

Figure~\ref{fig:res_plots_spectral_set_4} shows the resolutions for 
the reconstructed energy.  Here we see $\sim$~16\% for $dE/E$ in
all Cherenkov events, and $\sim$~9\% in Cherenkov events 
with $E>10^{16.7}$~eV and in Mixed events.  $dE/E$ only increases slightly to about 10\% for 
the case of fluorescence events.  In the last case, we manage to obtain 
a good energy resolution in spite of relatively large $d\psi$ and 
$dX_{\rm max}$, as we had seen in previous application of the 
profile-constrained monocular fit~\cite{Abbasi:2002ta}.  The reason for this is 
that reconstruction errors in $\psi$ and $X_{\rm max}$ are highly correlated 
along a direction orthogonal to errors in the reconstructed
$R_p$ and energy.  For the full range of the data set, we also see 
negligible bias in the reconstruction energy values.  These results show 
that our reconstruction procedure is clearly adequate for the spectrum 
study presented in this paper.

\begin{figure}[htb]
\centering
\includegraphics[height=1.45in]{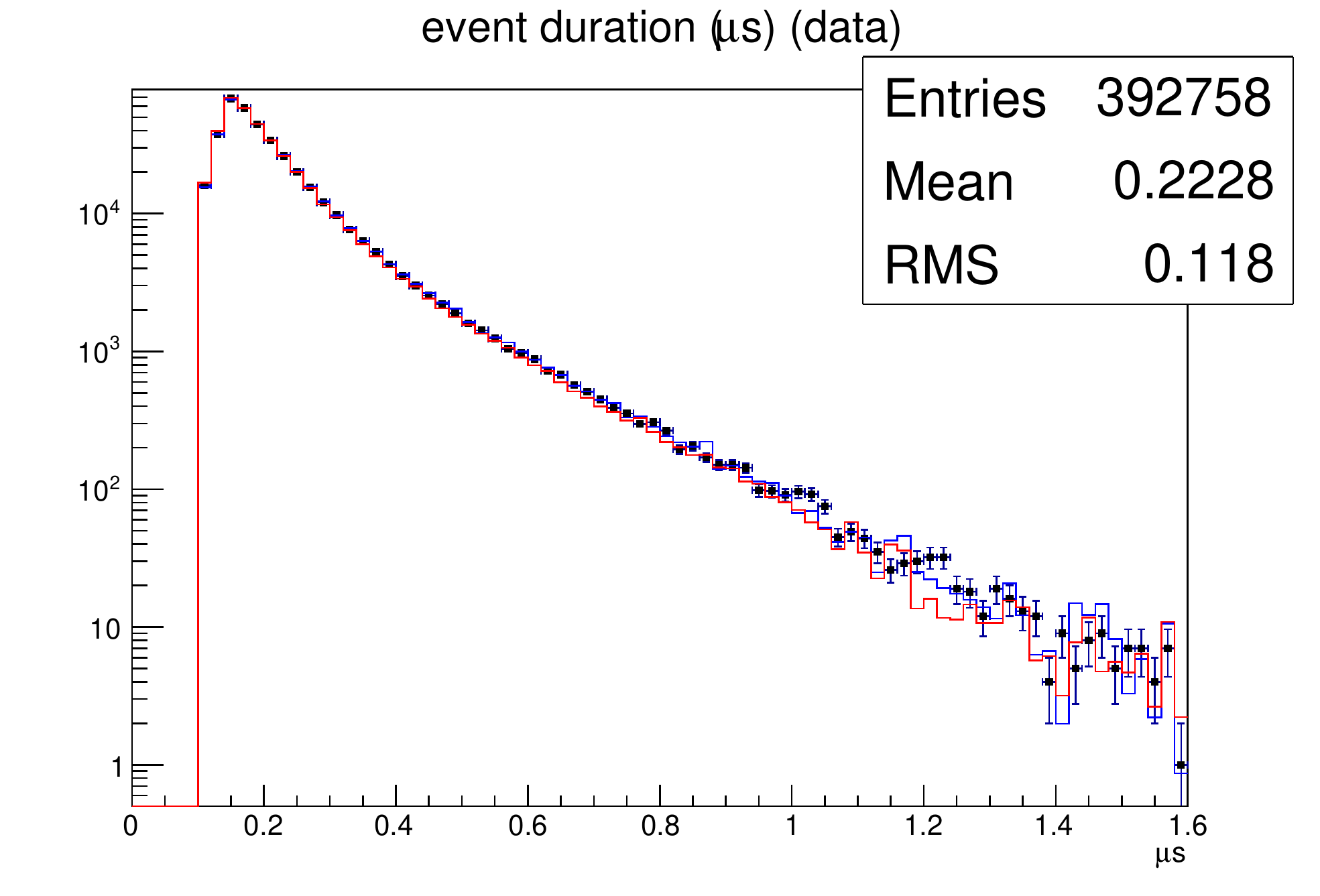}
\includegraphics[height=1.45in]{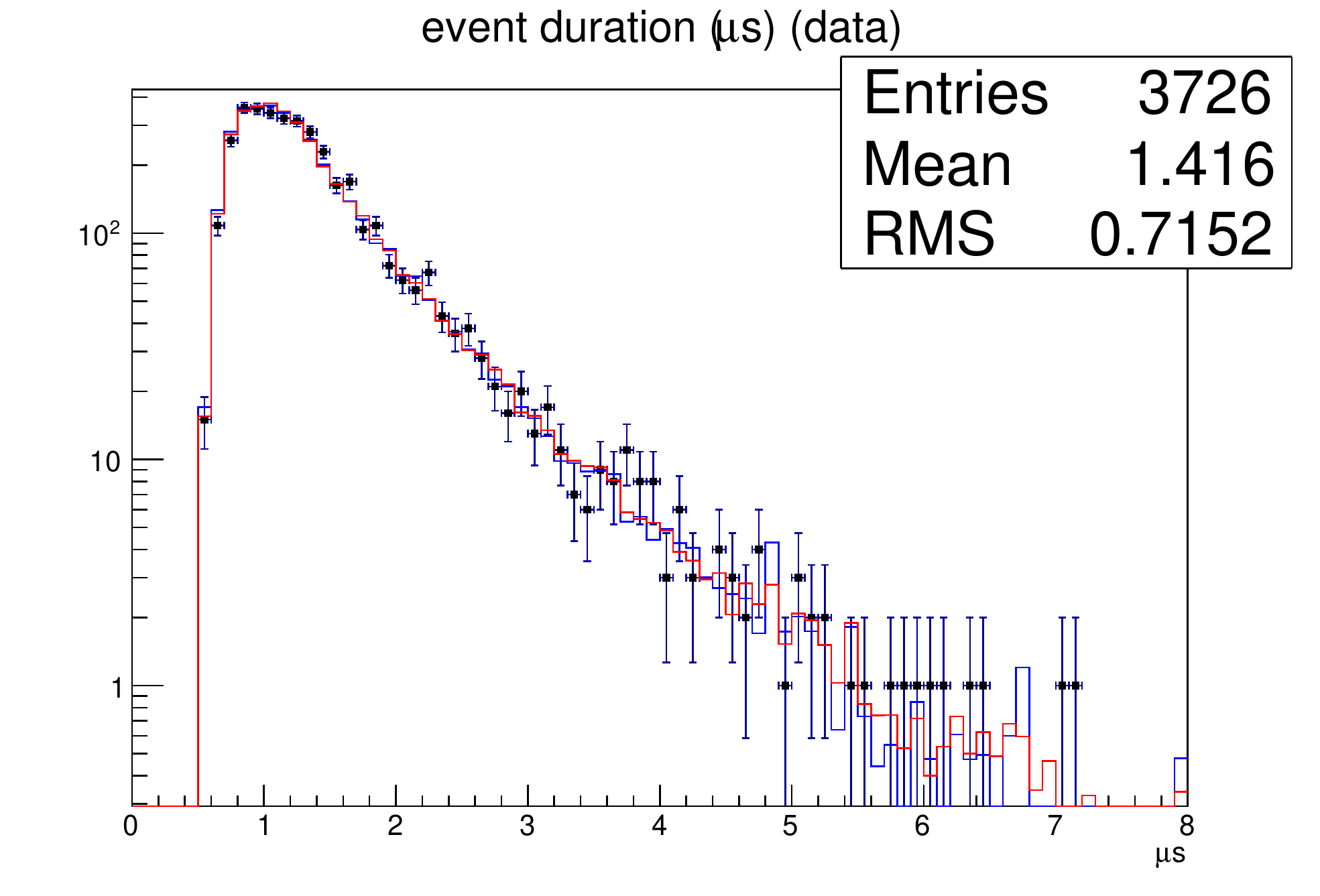}
\includegraphics[height=1.45in]{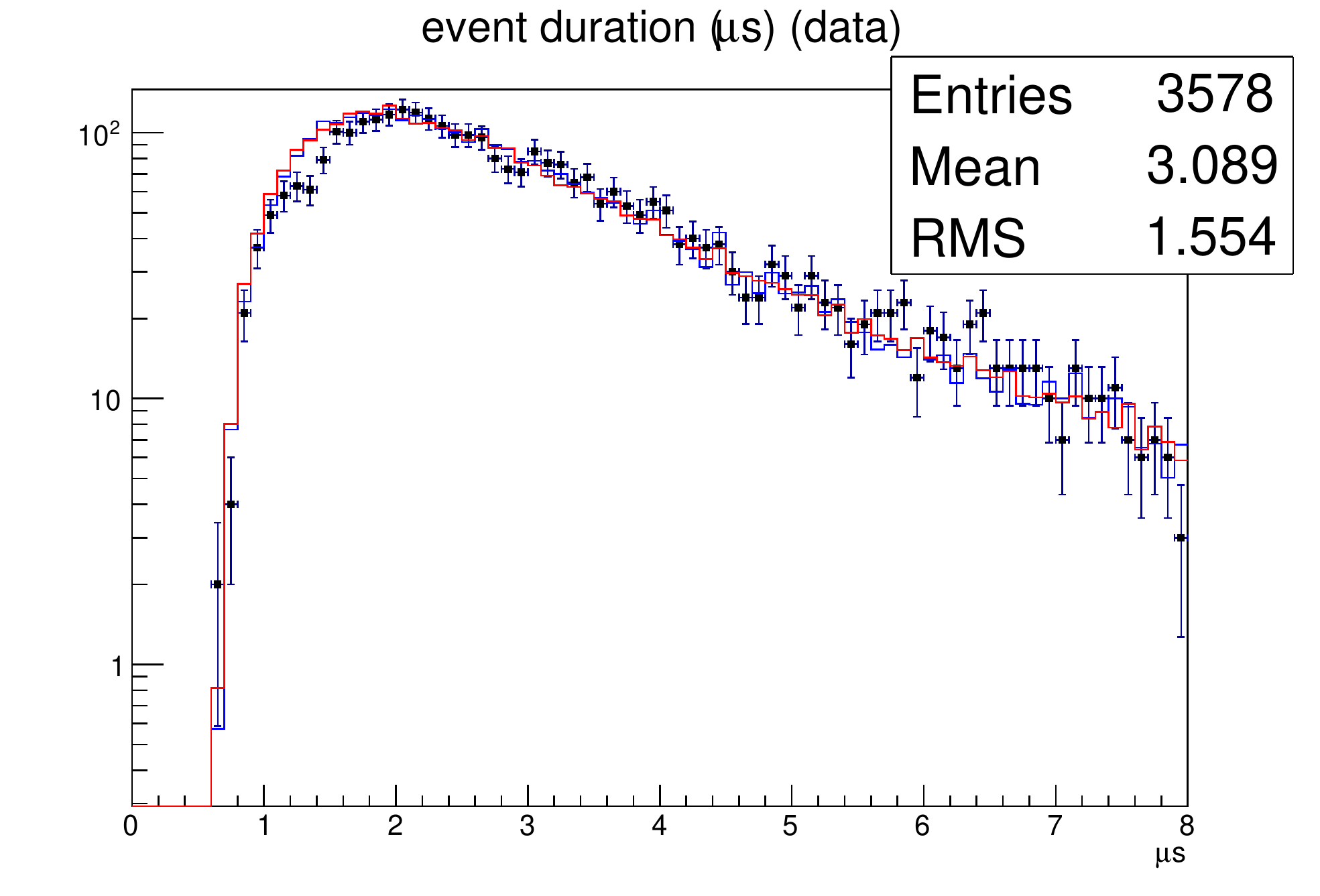}

\caption{Total event duration ($\mu s$), for Cherenkov (left), Mixed
  (center), and fluorescence events (right).  Black points are data,
  blue / red histograms are MC with mixed composition (H4a / TXF
  respectively).}
\label{fig:data_mc_duration}
\end{figure}

\begin{figure}[htb]
\centering
\includegraphics[height=1.45in]{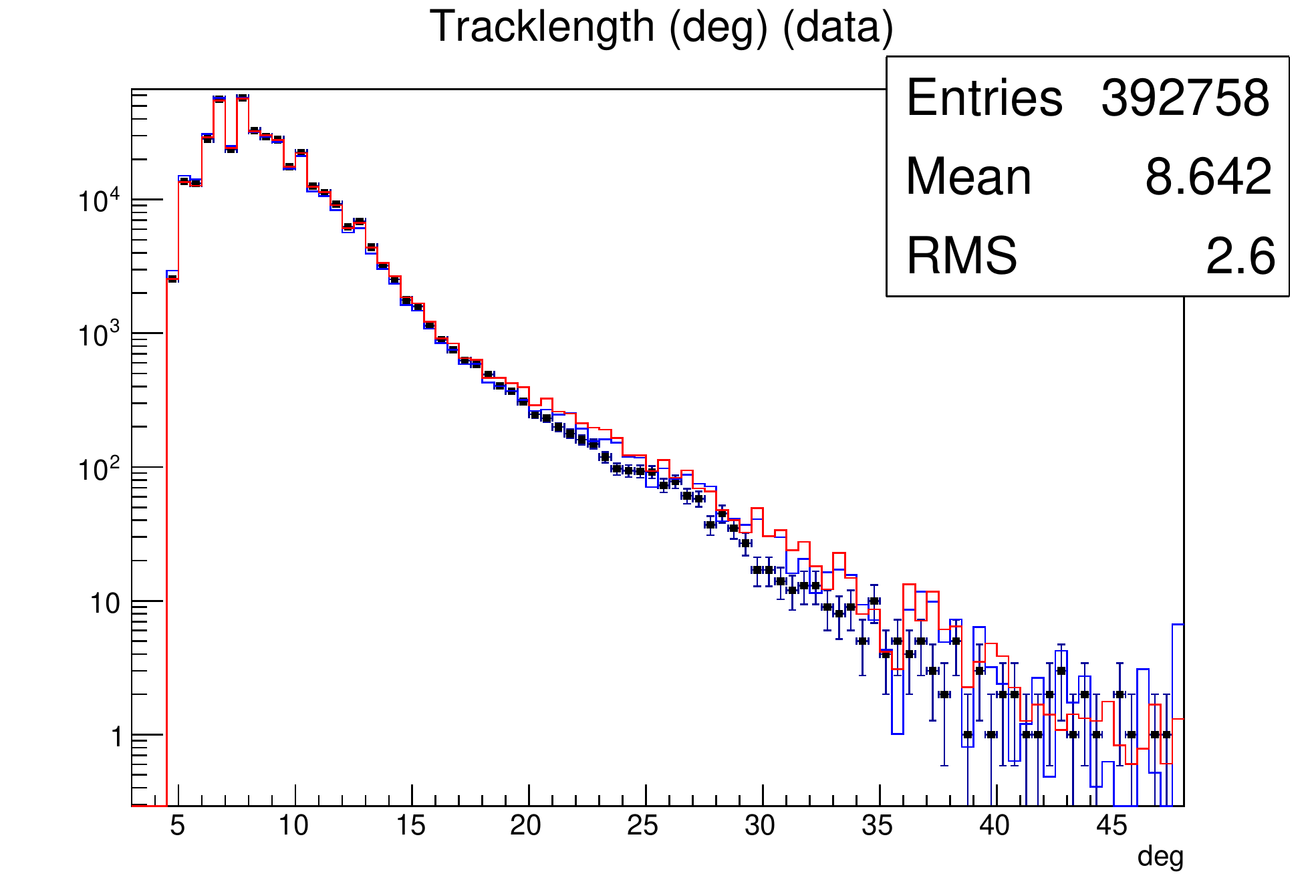}
\includegraphics[height=1.45in]{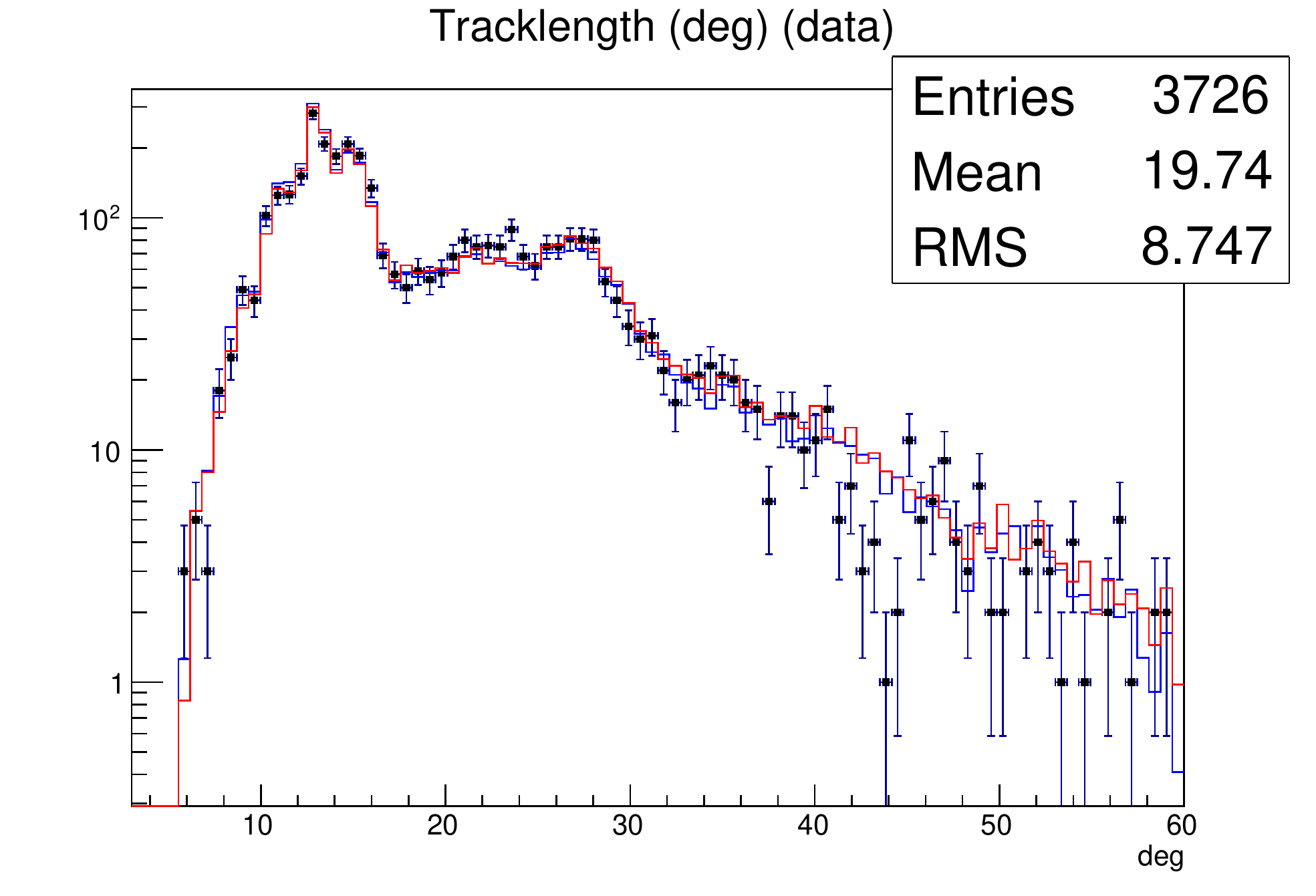}
\includegraphics[height=1.45in]{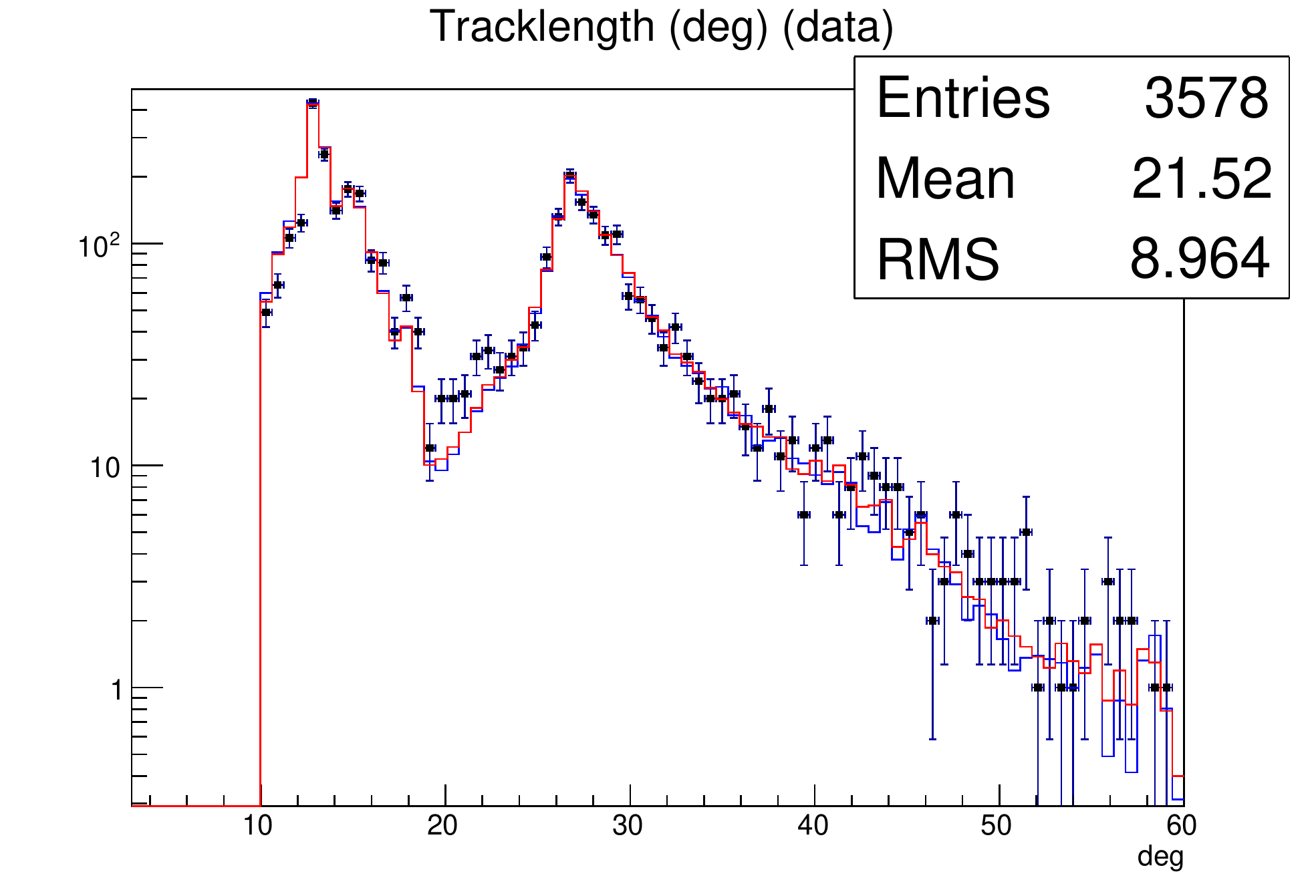}
\caption{Angular track-length (deg), for Cherenkov (left), Mixed
  (center), and fluorescence events (right).  Black points are data,
  blue / red histograms are MC with mixed composition (H4a / TXF
  respectively).}
\label{fig:data_mc_tracklength}
\end{figure}

The analysis depends on the MC simulation in two critical ways: (a)
The event quality cuts were chosen according to their effect on
the reconstruction resolution for simulated events, and (b) the
aperture of the detector is calculated from the acceptance of the
simulated events through the reconstruction and quality cuts.  It is
therefore important that the simulation gives an accurate account of
the data.  In Figure~\ref{fig:data_mc_duration}
and~\ref{fig:data_mc_tracklength} we show the comparisons between
distribution from the data to those of the MC for the event duration,
and for the track length. The MC distribution in each case has been
normalized to the number of data events, and the comparisons have been
split between Cherenkov, Mixed, and fluorescence events.
Two MC distributions are shown in each plot, one using the thrown MC
distribution (H4a composition), and the other being the re-weighted
distribution resulting from the TALE $X_{\rm max}$ distributions fits (TXF). 
All of these plots show good agreement between data and MC.

\begin{figure}[htb]
\centering
\includegraphics[height=1.45in]{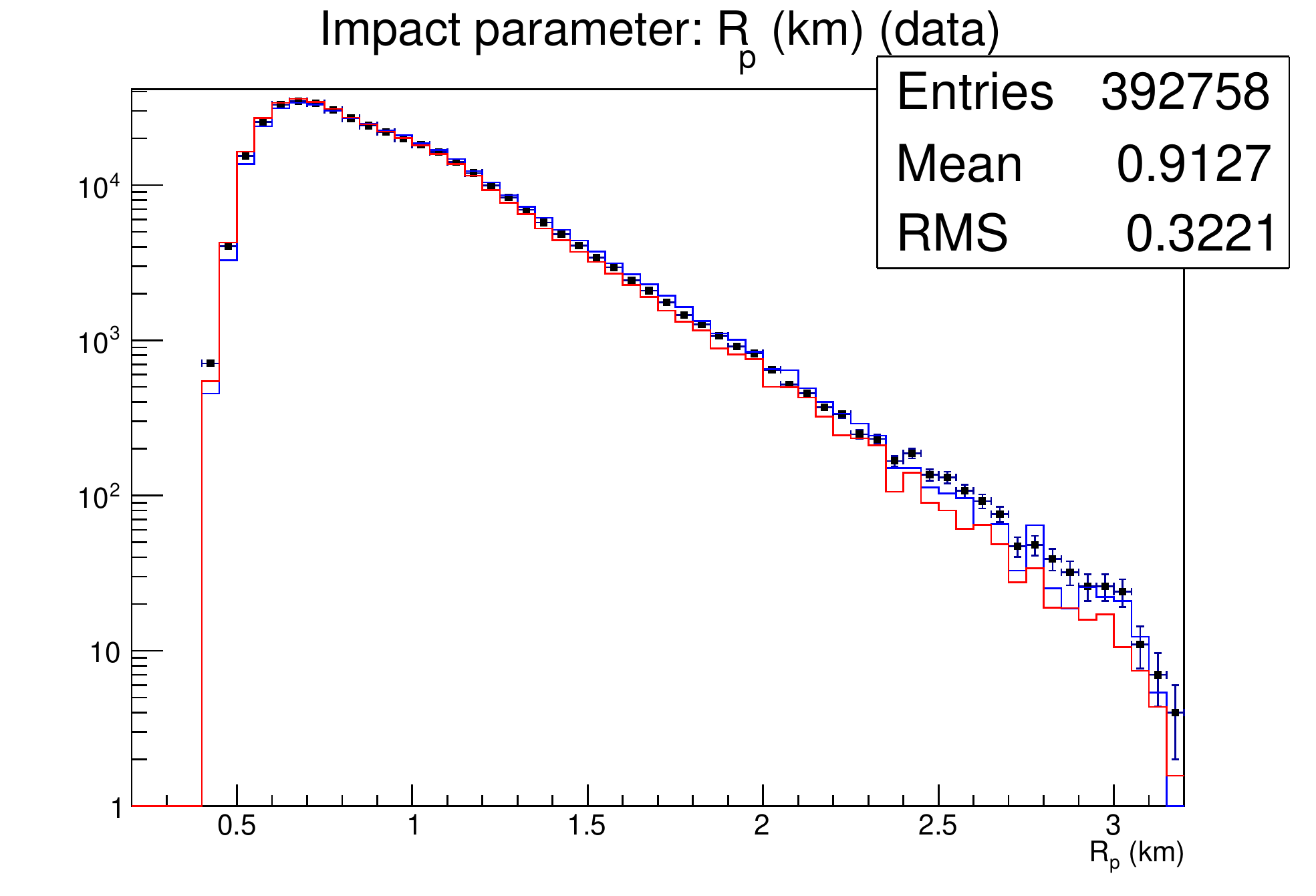}
\includegraphics[height=1.45in]{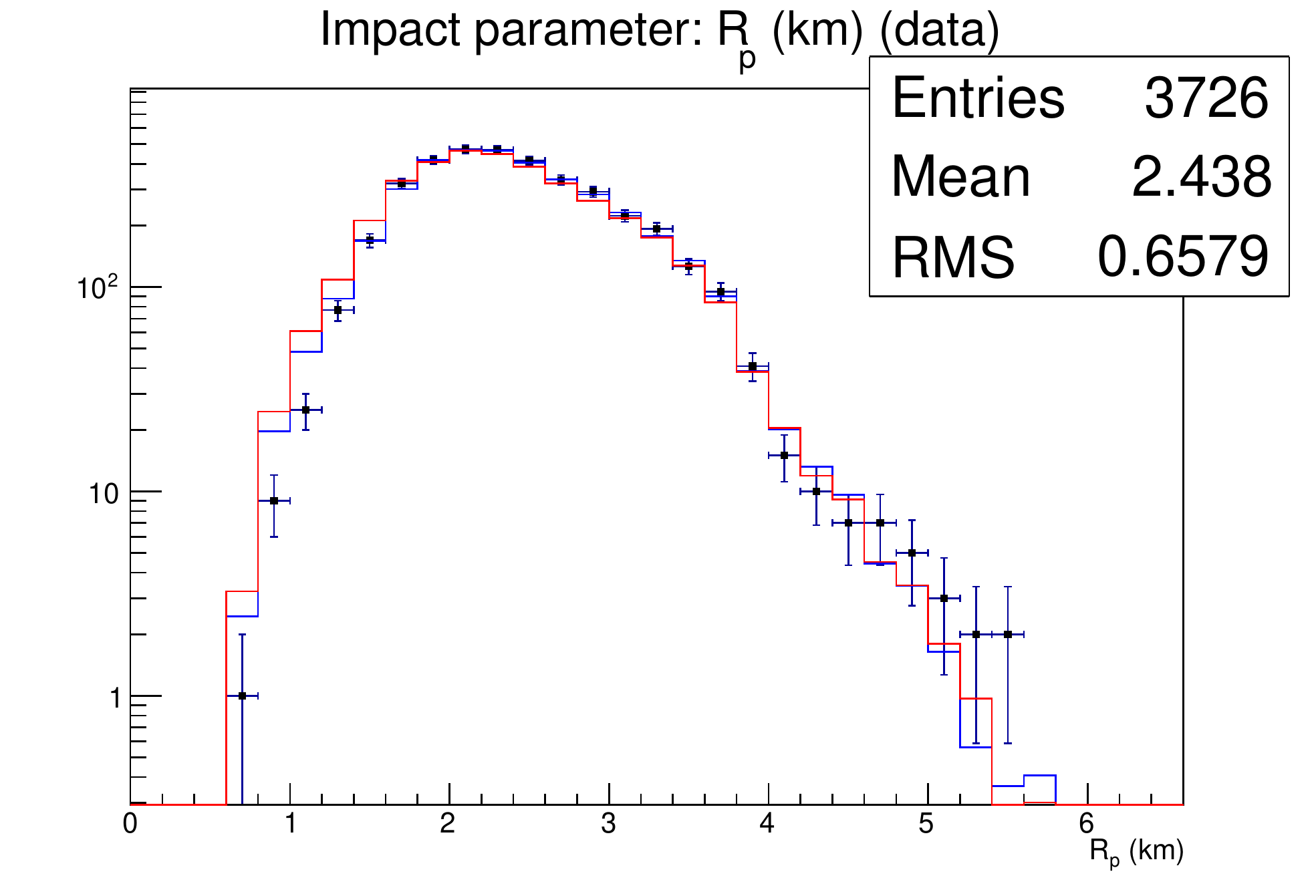}
\includegraphics[height=1.45in]{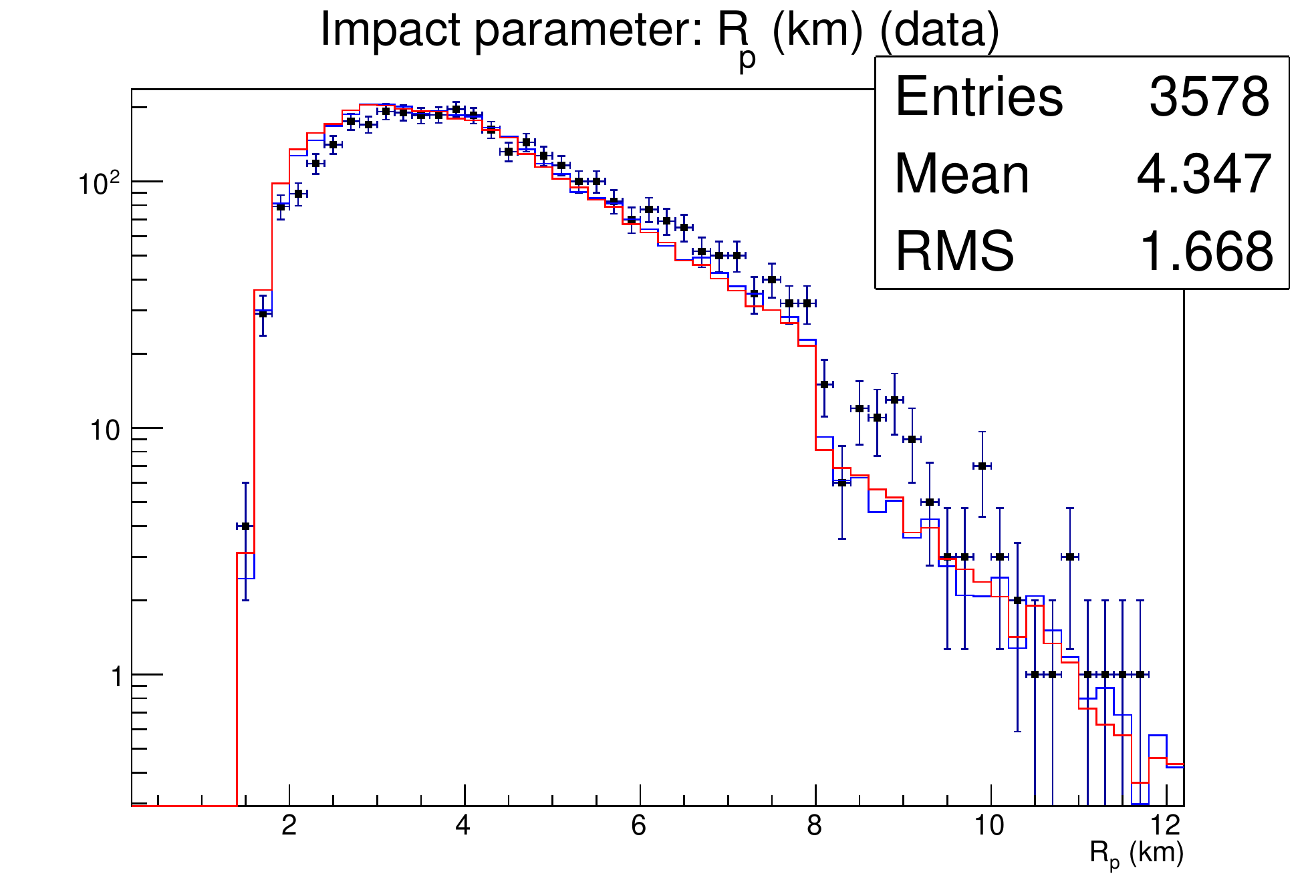}
\caption{Shower impact parameter (km), for Cherenkov (left), Mixed 
  (center), and fluorescence events (right).  Black points are data, 
  blue / red histograms are MC with mixed composition (H4a / TXF 
  respectively).}
\label{fig:data_mc_rp}
\end{figure}

\begin{figure}[htb]
\centering
\includegraphics[height=1.45in]{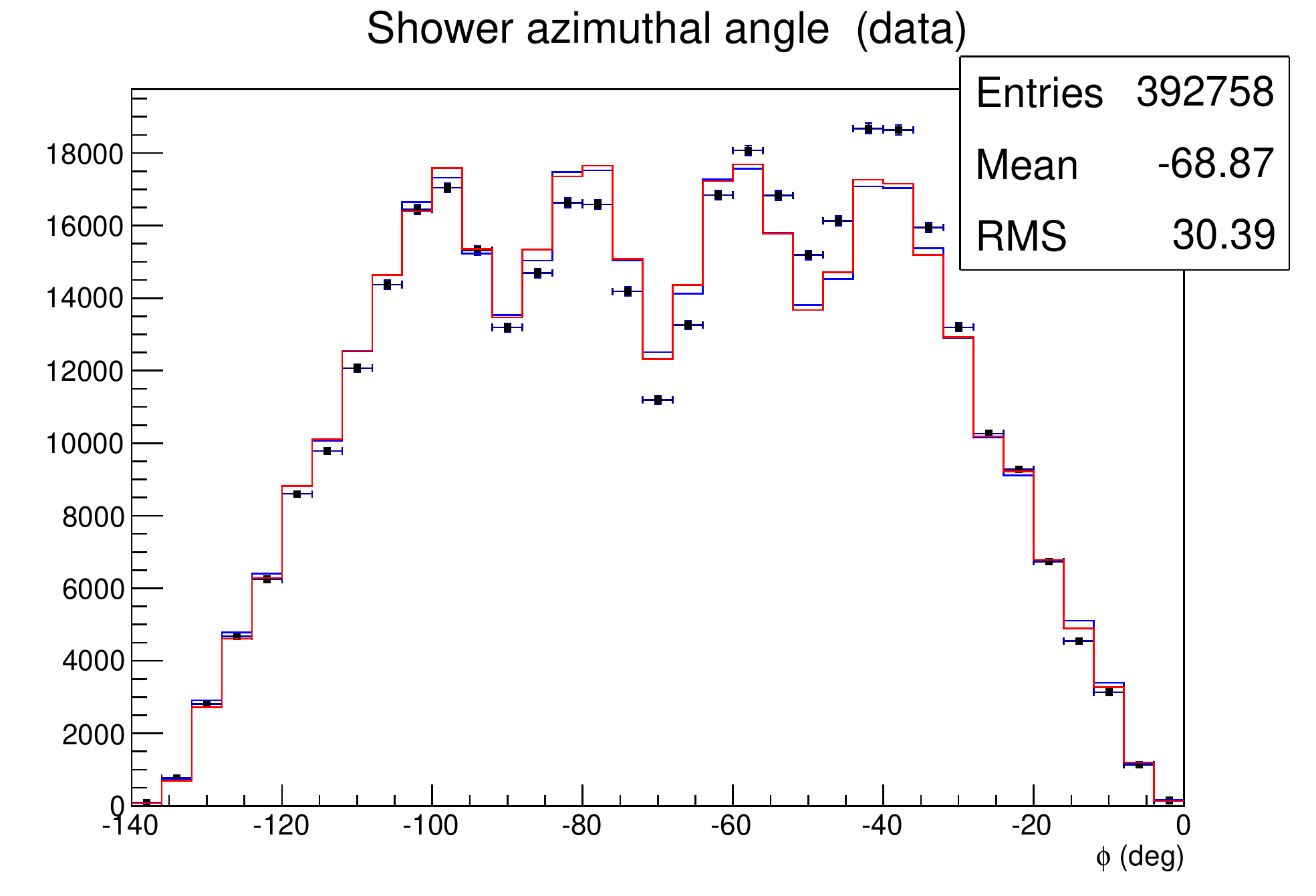}
\includegraphics[height=1.45in]{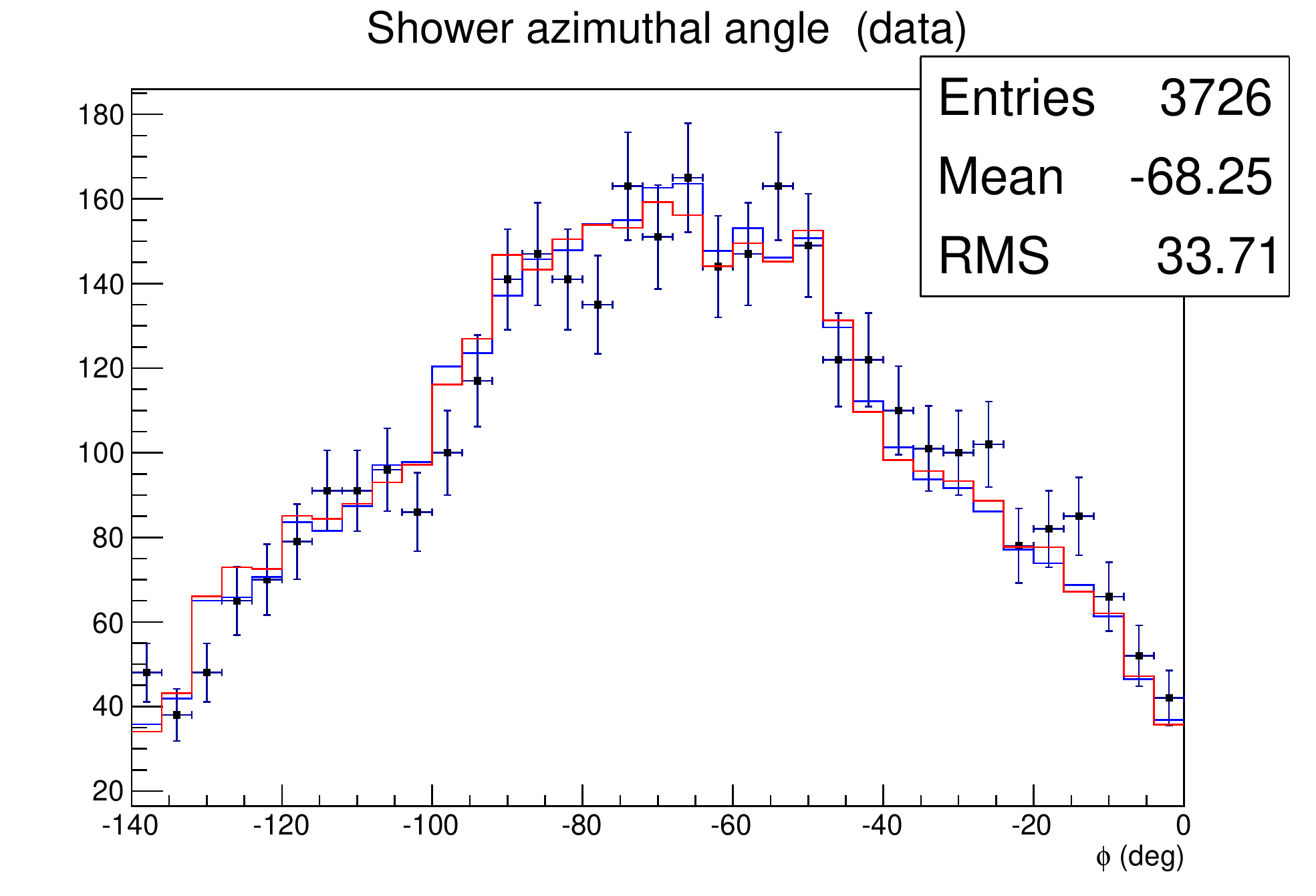}
\includegraphics[height=1.45in]{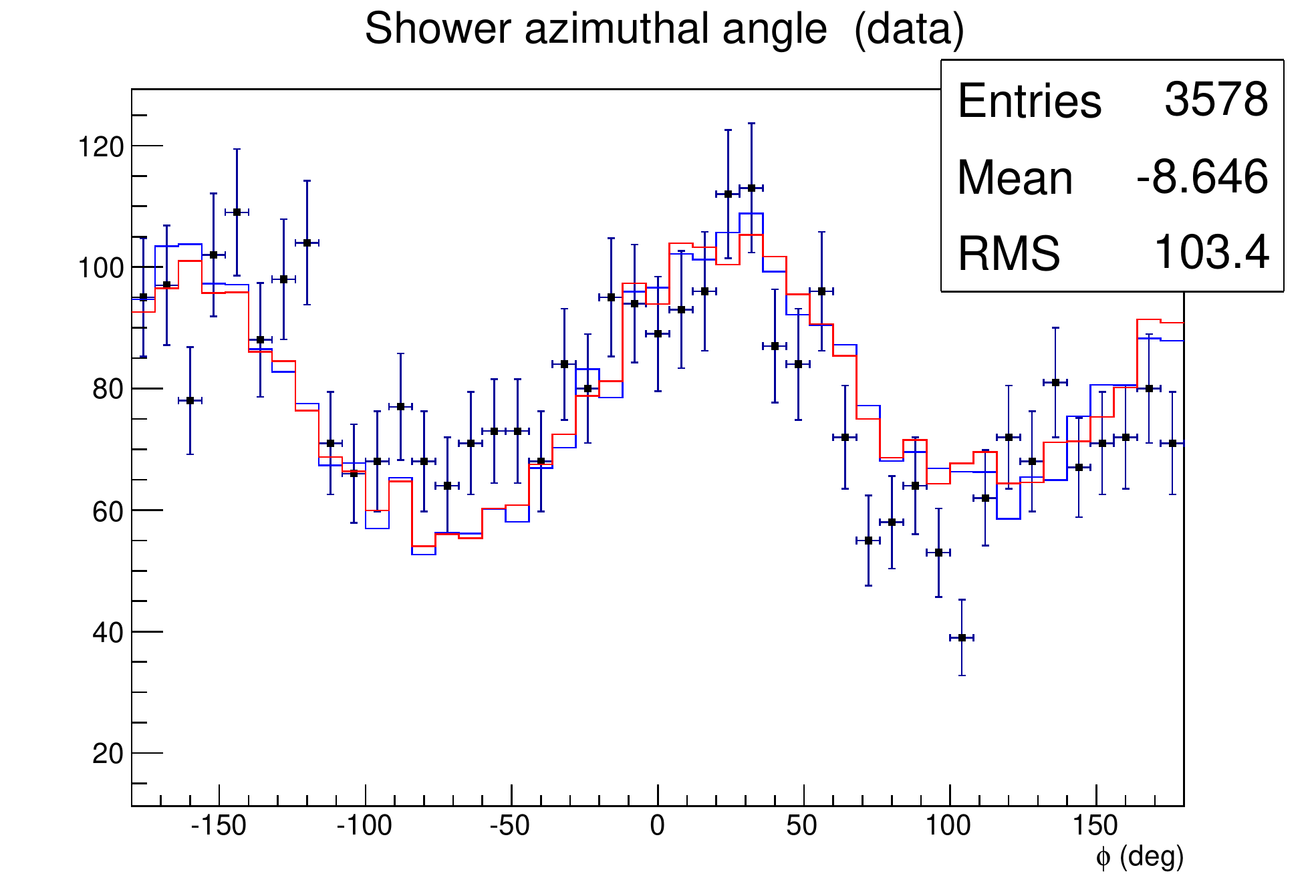}
\caption{Shower zenith angle (deg), for Cherenkov (left), Mixed
  (center), and fluorescence events (right).  Black points are data,
  blue / red histograms are MC with mixed composition (H4a / TXF
  respectively). }
\label{fig:data_mc_phi}
\end{figure}

The aperture calculation essentially measures the detection volume and
angular acceptance of the detector and analysis procedures.  The three
most important parameters in this regard are (a) the impact parameter
$R_{p}$, (b) the depth of shower maximum, $X_{\rm max}$, and (c) the
zenith angle of the shower, $\theta$.  The distributions of $R_{p}$
and $X_{\rm max}$ approximately define the fiducial volume of the atmosphere for
the fluorescence detector, and the zenith angle distribution gives the
acceptance solid angle.  For completeness we also include 
the azimuthal angle of the shower, $\phi$.  
The data-MC comparisons for $R_{p}$, $\theta$, and $\phi$, 
separately for Cherenkov, Mixed, and fluorescence events, are given in
figures~\ref{fig:data_mc_rp}~and~\ref{fig:data_mc_phi}. 
These distributions are again in good agreement between
data and MC.

\begin{figure}[htb]
\centering
\includegraphics[height=1.45in]{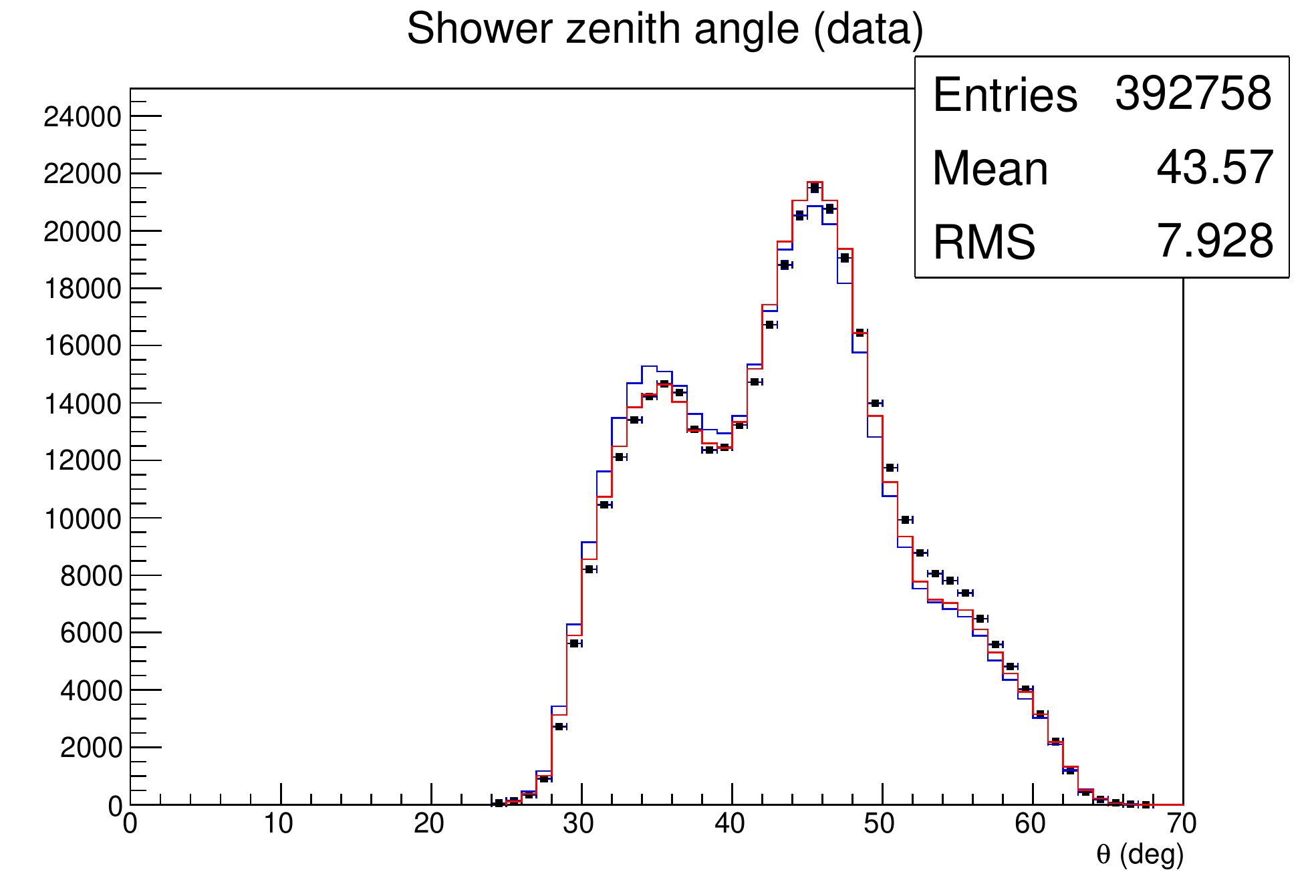}
\includegraphics[height=1.45in]{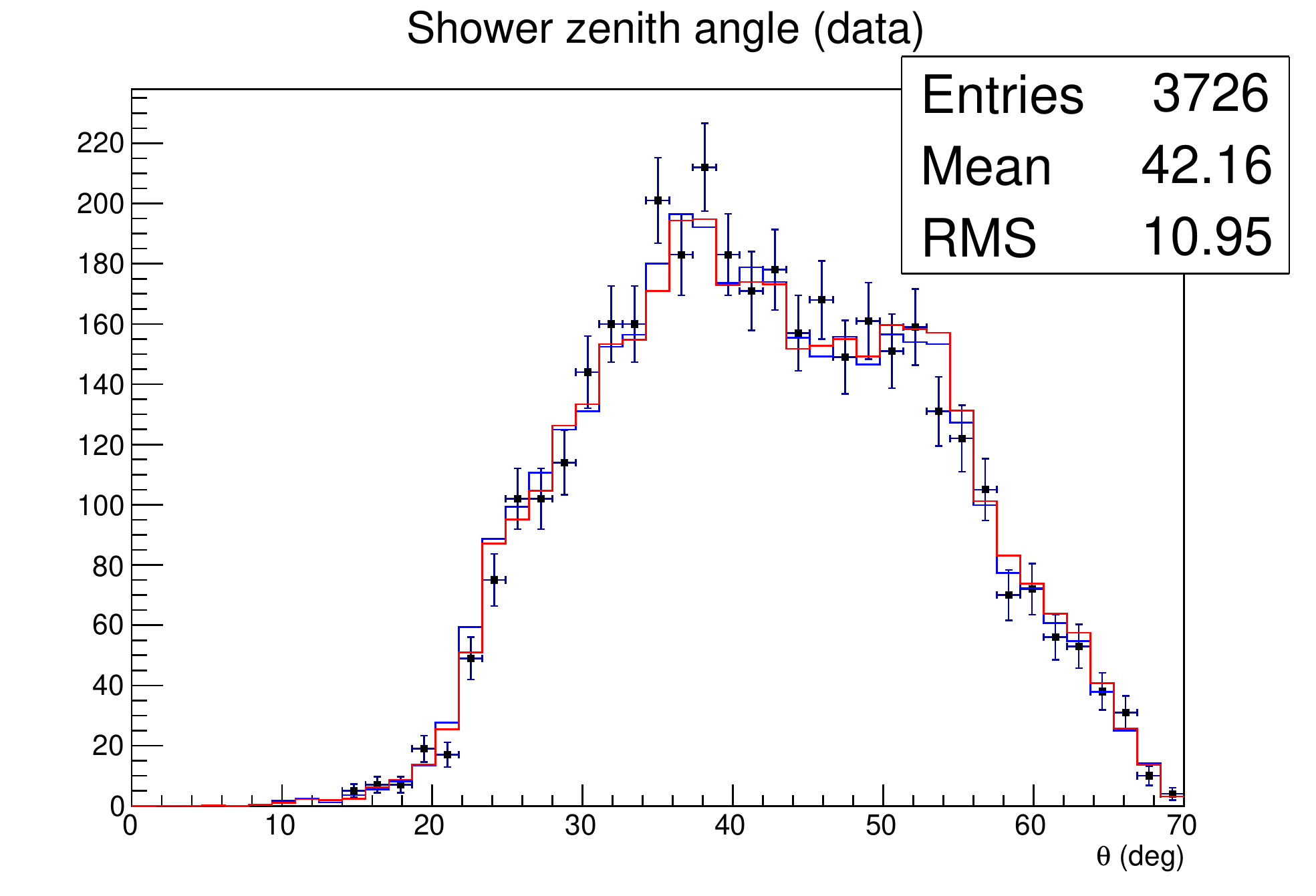}
\includegraphics[height=1.45in]{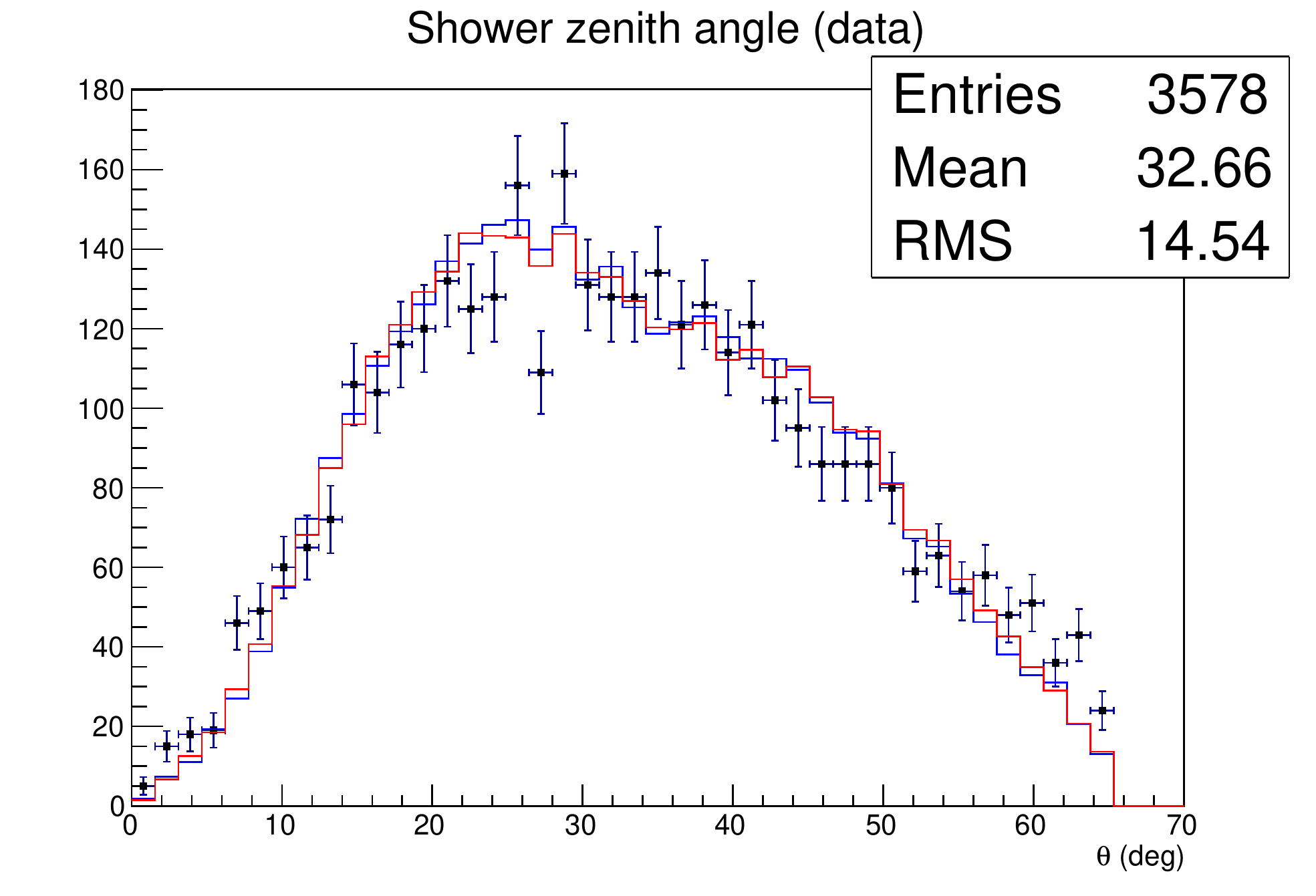}
\caption{Shower zenith angle (deg), for Cherenkov (left), Mixed
  (center), and fluorescence events (right).  Black points are data,
  blue / red histograms are MC with mixed composition (H4a / TXF
  respectively). }
\label{fig:data_mc_theta}
\end{figure}

\begin{figure}[htb]
\centering
\includegraphics[height=2.00in]{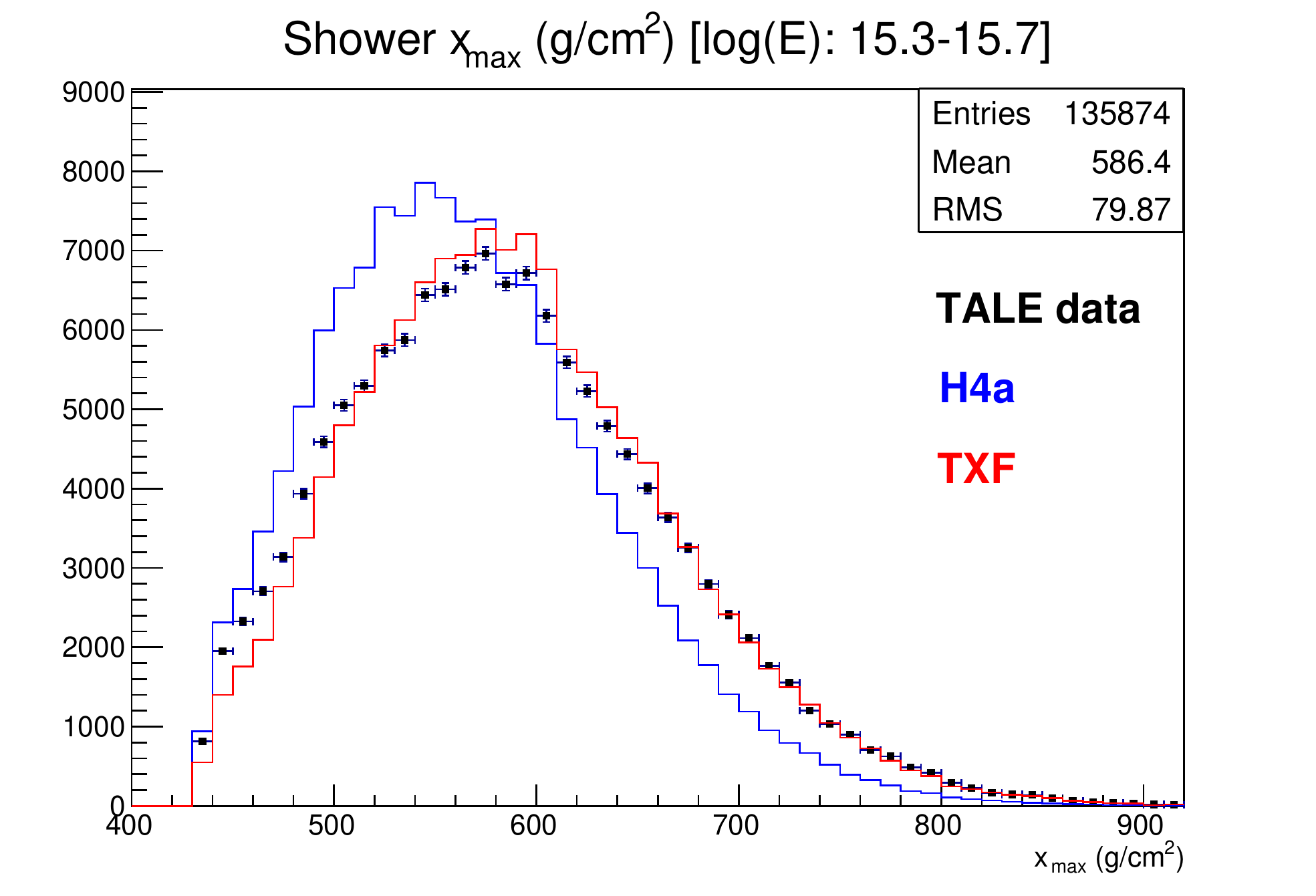}
\includegraphics[height=2.00in]{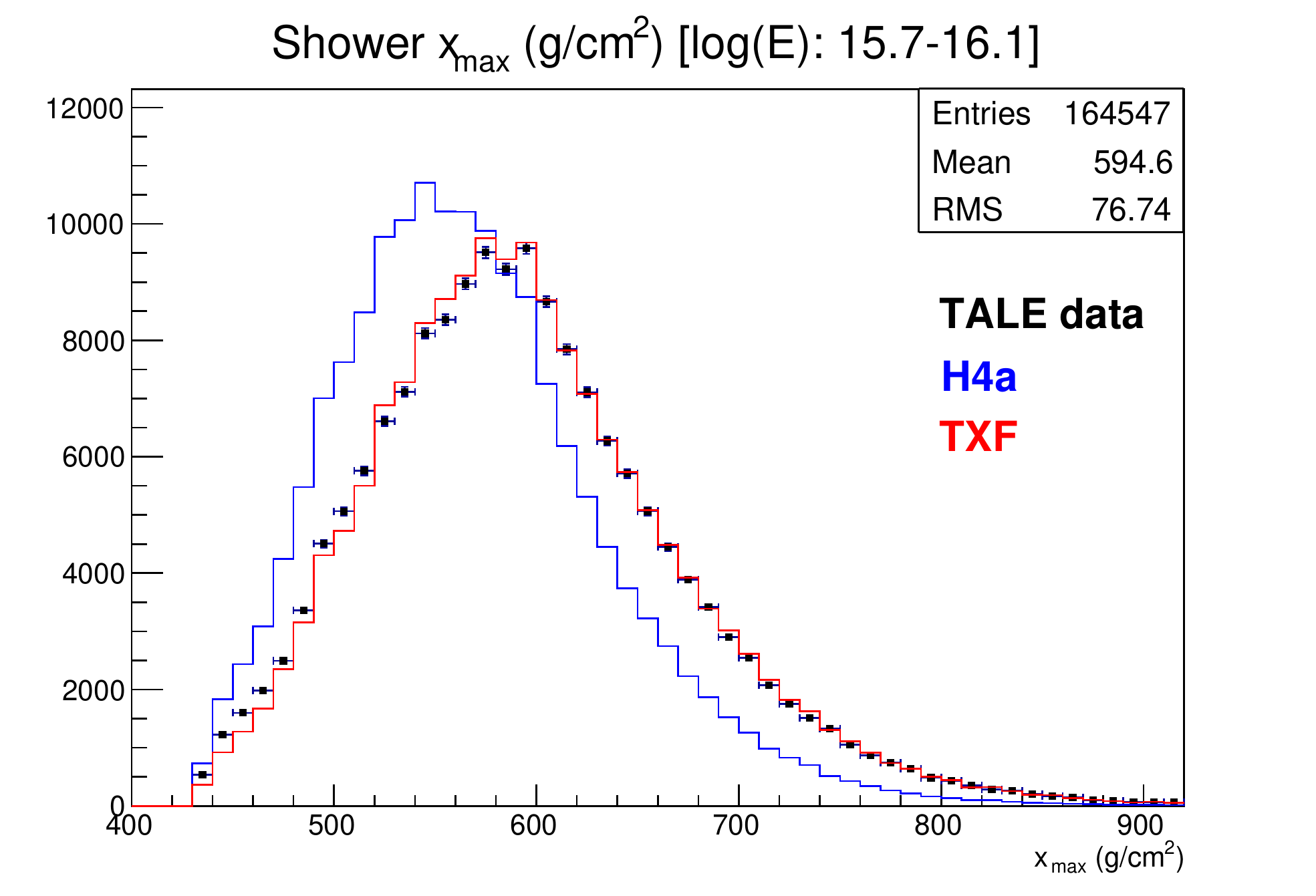}
\includegraphics[height=2.00in]{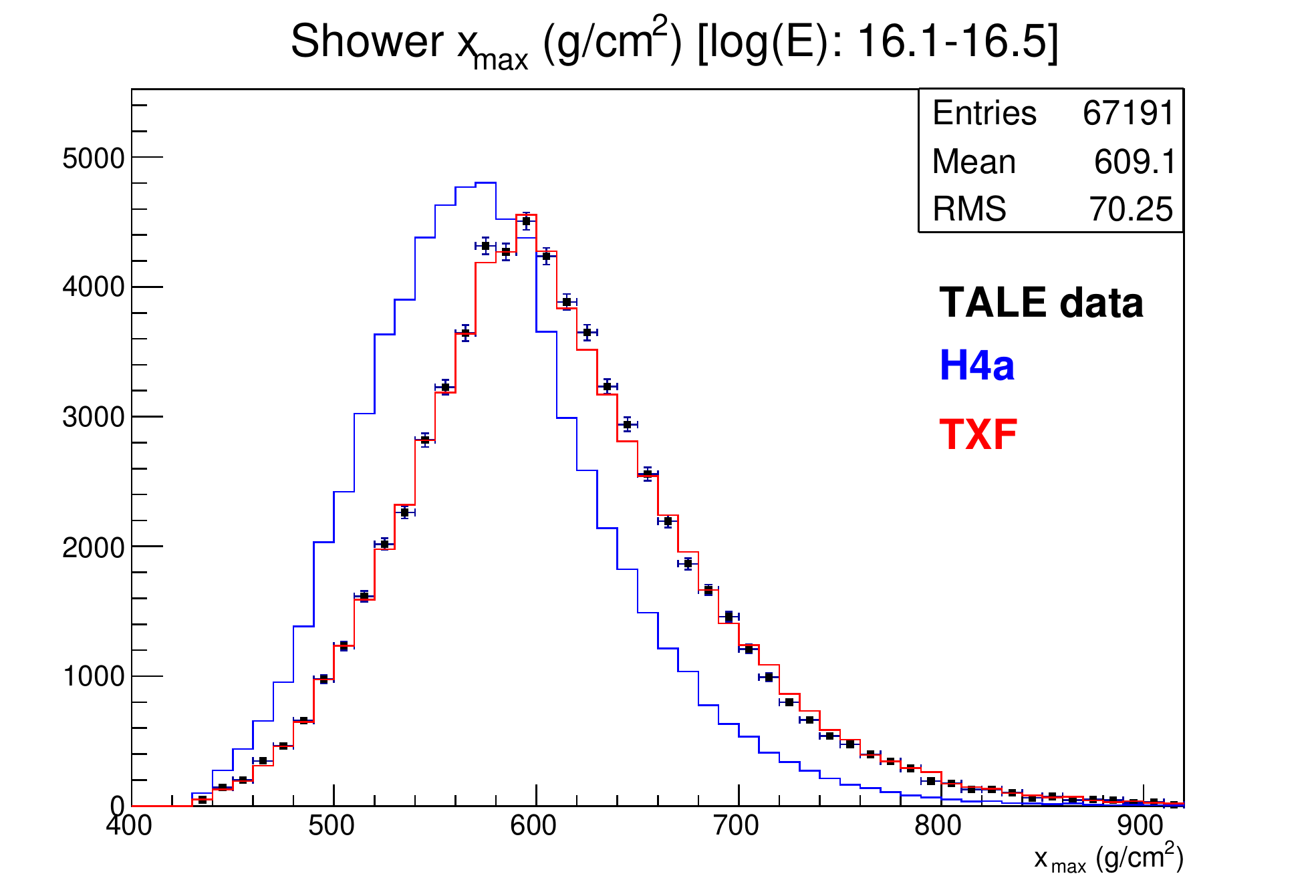}
\includegraphics[height=2.00in]{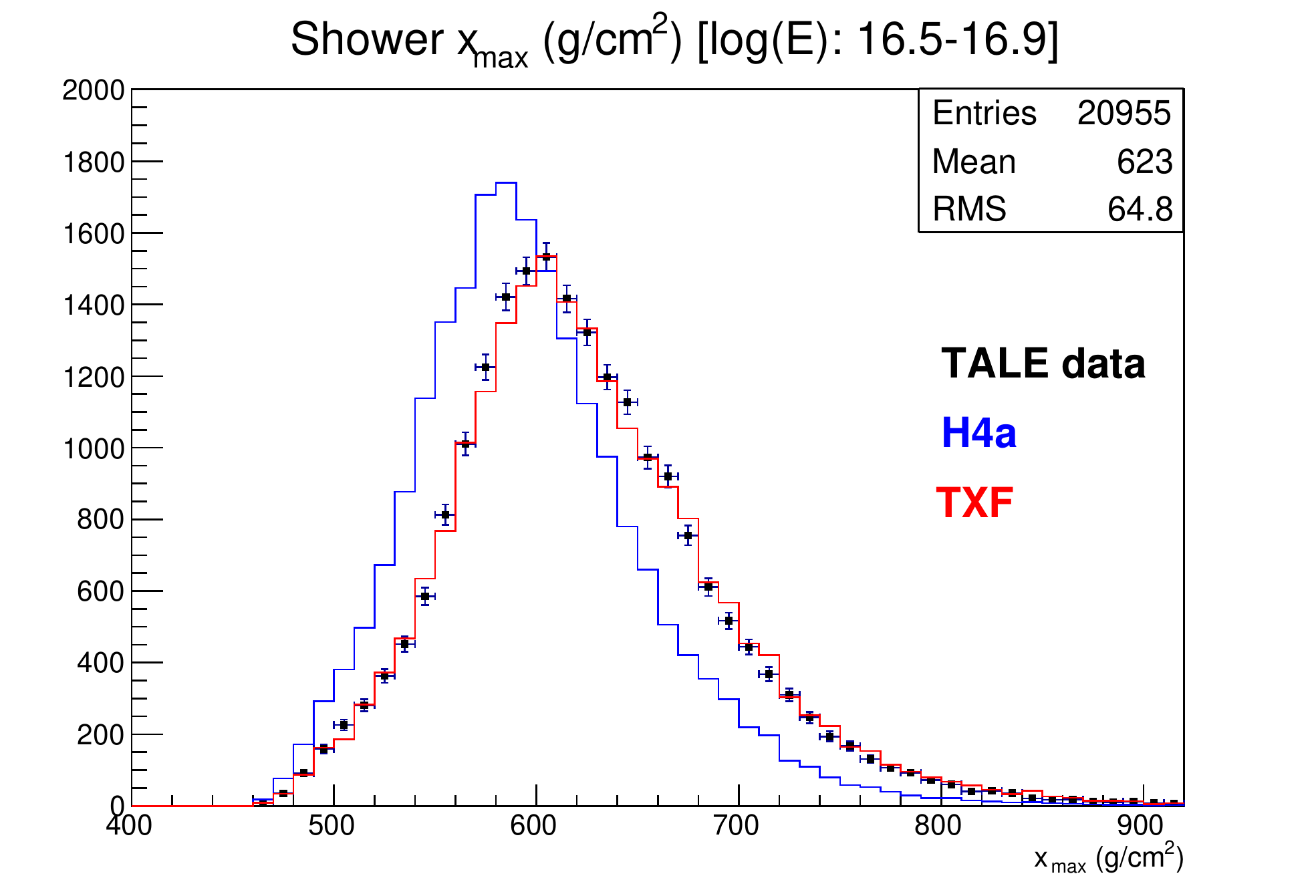}
\includegraphics[height=2.00in]{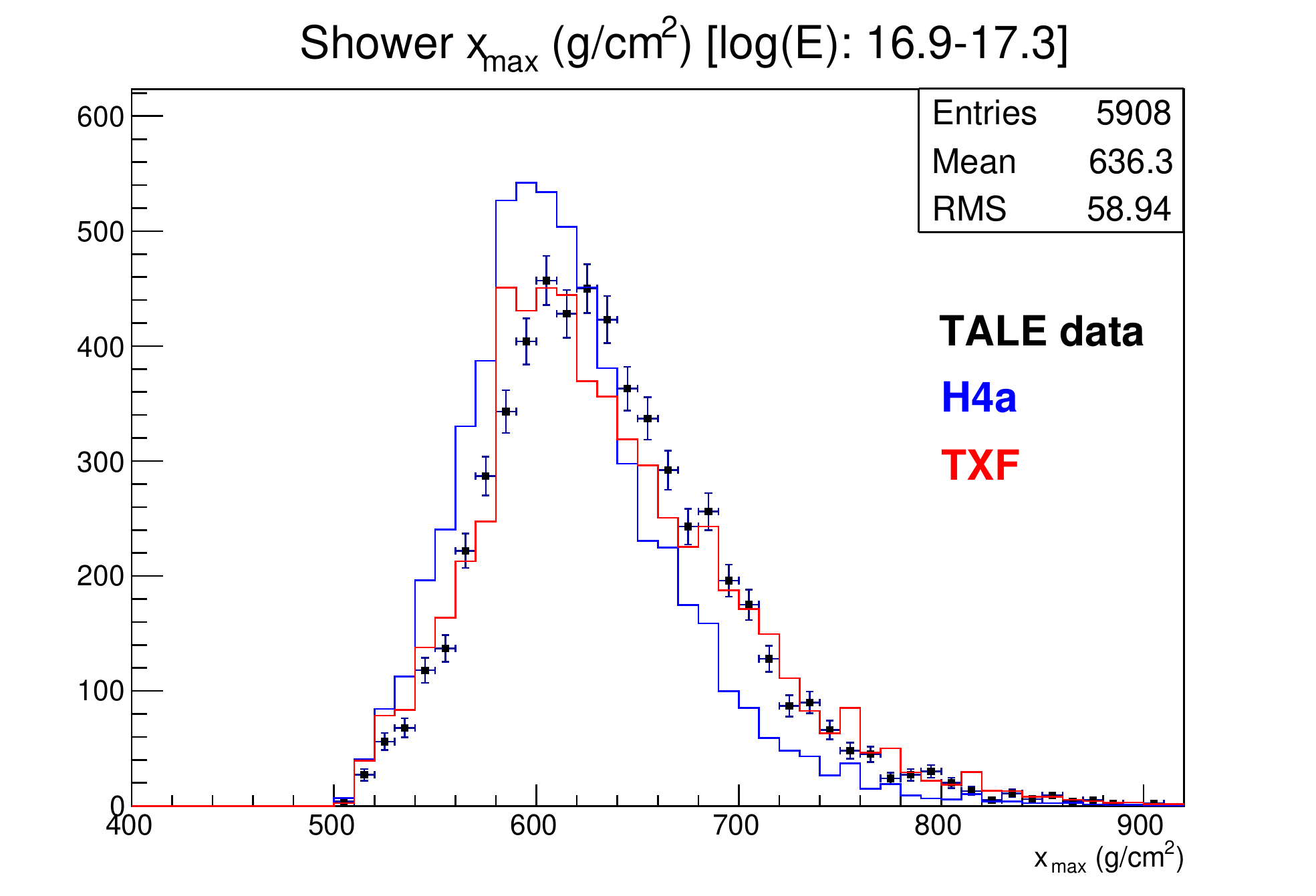}
\includegraphics[height=2.00in]{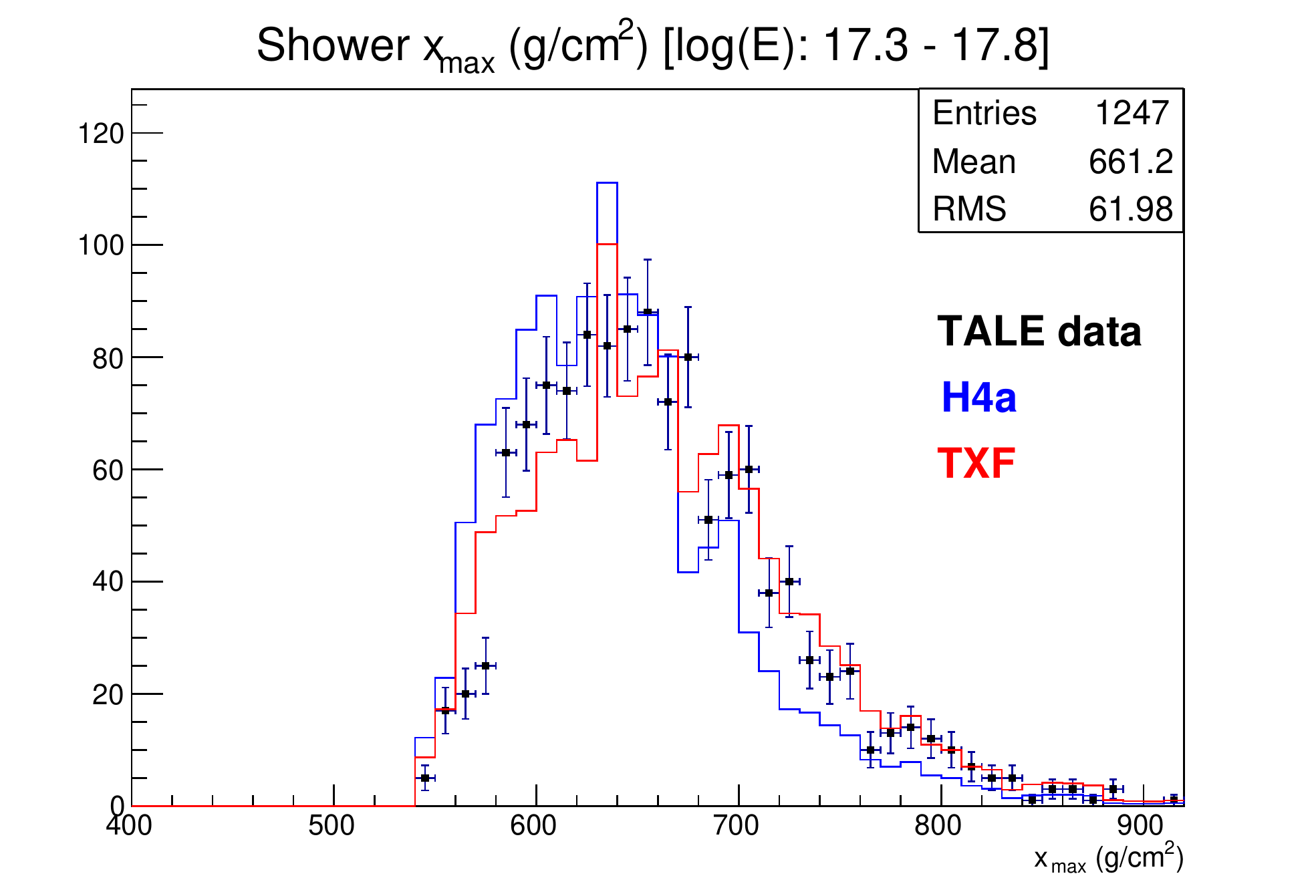}
\caption{Shower $X_{\rm max}$ (g/cm$^{2}$), in different energy ranges.
  Black points are data, blue / red histograms are MC with mixed composition
  (H4a / TXF respectively).  Note that the TXF histogram is a fit to the data.}
\label{fig:data_mc_xmax}
\end{figure}

The $X_{\rm max}$ distributions are shown in different energy ranges in 
figure~\ref{fig:data_mc_xmax}.  Here it should be noted that the TXF
histograms are the results of fitting the $X_{\rm max}$ distributions, and 
therefore ``match'' the data much better than the H4a composition model.
The data used for the TXF only includes Cherenkov events and Mixed events,
it does not include any fluorescence events.

Throughout this section we have presented the detector aperture, reconstruction 
resolution, and data/MC comparisons for a combined data set which includes epoch~2 
and epoch~3 of TALE operation.  This is justified by the fact that the only difference
between the two epochs is the removal, in epoch~3, of the online Cherenkov ``Blasts'' 
filter.  Even when allowed by the removal of the online filter, the vast majority of these
events get filtered out by the offline analysis, starting with the filtering procedure
described in ``Step 2'' of section~\ref{sec:data_process}, and also by later reconstruction
steps and the application of quality cuts to the final reconstructed events set.

We examined data from the two epochs separately and found only minor differences 
at energies below $\sim10^{15.5}$~eV. Since these blasts are mostly showers with 
energies of $\sim10^{15}$~eV or less, we expect that the difference to be due to 
some of these events getting through.  At energies above $\sim10^{15.5}$~eV the 
distributions obtained from the different epochs and the expected reconstruction 
resolutions are similar.

\FloatBarrier

\section{Systematic Uncertainties} 
\label{sec:systematics}

\begin{figure}[htb]
\centering
\includegraphics[height=2.0in]{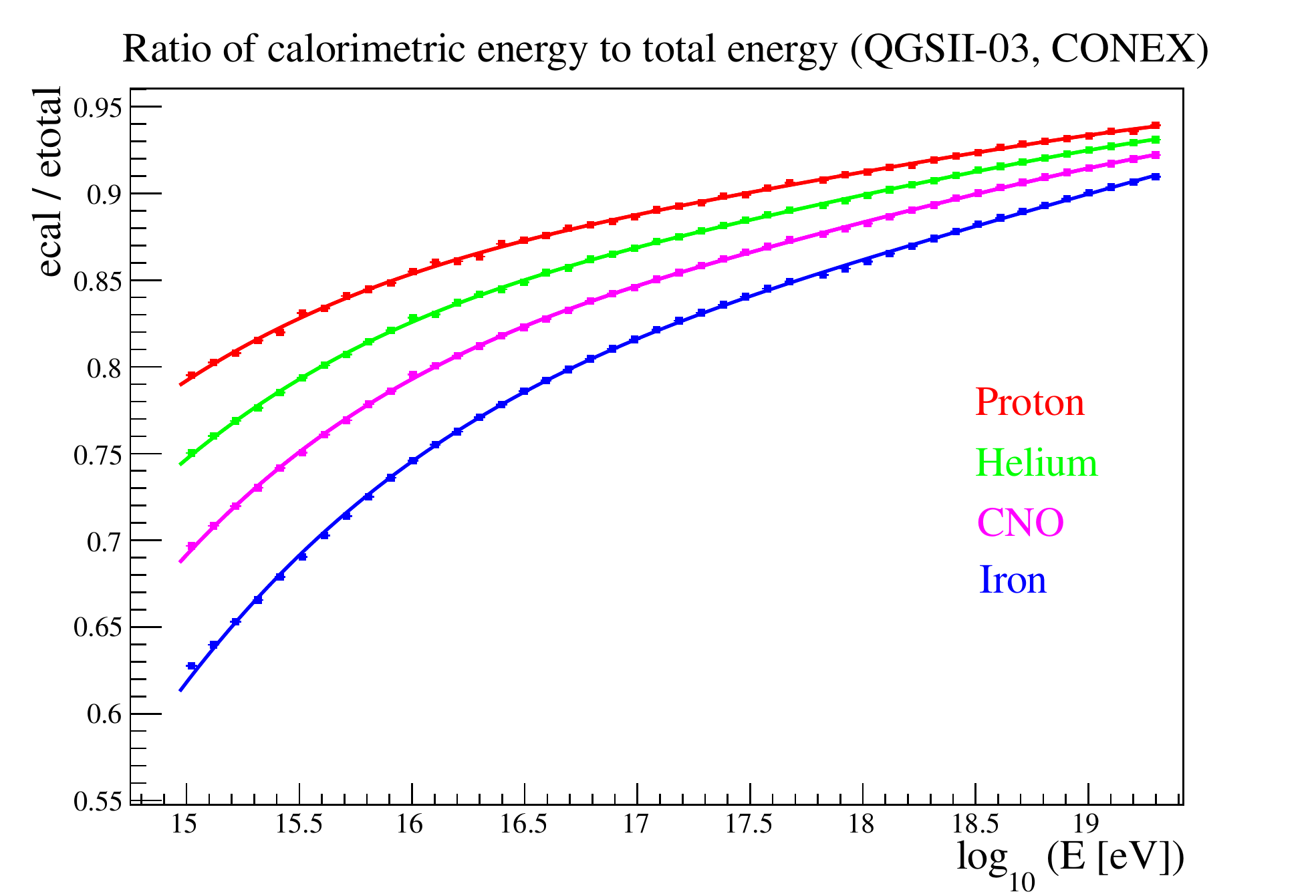}
\caption{Ratio of calorimetric energy to total shower energy as given by 
CONEX simulations.  Simulation sets of mono-energetic showers were used
to calculate the ratio.  Each point in the figure represents a simulation
set and the curves represent a 4-th degree polynomial fit to the point}
\label{fig:tale_ecal_to_etot}
\end{figure}

The main sources of systematic uncertainty in the measured flux 
come from the uncertainty in the energy scale which comes from 
a combination of uncertainties in the
TALE fluorescence detector photonic scale, atmospheric
corrections for scattering by aerosols of fluorescence and Cherenkov light, 
the fluorescence yield, and the correction for energy carried away
(mostly) by neutrinos and muons that does not appear calorimetrically.
The spectrum uncertainty consists of the energy scale uncertainty
squared plus the uncertainty in the aperture, which mostly arises from
the uncertainty in composition.  A brief description of these sources 
follows:

\begin{description}
\item [Detector photonic scale] This includes effects such as PMT gain, 
  UV filter transmission, telescope mirror reflectivity, etc.
  The estimate is based on previous studies by the HiRes 
  collaboration~\cite{Abbasi:2009ix}. As already mentioned, the TALE FD
  uses refurbished HiRes-II telescopes.
\item [Shower missing energy] Estimate is based on two factors: (1) the 
  dependence of the missing energy correction on the primary type,
  and (2) the dependence of the magnitude of the missing energy correction
  on the primary particle's energy.  The magnitude of the correction 
  decreases with increasing energy, and the absolute separation between 
  different primary types decreases as well, as can be seen in 
  figure~\ref{fig:tale_ecal_to_etot}.
\item [Atmosphere] The attenuation of the shower signal by aerosols is 
  modeled using an average aerosol density obtained from previous 
  studies by the HiRes collaboration~\cite{Abbasi:2009ix}.  TALE is located
  at $\sim 100$~km south of the original HiRes detectors and shares similar
  climate and weather patterns.  Aerosol measurements at the TA 
  sites~\cite{Tomida:2011cb} also confirm the aerosol density is similar to 
  that seen at HiRes.  Nightly fluctuations about the average aerosol density 
  have a small effect on the measured spectrum.
\item [Fluorescence yield] Several measurements of the absolute air fluorescence
  yield have been made, as summarized in the following 
  references~\cite{Rosado:2011qi,Cady:2011zz}.  Using an average yield based
  on multiple measurements leads to an estimate of the uncertainty of 6\% on 
  the energy due to the fluorescence yield, as described 
  in~\cite{Abbasi:2009ix} and used in~\cite{Abbasi:2015bha}.  Another 
  approach is to combine the errors reported by the two 
  experiments~\cite{Kakimoto:1995pr,Belz:2005pv}, as was done 
  in~\cite{TheTelescopeArray:2015mgw}, and this leads to an energy 
  uncertainty estimate of 11\%.  We use a value of 10\%..
\item [Composition] The distribution of $X_{max}$ values in the sky above 
  the detector determines the distances of the brightest parts of 
  the shower relative to the detector, and therefore affects how much
  light is collected by the telescope mirror.  This has a direct impact 
  on the detector aperture estimation.
\end{description}

Table~\ref{table:systematics} shows the contributions from the various 
sources, and shows that the systematic uncertainty in the spectrum is 
approximately constant as a function of energy. Events with energy 
$E < 10^{17}$ eV occur so close to the detector that there
is essentially no atmospheric effect.  They are seen almost
totally in Cherenkov light so are independent of the fluorescence
yield.  Events at $10^{18}$ eV occur within about 10 km of the
detector, and are also insensitive to atmospheric corrections.  In
both cases the composition-caused uncertainty in the aperture is
minimal.  For lower energy events we measure $X_{\rm max}$ accurately and
simulate the position of showers in the atmosphere in a
model-independent way.   For higher energy events we do the same, and in
addition the aperture is insensitive to composition.

\begin{table}[htb]
  \centering
\caption{Estimates of systematic uncertainties in the TALE FD energy
  scale and spectrum measurement.  This uncertainty is approximately
  constant as a function of energy}
\label{table:systematics}
\begin{tabular}{|l|c|c|c|}
  \hline
  Energy       &  Source                 &  value  &  contribution to spectrum \\ \hline
  $<10^{17}$ eV &  photonic scale         &  10\%   &  20\%                     \\ \hline
  $<10^{17}$ eV &  missing energy         &  10\%   &  20\%                     \\ \hline
  $<10^{17}$ eV &  atmosphere             &  0      &  0                        \\ \hline
  $<10^{17}$ eV &  fluorescence yield     &  0      &  0                        \\ \hline
  $<10^{17}$ eV &  composition ($X_{max}$) &  3\%    &  6\%        \\ \hline \hline

  $10^{18}$ eV  &  photonic scale         &  10\%   &  20\%                     \\ \hline
  $10^{18}$ eV  &  missing energy         &  5\%    &  10\%                     \\ \hline
  $10^{18}$ eV  &  atmosphere             &  2\%    &  4\%                      \\ \hline
  $10^{18}$ eV  &  fluorescence yield     &  10\%   &  20\%                       \\ \hline
  $10^{18}$ eV  &  composition ($X_{max}$) &  3\%    &  6\%        \\ \hline \hline
  
  $<10^{17}$ eV &  total                  &  14\%   & 29\%                      \\ \hline
  $10^{18}$ eV  &  total                  &  15\%   & 31\%                      \\ \hline
\end{tabular}
\end{table}

\FloatBarrier

\section{Results and Discussion}
\label{sec:results_and_discussion}

\begin{figure}[htb]
\centering
\includegraphics[height=3.2in]{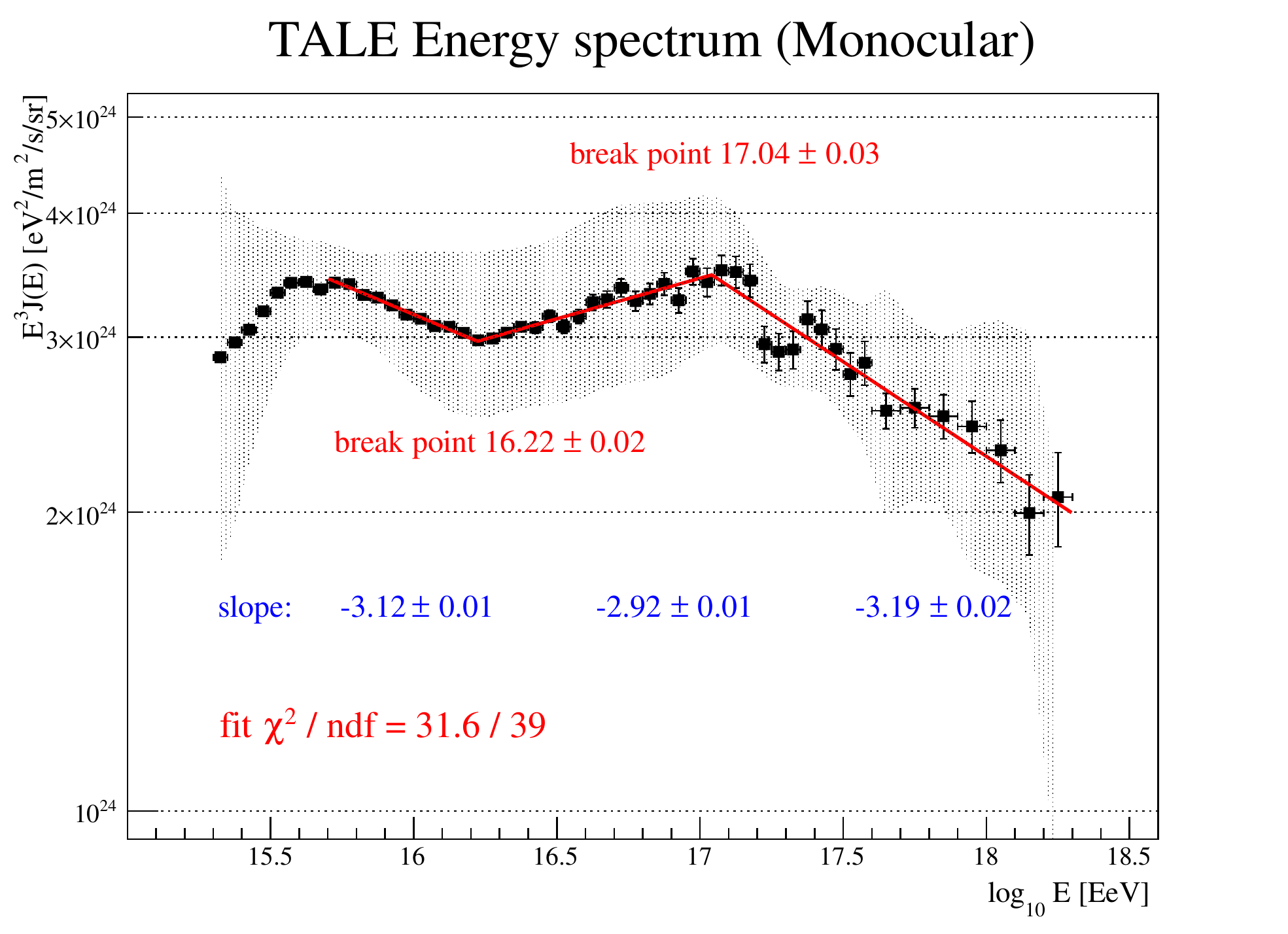}
\caption{TALE cosmic rays energy spectrum measured with 22 months of
  data.  A mixed primary composition given by the TXF is assumed.  The 
  gray band indicates the size of the systematic uncertainties.}
\label{fig:tale_spectrum_systematics}
\end{figure}

The measured spectrum using data from June of 2014 through  
March 2016 is shown in figure~\ref{fig:tale_spectrum_systematics}. 
The systematic errors shown with the gray band in the figure are calculated
by evaluating the effect on the spectrum of a systematic shift in the 
real data events energies of 15\% either up or down, see table~\ref{table:systematics}.

To calculate the systematic error band shown in the figure, reconstructed
real data event energies were shifted up or down by 15\%. 
A broken power-law fit, with two break points is also shown in the figure.
The fit was performed in the range~$10^{15.7}-10^{18.3}$~eV, 
with the locations of the break points being free parameters. 
A summary of the fit results is shown in table~\ref{table:spectrum_fit_results}.

\begin{figure}[htb]
\centering
\includegraphics[height=3.2in]{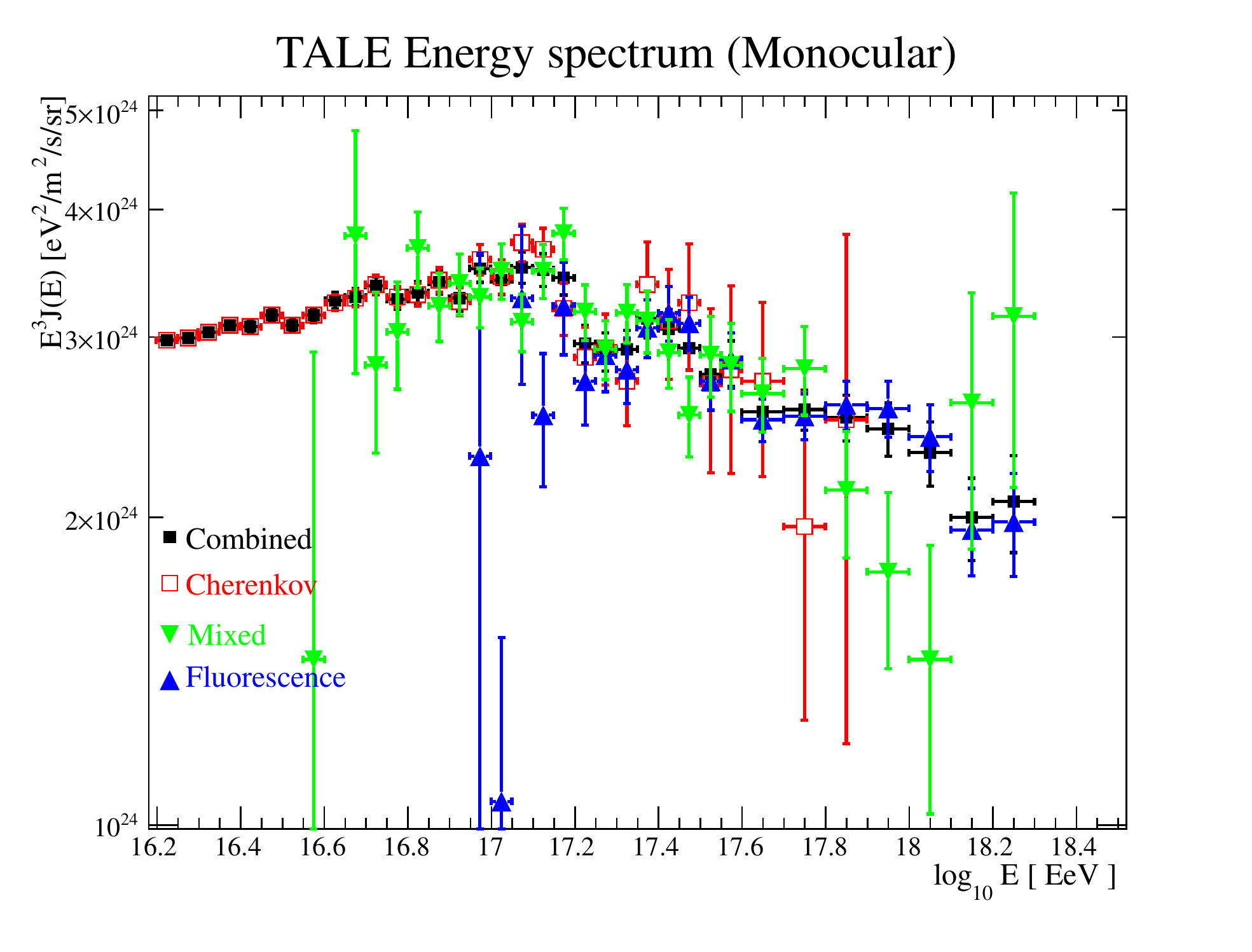}
\caption{TALE cosmic rays energy spectrum measured with 22 months of
  data.  Contributions from Cherenkov, mixed, and fluorescence events
  shown separately.  Note that only the Cherenkov subsets contributes
  to the spectrum below $10^{16.7}$ eV.}
\label{fig:tale_spectrum_all}
\end{figure}

The points in the spectrum shown in figure~\ref{fig:tale_spectrum_systematics}
are a combination of three separate subsets of events; 
each of which can be used to calculate an energy spectrum as shown in 
figure~\ref{fig:tale_spectrum_all}.  A comparison of the spectra measured 
by using the three event categories in the energy range where they overlap 
shows that, within statistics, there is good agreement between Cherenkov 
dominated, fluorescence dominated, and mixed signal events.

\begin{table}[htb]
  \centering
\caption{Fit parameters to a broken power law fit to TALE spectrum.}
\label{table:spectrum_fit_results}
\begin{tabular}{|l|r|}
  \hline

  Break point 1: $\log_{10}(E)$         & 16.22 $\pm$ 0.017 \\
  Break point 2: $\log_{10}(E)$         & 17.04 $\pm$ 0.035 \\ \hline
  Spectral index: $15.70 < \log_{10}(E) < 16.22$  &  3.12 $\pm$ 0.007  \\
  Spectral index: $16.22 < \log_{10}(E) < 17.04$  &  2.92 $\pm$ 0.008  \\
  Spectral index: $17.04 < \log_{10}(E) < 18.30$  &  3.19 $\pm$ 0.017  \\ \hline
\end{tabular}
\end{table}

While not part of the fit, it is easily seen from the spectrum plot that there
is a significant spectral index change at around $\sim10^{15.6}$~eV.  A
comparison of this energy and the second break point in 
table~\ref{table:spectrum_fit_results} shows that they are separated in energy 
by a factor of $\sim25$.  This is somewhat suggestive in that if we interpret 
the low energy break as the ``proton knee'', then the higher energy break occurs 
at where we would expect the ``iron knee'' to be according to the 
idea of rigidity dependent cutoffs in the spectra of individual nuclei as 
first proposed by Peters~\cite{Peters:1961}.

\begin{figure}[htb]
\centering
\includegraphics[height=3.2in]{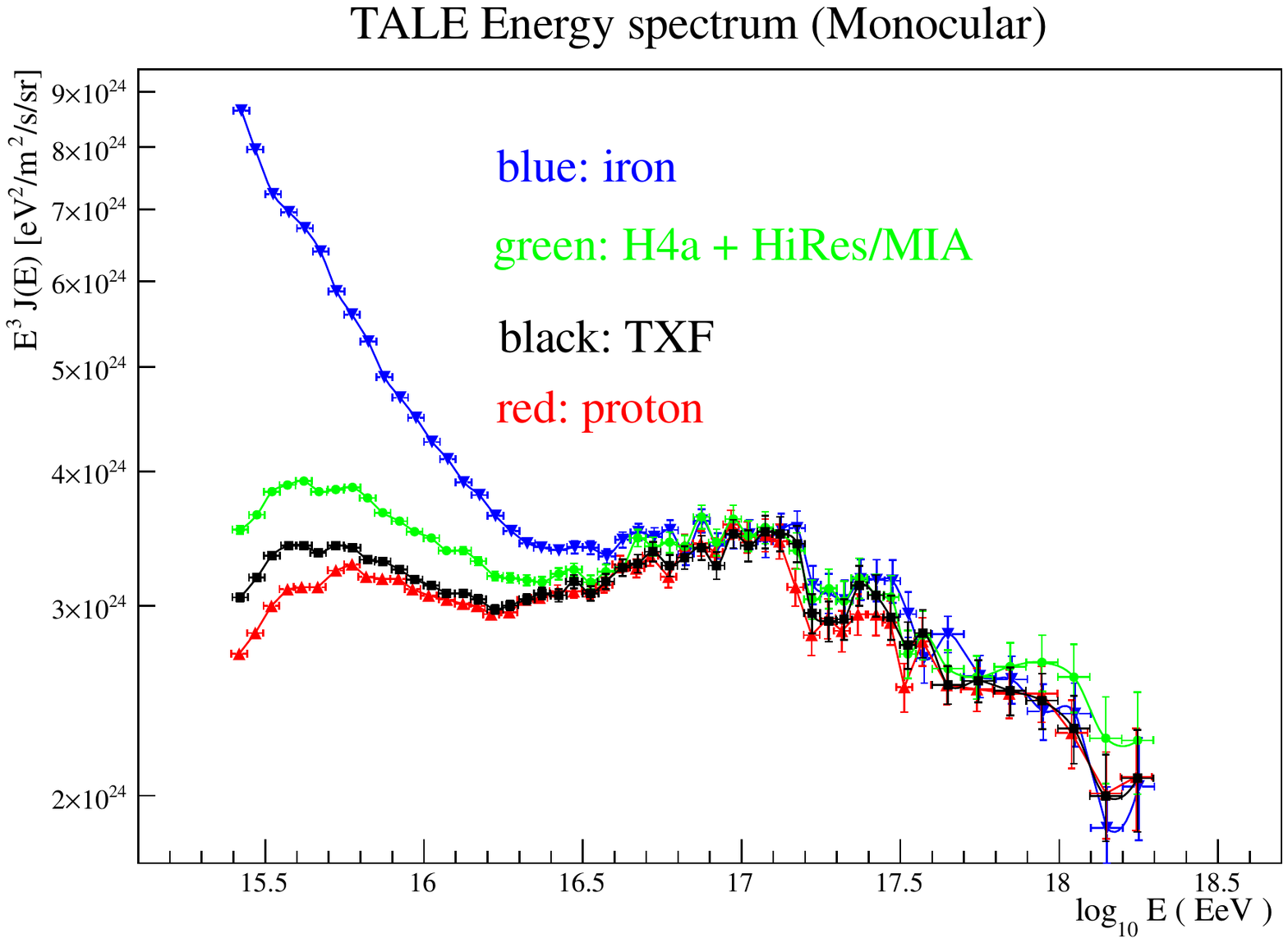}
\caption{TALE cosmic rays energy spectrum composition dependence.  A comparison
  of the spectrum calculation if we assume that cosmic rays are pure protons (red),
  pure iron (blue), follow the H4a composition (green), or the TXF result (black).
  The pure iron case is shown for reference only, at low energies it is excluded
  by previous measurements: e.g.~\cite{Budnev:2009mc, Apel:2013uni}
}
\label{fig:tale_spectrum_composition}
\end{figure}

To examine the composition dependence of the spectrum, and to demonstrate 
the applicability of the aperture calculation using a primary mixture from 
the TXF to the high energy end of the spectrum, we show in 
figure~\ref{fig:tale_spectrum_composition} a comparison of the spectrum
obtained with different compositions.  With respect to the energy spectrum 
for the case of pure iron composition assumption, note that composition 
measurements by other experiments, e.g.~\cite{Budnev:2009mc, Apel:2013uni}
exclude the possibility of iron dominated flux at energies below $10^{16}$~eV. 
The spectrum is included in the plot simply to demonstrate the extreme case 
of all heavy primaries.

\begin{figure}[htb]
\centering
\includegraphics[height=3.2in]{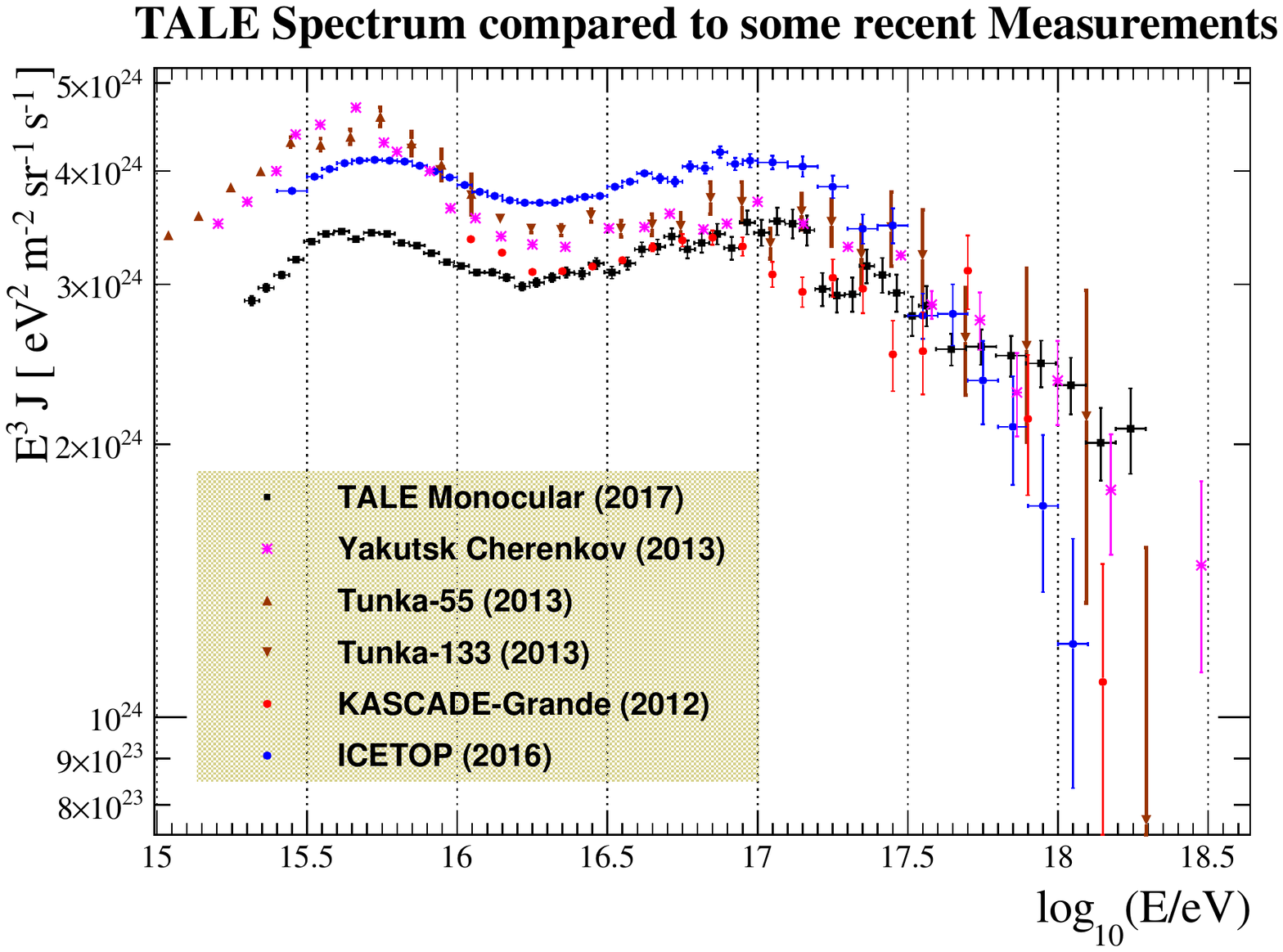}
\caption{TALE cosmic rays energy spectrum plotted along with measurements
  by Yakutsk~\cite{Knurenko:2013dia},
  TUNKA~\cite{Budnev:2013noa,Prosin:2014dxa},
  Kaskade-Grande~\cite{Apel:2012tda}, and
  IceTop~\cite{Rawlins:2016bkc}}
\label{fig:tale_spectrum_plus}
\end{figure}

Figure~\ref{fig:tale_spectrum_plus} compares the current result with some recent 
results from other experiments.  We note that qualitatively the spectra are in 
agreement.  The difference in normalization is within the systematics of the 
energy scales of the different experiments.  In particular, we note that a 6.5\%
downward shift in the IceTop energy scale, results in a spectrum that lies on
top of the TALE spectrum for energies below $10^{17}$~eV.

\begin{figure}[htb]
\centering
\includegraphics[height=3.2in]{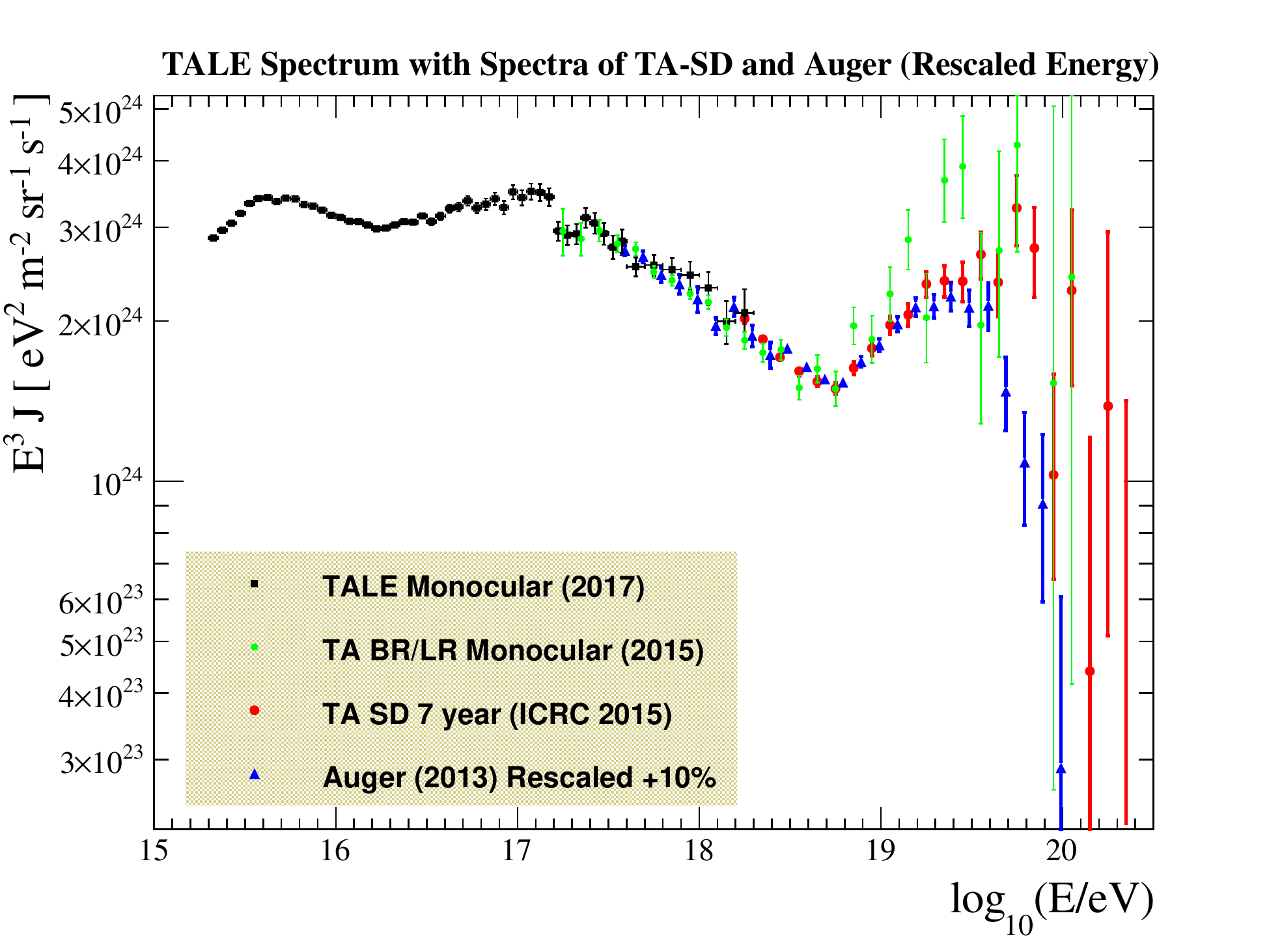}
\caption{TALE cosmic rays energy spectrum plotted along with measurements 
  by TA using the FD's at Black Rock and Long Ridge sites~\cite{TheTelescopeArray:2015mgw}, 
  and by the TA surface detector~\cite{Jui:2016amg}, also shown is the Auger spectrum~\cite{AugerSpectrum:2013}
  with a 10\% energy scaling applied to make it agree with the TA SD flux.}
\label{fig:tale_spectrum_plus_hi_energy}
\end{figure}

Figure~\ref{fig:tale_spectrum_plus_hi_energy} compares the current result with 
some recent results from TA Fluorescence~\cite{TheTelescopeArray:2015mgw} and surface 
detector~\cite{Jui:2016amg} measurements.  We note that above $10^{17}$~eV 
there is excellent agreement between the different results, demonstrating that 
the TALE spectrum can be seen as an extension of the measurements in the ultra-high 
energy regime down to lower energies.

\FloatBarrier

\section{Summary}
\label{sec:summary}

The Telescope Array Low Energy extension was constructed 
to extend TA's study of the spectrum and composition of
cosmic rays down to the $10^{16.5}-10^{18.5}$~eV regime. 
This is the energy range in which the transition from cosmic 
rays of galactic origin to those of extragalactic origin is 
thought to occur.  Although several experiments have already 
seen hints of a ``second knee'' structure in the $10^{17}$ decade, 
they have different energy scales and flux normalizations, so that
the actual energy of the feature was unknown.  TALE overlaps 
with other TA spectra, so that TALE measurements 
will share a single energy scale with TA.

TALE consists of 10 high-elevation fluorescence telescopes
and an in-fill array of 103 surface detectors.  The TALE FD has been
taking monocular data since 2013.  We analyzed 1080 hours of 
TALE monocular FD data taken between June 2014 and March 2016.
Events were reconstructed using a profile-constrained geometry fit
that reconstructs the spatial trajectory and the longitudinal
development of the extensive air shower in a single step.
We demonstrated that this technique gives energy resolutions
sufficient for spectrum measurement. In particular, we obtained
a $\sim 15\%$ energy reconstruction resolution of Cherenkov-dominated
showers pointing towards the detector. These events were
previously rejected for spectrum measurements.  Their inclusion
allowed us to extend the lower threshold of TALE FD by more than
another order of magnitude to $10^{15.3}$~eV.

Simulation based on CONEX, using the QGSJET-II.3 hadronic model
are used to calculate the aperture and to study the detector
resolutions. Monte Carlo vs. data comparisons showed good
agreement in the distributions of impact parameter($R_p$),
zenith angle and azimuthal angle ($\phi$). These quantities
are directly tied to the aperture calculation. 
However, the shower maximum depth ($X_{\rm max}$) distribution in the MC,
was adjusted to agree to the data. This last process removed
any dependence on the composition assumption related to the
dependence of $X_{\rm max}$ to the underlying hadronic model.

We present the resulting spectrum spanning $10^{15.3}-10^{18.3}$~eV.
The systematic error on the flux is estimated to be about 30\%, dominated
by the overall uncertainty in energy scale of about 14\%. The spectrum
shown in Figure~\ref{fig:tale_spectrum_systematics} shows a clear second 
knee near $10^{17.1}$~eV, along with an ankle-like structure 
at $10^{16.2}$~eV.  At the lowest reach of our measurement, we also see 
the ``first knee'' at $\sim10^{15.6}$~eV.
The ratio of the energies of the two knees is tantalizingly close
to the factor of 26 in the charge ratio of iron to protons.  This
suggests that the two ``knee'' structures
represent the ends of the galactic iron 
(second knee) and proton/helium (first knee) fluxes.

\section*{Acknowledgements}

The Telescope Array experiment is supported by the Japan Society for
the Promotion of Science(JSPS) through 
Grants-in-Aid
for Specially Promoted Research 
JP21000002, 
for Scientific  Research (S) 
JP19104006, 
for Specially Promoted Research 
JP15H05693, 
for Scientific  Research (S)
JP15H05741 and
for Young Scientists (A)
JPH26707011; 
by the joint research program of the Institute for Cosmic Ray Research (ICRR), The University of Tokyo; 
by the U.S. National Science
Foundation awards PHY-0601915,
PHY-1404495, PHY-1404502, and PHY-1607727; 
by the National Research Foundation of Korea
(2015R1A2A1A01006870, 2015R1A2A1A15055344, 2016R1A5A1013277,
2007-0093860, 2016R1A2B4014967); by the Russian Academy of
Sciences, RFBR grant 16-02-00962a (INR), IISN project No. 4.4502.13,
and Belgian Science Policy under IUAP VII/37 (ULB). The foundations of
Dr. Ezekiel R. and Edna Wattis Dumke, Willard L. Eccles, and George
S. and Dolores Dor\'e Eccles all helped with generous donations. The
State of Utah supported the project through its Economic Development
Board, and the University of Utah through the Office of the Vice
President for Research. The experimental site became available through
the cooperation of the Utah School and Institutional Trust Lands
Administration (SITLA), U.S. Bureau of Land Management (BLM), and the
U.S. Air Force. We appreciate the assistance of the State of Utah and
Fillmore offices of the BLM in crafting the Plan of Development for
the site.  Patrick Shea assisted the collaboration with valuable advice 
on a variety of topics. The people and the officials of Millard County, 
Utah have been a source of
steadfast and warm support for our work which we greatly appreciate. 
We are indebted to the Millard County Road Department for their efforts 
to maintain and clear the roads which get us to our sites. 
We gratefully acknowledge the contribution from the technical staffs of
our home institutions. An allocation of computer time from the Center
for High Performance Computing at the University of Utah is gratefully
acknowledged.


\section*{Appendix A}
Spectrum data listed in table \ref{table_spectrum}
\begin{table}[htb]
\centering
\caption{Spectrum data}.
\label{table_spectrum}
\begin{tabular}{|l|l|l|}
\hline
energy-bin  & Num. Events &   $E^{3}j(E) \pm \sigma_{stat.} \pm \sigma_{sys.}$  \\
15.30-15.35 &     7207 &  2.864 $\pm$ 0.034 + 1.585 - 1.132 \\ 
15.35-15.40 &    10663 &  2.965 $\pm$ 0.029 + 1.073 - 1.056 \\ 
15.40-15.45 &    13829 &  3.052 $\pm$ 0.026 + 0.897 - 0.811 \\ 
15.45-15.50 &    17040 &  3.187 $\pm$ 0.024 + 0.705 - 0.724 \\ 
15.50-15.55 &    19821 &  3.327 $\pm$ 0.024 + 0.485 - 0.603 \\ 
15.55-15.60 &    21628 &  3.398 $\pm$ 0.023 + 0.341 - 0.432 \\ 
15.60-15.65 &    22664 &  3.411 $\pm$ 0.023 + 0.311 - 0.311 \\ 
15.65-15.70 &    22553 &  3.352 $\pm$ 0.022 + 0.463 - 0.463 \\ 
15.70-15.75 &    24448 &  3.405 $\pm$ 0.022 + 0.413 - 0.413 \\ 
15.75-15.80 &    24241 &  3.393 $\pm$ 0.022 + 0.200 - 0.277 \\ 
15.80-15.85 &    23375 &  3.310 $\pm$ 0.022 + 0.294 - 0.301 \\ 
15.85-15.90 &    22309 &  3.288 $\pm$ 0.022 + 0.329 - 0.400 \\ 
15.90-15.95 &    20576 &  3.231 $\pm$ 0.023 + 0.447 - 0.456 \\ 
15.95-16.00 &    18558 &  3.161 $\pm$ 0.023 + 0.519 - 0.454 \\ 
16.00-16.05 &    16685 &  3.132 $\pm$ 0.024 + 0.526 - 0.482 \\ 
16.05-16.10 &    14820 &  3.079 $\pm$ 0.025 + 0.561 - 0.471 \\ 
16.10-16.15 &    13181 &  3.074 $\pm$ 0.027 + 0.555 - 0.536 \\ 
16.15-16.20 &    11428 &  3.030 $\pm$ 0.028 + 0.640 - 0.558 \\ 
16.20-16.25 &     9788 &  2.978 $\pm$ 0.030 + 0.673 - 0.526 \\ 
16.25-16.30 &     8462 &  2.992 $\pm$ 0.033 + 0.642 - 0.513 \\ 
16.30-16.35 &     7363 &  3.032 $\pm$ 0.035 + 0.627 - 0.510 \\ 
16.35-16.40 &     6432 &  3.074 $\pm$ 0.038 + 0.621 - 0.536 \\ 
16.40-16.45 &     5576 &  3.068 $\pm$ 0.041 + 0.648 - 0.478 \\ 
16.45-16.50 &     4942 &  3.150 $\pm$ 0.045 + 0.582 - 0.620 \\ 
16.50-16.55 &     4168 &  3.075 $\pm$ 0.048 + 0.753 - 0.434 \\ 
16.55-16.60 &     3759 &  3.149 $\pm$ 0.051 + 0.517 - 0.733 \\ 
16.60-16.65 &     3028 &  3.254 $\pm$ 0.059 + 0.987 - 0.505 \\ 
16.65-16.70 &     2686 &  3.275 $\pm$ 0.063 + 0.602 - 0.565 \\ 
16.70-16.75 &     2334 &  3.365 $\pm$ 0.070 + 0.701 - 0.737 \\ 
16.75-16.80 &     1914 &  3.262 $\pm$ 0.075 + 0.915 - 0.568 \\ 
16.80-16.85 &     1660 &  3.317 $\pm$ 0.081 + 0.699 - 0.546 \\ 
16.85-16.90 &     1456 &  3.395 $\pm$ 0.089 + 0.669 - 0.737 \\ 
16.90-16.95 &     1197 &  3.269 $\pm$ 0.094 + 0.906 - 0.452 \\ 
16.95-17.00 &     1083 &  3.495 $\pm$ 0.106 + 0.561 - 0.686 \\ 
17.00-17.05 &      914 &  3.409 $\pm$ 0.113 + 0.832 - 0.482 \\ 
17.05-17.10 &      824 &  3.504 $\pm$ 0.122 + 0.577 - 0.511 \\ 
17.10-17.15 &      739 &  3.491 $\pm$ 0.128 + 0.596 - 0.445 \\ 
17.15-17.20 &      677 &  3.422 $\pm$ 0.132 + 0.500 - 0.683 \\ 
17.20-17.25 &      569 &  2.952 $\pm$ 0.124 + 0.736 - 0.264 \\ 
17.25-17.30 &      544 &  2.901 $\pm$ 0.124 + 0.285 - 0.240 \\ 
17.30-17.35 &      524 &  2.916 $\pm$ 0.127 + 0.263 - 0.263 \\ 
17.35-17.40 &      540 &  3.128 $\pm$ 0.135 + 0.381 - 0.381 \\ 
17.40-17.45 &      498 &  3.055 $\pm$ 0.137 + 0.423 - 0.438 \\ 
17.45-17.50 &      448 &  2.921 $\pm$ 0.138 + 0.488 - 0.437 \\ 
17.50-17.55 &      400 &  2.755 $\pm$ 0.138 + 0.485 - 0.236 \\ 
17.55-17.60 &      384 &  2.828 $\pm$ 0.144 + 0.265 - 0.387 \\ 
17.60-17.70 &      614 &  2.530 $\pm$ 0.103 + 0.865 - 0.752 \\ 
17.70-17.80 &      490 &  2.548 $\pm$ 0.115 + 0.611 - 0.518 \\ 
17.80-17.90 &      380 &  2.498 $\pm$ 0.128 + 0.449 - 0.586 \\ 
17.90-18.00 &      282 &  2.440 $\pm$ 0.146 + 0.710 - 0.659 \\ 
18.00-18.10 &      191 &  2.308 $\pm$ 0.167 + 0.824 - 0.646 \\ 
18.10-18.20 &      117 &  1.996 $\pm$ 0.185 + 0.794 - 0.309 \\ 
18.20-18.30 &       85 &  2.071 $\pm$ 0.225 + 0.832 - 1.067 \\ \hline

\end{tabular}
\end{table}


\clearpage
\begin{thebibliography}{99}

\bibitem{Hill:1983mk}
  C.~T.~Hill and D.~N.~Schramm,
  \emph{The Ultrahigh-Energy Cosmic Ray Spectrum},
  \emph{Phys. Rev. D} {\bf 31} (1985) 564.
  
\bibitem{Berezinsky:1988wi}
  V.~S.~Berezinsky and S.~I.~Grigor'eva,
  \emph{A Bump in the ultrahigh-energy cosmic ray spectrum},
  \emph{Astron. Astrophys.}  {\bf 199} (1988) 1.

\bibitem{Berezinsky:2005qe} 
  V.~Berezinsky,
  \emph{Dip in uhecr and transition from galactic to extragalactic cosmic rays},
  \emph{astro-ph/0509069}.

\bibitem{Abbasi:2004nz} 
  R.~U.~Abbasi {\it et al.} [HiRes Collaboration],
  \emph{A Study of the composition of ultrahigh energy cosmic rays using the High Resolution Fly's Eye},
  \emph{Astrophys. J.}  {\bf 622}, 910 (2005)


\bibitem{Abbasi:2014sfa} 
  R.~U.~Abbasi {\it et al.},
  \emph{Study of Ultra-High Energy Cosmic Ray composition using Telescope Array’s Middle Drum detector and surface array in hybrid mode},
  \emph{Astropart. Phys.}  {\bf 64}, 49 (2015)

\bibitem{Aloisio:2013hya} 
  R.~Aloisio, V.~Berezinsky and P.~Blasi,
  \emph{Ultra high energy cosmic rays: implications of Auger data for source spectra and chemical composition},
  \emph{JCAP} {\bf 1410}, no. 10, 020 (2014)

\bibitem{Bird:1994wp} 
  D.~J.~Bird {\it et al.} [HiRes Collaboration], 
  \emph{The Cosmic ray energy spectrum observed by the Fly's Eye}, 
  \emph{Astrophys. J.}  {\bf 424}, 491 (1994).

\bibitem{AbuZayyad:2000ay} 
  Abu-Zayyad, T. {\it et al.}  [HiRes-MIA Collaboration],
  \emph{Measurement of the cosmic ray energy spectrum and composition from 10$^{17}$ eV to 10$^{18.3}$ eV using a hybrid fluorescence technique}, 
  \emph{Astrophys. J.}  {\bf 557}, 686-699 (2001)

\bibitem{Nagano:1991jz} 
  M.~Nagano {\it et al.},
  \emph{Energy spectrum of primary cosmic rays above 10**17-eV determined from the extensive air shower experiment at Akeno},
  \emph{J. Phys. G} {\bf 18}, 423 (1992).

\bibitem{Egorova:2004cm} 
  V.~P.~Egorova {\it et al.},
  \emph{The Spectrum features of UHECRs below and surrounding GZK},
  \emph{Nucl. Phys. Proc. Suppl.}  {\bf 136}, 3 (2004)


\bibitem{tasd-nim-a}
  T. Nonaka {\em et al.},
  \emph{The Surface Detector Array of the Telescope Array Experiment},
  \emph{Nucl.Instrum.Meth.} {\bf A689}, 87-97 (2012)

\bibitem{AbuZayyad:2012qk}
  Abu-Zayyad, T. and Aida, R. and Allen, M. and Anderson, R. and Azuma, R. {\em et al.},
  \emph{The Energy Spectrum of Telescope Array's Middle Drum Detector and the Direct Comparison to the High Resolution Fly's Eye Experiment},
  \emph{Astropart.Phys.} {\bf 39-40}, 109-119 (2012)

\bibitem{tafd-nim-a}
  Tokuno, H. and Tameda, Y. and Takeda, M. and Kadota, K. and Ikeda, D. {\em et al.}, 
  \emph{New air fluorescence detectors employed in the Telescope Array experiment},
  \emph{Nucl.Instrum.Meth.} {\bf A676}, 54-65 (2012)

\bibitem{Teshima:1985vs}
  Teshima, M. and Ohoka, H. and Matsubara, Y. and Hara, T. and Hatano, Y. {\em et al.}, 
  \emph{Expanded Array for Giant Air Shower Observation at Akeno}, 
  \emph{Nucl.Instrum.Meth.} {\bf A247}, 399 (1986)

\bibitem{Sokolsky:2011zz}
  Sokolsky, P., 
  \emph{Final Results from the High Resolution Fly's Eye (HiRes) Experiment}, 
  \emph{Nucl.Phys.Proc.Suppl.} {\bf 212-213}, 74-78 (2011) 

\bibitem{AbuZayyad:2012ru}
  T.~Abu-Zayyad {\it et al.} [Telescope Array Collaboration],
  \emph{The Cosmic Ray Energy Spectrum Observed with the Surface Detector of the Telescope Array Experiment},
  \emph{Astrophys. J.}  {\bf 768} (2013) L1

\bibitem{Abbasi:2014lda} 
  R.~U.~Abbasi {\it et al.} [Telescope Array Collaboration],
  \emph{Indications of Intermediate-Scale Anisotropy of Cosmic Rays with Energy Greater Than 57 EeV in the Northern Sky Measured with the Surface Detector of the Telescope Array Experiment},
  \emph{Astrophys. J.}  {\bf 790}, L21 (2014)


\bibitem{tale_icrc2011}
  G.B. Thomson {\em et al.}, 
  \emph{The Telescope Array Low Energy Extension ({TALE})}, 
  in proceedings of \emph{International Cosmic Ray Conference} {\bf 3}, 337-339 (2011)

\bibitem{Boyer:2002zz}
  Boyer, J. H. and Knapp, B. C. and Mannel, E. J. and Seman, M., 
  \emph{FADC-based DAQ for HiRes Fly's Eye}, 
  \emph{Nucl.Instrum.Meth.} {\bf A482}, 457-474 (2002)


\bibitem{Zundel_thesis_ch4}
  Zundel, Z. J., 
  \emph{Spectrum Measurement with the Telescope Array Low Energy Extension (TALE) Fluorescence Detector}, 
  \emph{ Ph.D. Thesis, University of Utah}, (2016), Chapter 4.

\bibitem{hires_gpsy}
  J. D. Smith {\em et al.},
  \emph{Absolute GPS time event generation and capture for remote locations},
  in proceedings of \emph{International Cosmic Ray Conference}{\bf}, 825-827 (2001)

\bibitem{jNICHE}
  D. Bergman, Y. Tsunesada, J.F. Krizmanic, Y. Omura
  \emph{jNICHE: Prototype detectors of a non-imaging Cherenkov array at the TA site},
  PoS ICRC {\bf 2017}, 415 (2017)
  in proceedings of \emph{International Cosmic Ray Conference}{\bf}, 825-827 (2001)

\bibitem{Abuzayyad:2015fvm}
  Abu-Zayyad, T. {\it et al.}  [Telescope Array Collaboration],
  \emph{Cosmic Rays Energy Spectrum observed by the TALE detector using Cerenkov light}, 
  PoS ICRC {\bf 2015}, 422 (2016)

\bibitem{Baltrusaitis:1985mx} 
  R.~M.~Baltrusaitis {\it et al.},
  \emph{The Utah Fly's Eye Detector},
  \emph{Nucl. Instrum. Meth. A} {\bf 240}, 410 (1985)

\bibitem{AbuZayyad_thesis}
  AbuZayyad, T. {\em The Energy Spectrum of Ultra High Energy Cosmic Rays}, 
  PhD thesis, University of Utah (2000)

\bibitem{Abbasi:2007sv}
  Abbasi, R.U. {\em et al.}, 
  \emph{First Observation of the Greisen-Zatsepin-Kuzmin suppression}, 
  \emph{Phys.Rev.Lett.} {\bf 100}, 101101 (2008)



\bibitem{Bergmann:2006yz} 
  Bergmann, T.  {\it et al.}, 
  \emph{One-dimensional Hybrid Approach to Extensive Air Shower Simulation}, 
  \emph{Astropart.Phys.} {\bf 26}, 420-432 (2007)

\bibitem{Heck:1998vt}
  Heck, D. {\it et al.}, 
  \emph{CORSIKA: A Monte Carlo code to simulate extensive air showers}, 
  \emph{FZKA-6019} (1998)

\bibitem{Bernlohr:2008kv}
  Bernlohr, Konrad, 
  \emph{Simulation of Imaging Atmospheric Cherenkov Telescopes with CORSIKA and sim\_telarray}, 
  \emph{Astropart.Phys.} {\bf 30}, 149-158 (2008)

\bibitem{Lafebre:2009en}
  Lafebre, S. {\em et al.},
  \emph{Universality of electron-positron distributions in extensive air showers},
  \emph{Astropart.Phys.} {\bf 31} 243-254 (2009)

\bibitem{NKG_G}
  Greisen, K.,
  \emph{},
  \emph{Prog. Cosm. Ray Phys.} {\bf 3} 1 (1956)

\bibitem{NKG_NK}
  Kamata, K., and Nishimura, J.,
  \emph{},
  \emph{Prog. Theor. Phys. Suppl.} {\bf 6} 93 (1958)

\bibitem{Kakimoto:1995pr} 
  F.~Kakimoto, E.~C.~Loh, M.~Nagano, H.~Okuno, M.~Teshima and S.~Ueno,
  \emph{A Measurement of the air fluorescence yield},
  \emph{Nucl. Instrum. Meth. A} {\bf 372}, 527 (1996).

\bibitem{Belz:2005pv} 
  J.~W.~Belz {\it et al.} [FLASH Collaboration],
  \emph{Measurement of pressure dependent fluorescence yield of air: Calibration factor for UHECR detectors},
  \emph{Astropart. Phys.}  {\bf 25}, 129 (2006)

\bibitem{Nerling:2005fj} 
  F.~Nerling, J.~Bluemer, R.~Engel and M.~Risse,
  \emph{Universality of electron distributions in high-energy air showers: Description of Cherenkov light production},
  \emph{Astropart. Phys.} {\bf 24}, 421 (2006)

\bibitem{GDAS}
  \url{http://ready.arl.noaa.gov/gdas1.php}

\bibitem{Gaisser:2012cc}
  Gaisser, T.K., 
  \emph{Spectrum of cosmic-ray nucleons, kaon production, and the atmospheric muon charge ratio}, 
  \emph{Astropart.Phys.} {\bf 35} 801-806 (2012)

\bibitem{TFractionFitter}
  \url{https://root.cern/doc/master/classTFractionFitter.html}

\bibitem{Barlow:1993dm} 
  R.~J.~Barlow and C.~Beeston,
  \emph{Fitting using finite Monte Carlo samples},
  \emph{Comput. Phys. Commun.}  {\bf 77}, 219 (1993).


\bibitem{Fujii:2017} 
  Toshihiro Fujii {\it et al.}
  \emph{A systematic uncertainty on the energy scale of the
    Telescope Array fluorescence detectors},
  PoS ICRC {\bf 2017}, 524 (2017).


\bibitem{Abbasi:2002ta} 
  R.~U.~Abbasi {\it et al.} [HiRes Collaboration],
  \emph{Measurement of the flux of ultrahigh energy cosmic rays from monocular observations by the High Resolution Fly's Eye experiment},
  \emph{Phys. Rev. Lett.}  {\bf 92}, 151101 (2004)

\bibitem{Abbasi:2009ix} 
  R.~U.~Abbasi {\it et al.} [HiRes Collaboration],
  \emph{Measurement of the Flux of Ultra High Energy Cosmic Rays by the Stereo Technique},
  \emph{Astropart. Phys.}  {\bf 32}, 53 (2009)

\bibitem{Tomida:2011cb} 
  T.~Tomida {\it et al.},
  \emph{The atmospheric transparency measured with a LIDAR system at the Telescope Array experiment},
  \emph{Nucl. Instrum. Meth. A} {\bf 654}, 653 (2011)
  doi:10.1016/j.nima.2011.07.012
  [arXiv:1109.1196 [astro-ph.IM]].

\bibitem{Rosado:2011qi} 
  J.~Rosado, F.~Blanco and F.~Arqueros,
  \emph{Comparison of available measurements of the absolute air-fluorescence yield and determination of its global average value},
  \emph{AIP Conf. Proc.}  {\bf 1367}, 34 (2011)
  doi:10.1063/1.3628711
  [arXiv:1104.0894 [astro-ph.IM]].

\bibitem{Cady:2011zz} 
  R.~Cady,
  \emph{A world average of fluorescence yield measurements},
  \emph{AIP Conf. Proc.}  {\bf 1367}, 40 (2011).


\bibitem{Abbasi:2015bha} 
  R.~U.~Abbasi {\it et al.},
  \emph{The hybrid energy spectrum of Telescope Array’s Middle Drum Detector and surface array},
  \emph{Astropart. Phys.}  {\bf 68}, 27 (2015).

\bibitem{Peters:1961}
  B. Peters,
  \emph{Primary cosmic radiation and extensive air showers},
  \emph{Il Nuovo Cimento} {\bf 22} 800-819 (1961)

\bibitem{Budnev:2009mc}
  N.~M.~Budnev {\it et al.},
  \emph{The Cosmic Ray Mass Composition in the Energy Range $10^{15}$ - $10^{18}$~eV 
    measured with the Tunka Array: Results and Perspectives},
  \emph{Nucl. Phys. Proc. Suppl.}  {\bf 190} (2009) 247

\bibitem{Apel:2013uni} 
  W.~D.~Apel {\it et al.},
  \emph{KASCADE-Grande measurements of energy spectra for elemental groups of cosmic rays},
  \emph{Astropart. Phys.}  {\bf 47}, 54 (2013)

\bibitem{Knurenko:2013dia} 
  S.~P.~Knurenko {\it et al.},
  \emph{Cosmic ray spectrum in the energy range 1.0E15-1.0E18 eV and the second knee according to the small Cherenkov setup at the Yakutsk EAS array},
  \emph{arXiv:1310.1978 [astro-ph.HE].}

\bibitem{Budnev:2013noa} 
  N.~Budnev {\it et al.},
  \emph{Tunka-25 Air Shower Cherenkov array: The main results},
  \emph{Astropart. Phys.}  {\bf 50-52}, 18 (2013).

\bibitem{Prosin:2014dxa} 
  V.~V.~Prosin {\it et al.},
  \emph{Tunka-133: Results of 3 year operation},
  \emph{Nucl. Instrum. Meth. A} {\bf 756}, 94 (2014).

\bibitem{Apel:2012tda}
  W.~D.~Apel {\it et al.},
  \emph{The spectrum of high-energy cosmic rays measured with KASCADE-Grande},
  \emph{Astropart. Phys.}  {\bf 36} (2012) 183.

\bibitem{Rawlins:2016bkc} 
  K.~Rawlins [IceCube Collaboration],
  \emph{Cosmic ray spectrum and composition from three years of IceTop and IceCube},
  \emph{J. Phys. Conf. Ser.}  {\bf 718}, no. 5, 052033 (2016).


\bibitem{TheTelescopeArray:2015mgw} 
  R.~U.~Abbasi {\it et al.} [Telescope Array Collaboration],
  \emph{The energy spectrum of cosmic rays above 10$^{17.2}$ eV measured by the fluorescence detectors of the Telescope Array experiment in seven years},
  \emph{Astropart. Phys.}  {\bf 80}, 131 (2016)

\bibitem{Jui:2016amg} 
  C.~Jui,
  \emph{Summary of Results from the telescope Array Experiment},
  \emph{PoS ICRC} {\bf 2015}, 035 (2016).

\bibitem{AugerSpectrum:2013} 
  A.~Schulz [Pierre Auger Collaboration],
  \emph{The measurement of the energy spectrum of cosmic rays above 3x10$^{17}$ eV with the Pierre Auger Observatory},
  \emph{PoS ICRC} {\bf 2013}, 0769 (2013).

\end{thebibliography}
\end{document}